\documentclass{aa}

\newcommand{\Bnabla}{\boldsymbol{\nabla}}
\newcommand{\rr}{\boldsymbol{r}}

\def\EM{E_{\rm M}}

\def\MATINS{\texttt{MATINS}\xspace}
\def\CCSN{\texttt{CCSN}\xspace}
\usepackage{color}
\usepackage{amsmath}
\usepackage{graphicx}
\usepackage{txfonts}
\graphicspath{{./fig/}{./png/}}
\usepackage[colorlinks=true,linkcolor=blue,citecolor=blue,urlcolor=blue]{hyperref}
\begin{document} 

\title{Spontaneous wandering of the magnetic axis in pulsars: 3D magneto-thermal simulations and the imprint on braking indices}

\author{Clara~Dehman\inst{1}
\and Daniele~Vigan\`o\inst{2,3}}

\institute{Departament de Física, Universitat d'Alacant, 03690 Alicante, Spain
\and Institut de Ci\`encies de l'Espai (ICE-CSIC), Campus UAB, Carrer de Can Magrans s/n, 08193 Cerdanyola del Vall\`es, Barcelona, Catalonia, Spain
\and Institut d'Estudis Espacials de Catalunya (IEEC), 08860 Castelldefels, Barcelona, Catalonia, Spain}

\date{Received; accepted}

\abstract
{The braking index measurements of hundreds of pulsars, despite the observational caveats, show a clear trend: $n<3$ at early ages, $n>3$ at middle ages, and increasingly large, fluctuating values at larger characteristic ages. Crustal magnetic fields in isolated neutron stars evolve through Ohmic dissipation and non-linear Hall redistribution across scales, with additional early-time contributions from the chiral instability and from post-burial re-emergence. Initial conditions are usually assumed dipole-dominated, whereas more complex and realistic fields, with energy spread more evenly over a broad range of scales, have not been explored in the context of the braking index.}
{We assess how the internal dynamics of initially non-trivial crustal fields change the dipolar field strength and obliquity, and compare their combined contribution to the braking index $n$, together with that of the expected alignment, with the observed values.}
{We perform long-term 3D magneto-thermal simulations, for initial configurations ranging from the standard dipole-dominated case to tangled, small-scale-dominated fields.}
{Small-scale dominated configurations with relatively weak dipolar components tend to show rich dynamics: the dipolar mode is continuously fed by the small scales and may grow or decay depending on the initial spectrum, driving $n<3$ or $n>3$, while the obliquity varies on timescales as short as ${\cal O}$(kyr). These spontaneous changes often dominate the alignment torque, and can account for the large, sign-alternating braking indices observed at characteristic ages $\tau_c\gtrsim10^{5}$~yr, unlike the dipole-dominated cases.}
{Small-scale initial conditions can reproduce the large values of $n-3$, of either sign, observed at late ages. However, short-term magnetospheric fluctuations or superfluid-driven variations of the torque may still be needed to explain the available measurements and the structured timing residuals quantitatively. Future work comparing the observed timing properties in detail could fine-tune the most promising initial configurations.}

\keywords{pulsars: general -- magnetic field -- stars: neutron}

\titlerunning{Pulsar obliquity and braking index}
\authorrunning{Dehman \& Vigan\`o}
\maketitle
\nolinenumbers

\section{Introduction}
\label{sec: intro}

Recent magnetohydrodynamic (MHD) simulations mimicking the post-core-collapse supernova environment have substantially improved our understanding of magnetic-field amplification in proto-neutron stars (e.g. \citealt{obergaulinger2014,aloy2021,reboul2021,masada2022,matsumoto2022,barrere2025}). These studies indicate that neutron stars are born with complex magnetic configurations, in which the magnetic energy is distributed over a broad range of spatial scales and is dominated by non-axisymmetric and toroidal components, rather than by large-scale dipoles or similarly simple configurations commonly adopted as initial conditions in long-term evolution studies of neutron stars.

The magneto-thermal evolution of isolated neutron stars (see \citealt*{pons26} for a recent review) is shaped by two distinct processes acting on the crustal magnetic field: the Hall effect and Ohmic dissipation \citep{goldreich92}. In addition, at early times, and when small-scale structures are well resolved, the chiral magnetic effect (CME) can efficiently transfer magnetic energy from small to large scales \citep{dehmanpons25,dehman26}. Thereafter, the secular evolution is governed by Hall--Ohmic dynamics, under which the redistribution of magnetic energy occurs predominantly on small scales \citep{wood15,gourgouliatos16,dehman23ccsn}: small-scale structures are particularly susceptible to Ohmic dissipation and are continuously regenerated through the Hall cascade, whereas large-scale current loops (i.e. low-$\ell$ magnetic multipoles) are comparatively long-lived. Ohmic and Hall contributions therefore reach a balance at high $\ell$ (small scales), producing a spectral slope of $\ell^{-2}$ (where $\ell$ denotes the spherical harmonic degree), while the low-$\ell$ part of the spectrum tends to preserve the imprint of the field configuration at the onset of this Hall--Ohmic phase -- not necessarily the configuration present at birth, since the CME may have already reshaped the large-scale field beforehand.

This low-$\ell$ persistence does not imply that any particular low-$\ell$ configuration is universal. The so-called Hall attractor -- a poloidal field dominated by $\ell=1,3$ plus an $\ell=2$ toroidal component, found in several simulations (e.g. \citealt{gourgouliatos14b,gourgouliatos14,wood15,bransgrove18}) -- is itself the outcome of a specific choice of large-scale configuration at the start of the Hall--Ohmic phase: typically a dipolar ($\ell=1$) poloidal field tightly Hall-coupled to a quadrupolar ($\ell=2$) toroidal one, whose mutual coupling generates the additional $\ell=3$ poloidal component characteristic of the attractor. By contrast, 3D pseudo-equilibrium (Ohmic+Hall) configurations starting from complex, non-axisymmetric, non-dipole-dominated fields \citep{gourgouliatos16,dehman23ccsn} are considerably more dynamic and tangled: poloidal and toroidal multipoles continuously exchange energy, and the resulting configuration does not necessarily resemble this Hall attractor.

Given this sensitivity to the field entering the Hall--Ohmic phase, the configuration at that stage has been the main multi-dimensional degree of freedom explored by previous studies -- most of which neglect the CME and instead prescribe, by hand, specific large-scale-dominated configurations chosen to capture a range of evolutionary scenarios. These include: configurations dominated by a large-scale toroidal dipolar component, which favour specific Hall instability modes \citep{gourgouliatos19} and produce magnetic spots at the poles, as required by some radio-emission models \citep{geppert14}; toroidal quadrupolar-dominated setups leading to dipolar-field growth compatible with the braking indices observed in young pulsars \citep{gourgouliatos15}; small-scale fields in the context of central compact objects \citep{gourgouliatos20,igoshev21,dehman23ccsn}; and dipolar-dominated configurations with a quadrupolar toroidal component, used as a practical benchmark scenario \citep{vigano12code,vigano21,dehman23matins}. The latter has been used to test different background electron velocity profiles \citep{gourgouliatos14b,gourgouliatos14}, to compare with observed luminosity and timing properties \citep{vigano12,vigano13,marino24}, and to study the corresponding temperature maps, hotspots, and thermal X-ray light curves \citep{igoshev21b,ascenzi24}.

The memory of the initial configuration has a clear physical origin. Since the magnetic field is not replenished by dynamo action, and Hall redistribution is limited both by the small crustal thickness ($\sim1$~km, i.e. $\sim10\%$ of the stellar radius) and by the non-zero resistivity, the choice of large-scale configuration at the onset of the Hall--Ohmic phase largely determines to what extent large-scale current structures dominate throughout the subsequent evolution. Indeed, while the CME-driven inverse cascade has been shown to operate effectively \citep{dehmanpons25,dehman26}, the Hall inverse cascade itself is confined to scales of order a few tens in $\ell$, unable to reach lower-$\ell$ (larger-scale) modes -- a limit set by the extreme aspect ratio of the neutron-star crust \citep{dehmanbrande25}.

One so-far scarcely studied consequence of such realistic configurations is that a non-dominant poloidal dipolar component, which nonetheless determines the spin-down evolution, can spontaneously change direction, similarly to what happens in convective dynamo-hosting objects (e.g. \citealt{jackson00} for the Earth). This drives an evolution of the angle $\alpha$ (known as the obliquity or inclination angle) between the spin axis and the dipolar magnetic moment. Such magnetic-axis drift has indeed been observed by \citet{gourgouliatos18} in 3D simulations of the early evolution of initial configurations dominated by a large-scale toroidal field, who related it to the secular changes observed in the pulse profile of the Crab pulsar, and to the braking indices measured in young pulsars \citep{espinoza11}.

Motivated by the lack of systematic studies in this regard, we investigate a small but representative variety of initial magnetic-field configurations, ranging from a small-scale-dominated case whose dipolar component is initially negligible, through the typical pulsar range, up to a single higher-field case probing the magnetar regime. In contrast to \citet{gourgouliatos18}, who considered a single, large-scale initial configuration, we span two opposite extremes -- nearly pure dipoles, and strongly non-axisymmetric, small-scale-dominated configurations motivated by realistic core-collapse simulations -- allowing us to assess how robust this spontaneous obliquity drift is to the assumed initial magnetic structure. Other previous efforts along these lines mostly focused on the evolution of the surface dipolar intensity only, $B_d$, in either axisymmetric \citep{pons12,vigano12} or 3D \citep{gourgouliatos15} simulations, or relied on semi-analytical models \citep{igoshev20}.

For each case, we follow the long-term evolution of the magnetic field and obliquity using the publicly available, state-of-the-art 3D magneto-thermal code \MATINS\ (MAgneto-Thermal evolution of Isolated Neutron Stars)\footnote{\url{https://github.com/ice-csic-astroexotic/MATINS}} \citep{dehman23matins,ascenzi24}. We aim to assess how short the spontaneous obliquity-evolution timescale can be, and to compare it with both the $B_d$-evolution timescale and the timescale of magnetospheric torque feedback \citep{philippov14}, which is often invoked instead to prescribe the long-term obliquity evolution. Finally, we compare our results with the measured braking indices, which deviate from the canonical $n=3$ in an age-dependent manner, keeping in mind that short-term variations arising from purely magnetospheric dynamics, rather than from the mechanism studied here, are expected to be the main contributors to the observed timing noise.

In Sect.~\ref{sec: Timing}, we summarise the essential observational constraints available from pulsar timing; in Sect.~\ref{sec: braking index}, we introduce the different physical contributions to the timing properties, and to the braking index in particular; in Sect.~\ref{sec: 3D simulations}, we present the numerical setup; in Sect.~\ref{sec: results}, we show the results of the simulations; and in Sect.~\ref{sec: discussion}, we summarise their implications.

\section{Braking index: measurements and caveats}
\label{sec: Timing}

\begin{figure*}
\includegraphics[width=0.33\linewidth]{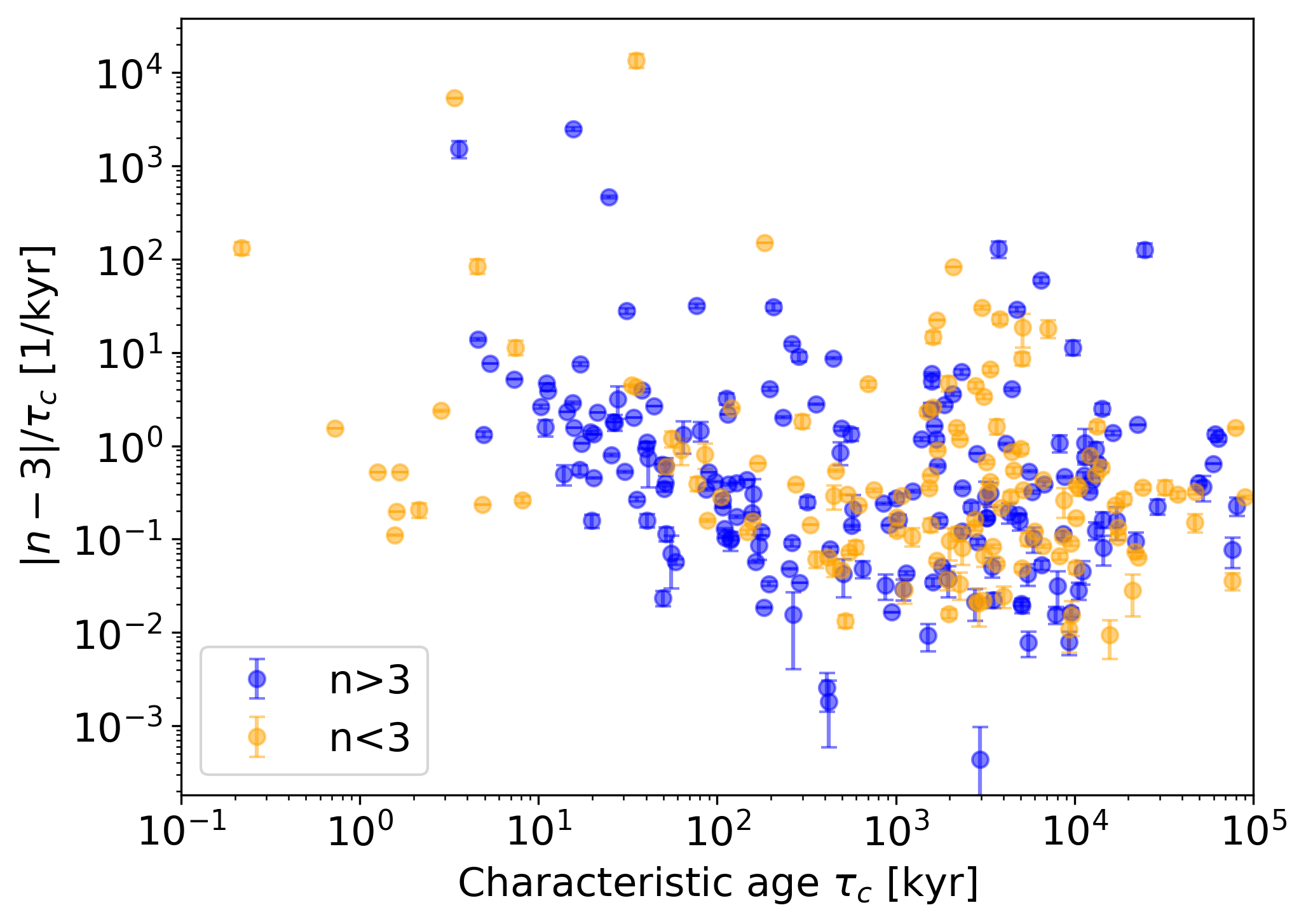}
\includegraphics[width=0.33\linewidth]{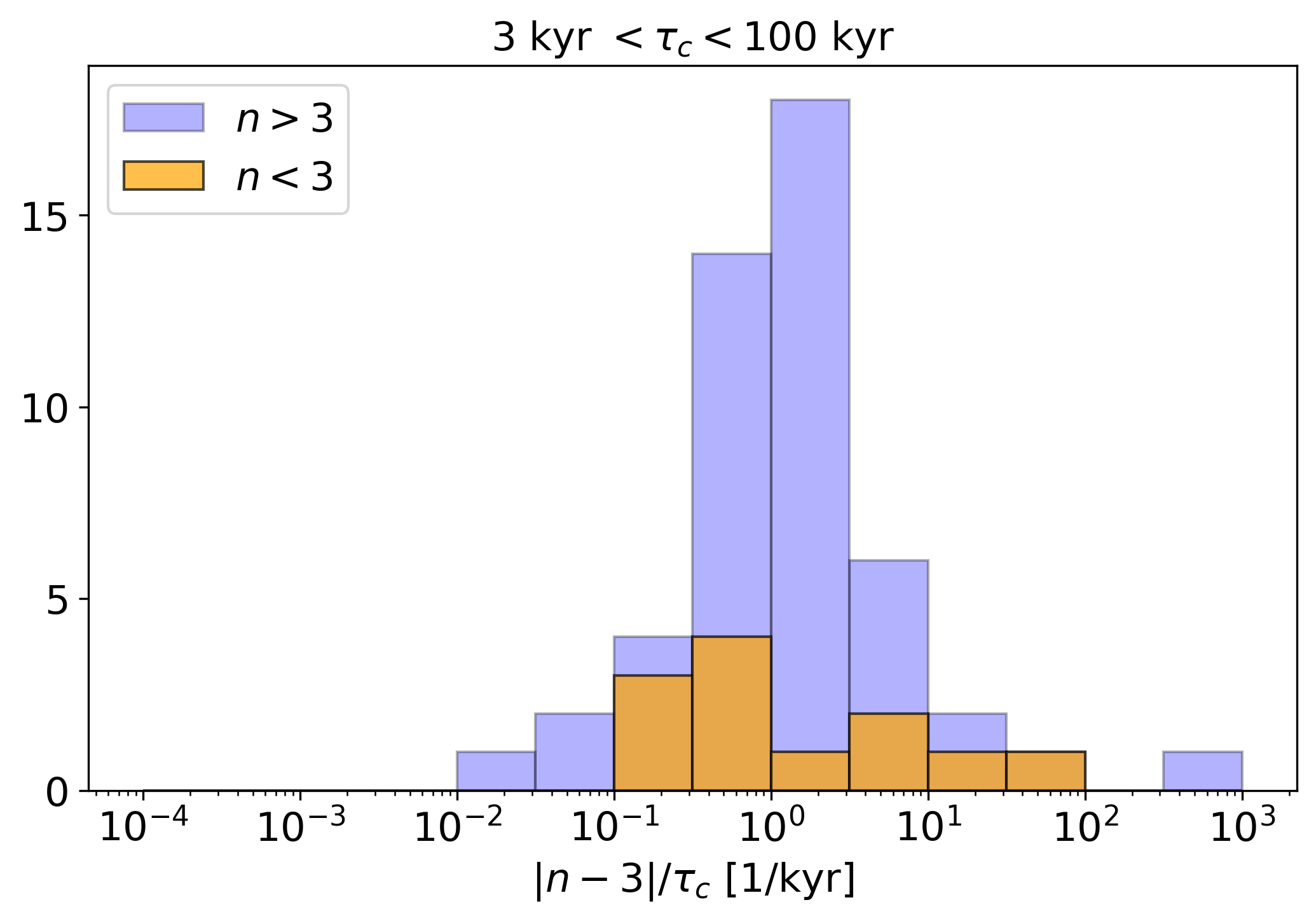}
\includegraphics[width=0.33\linewidth]{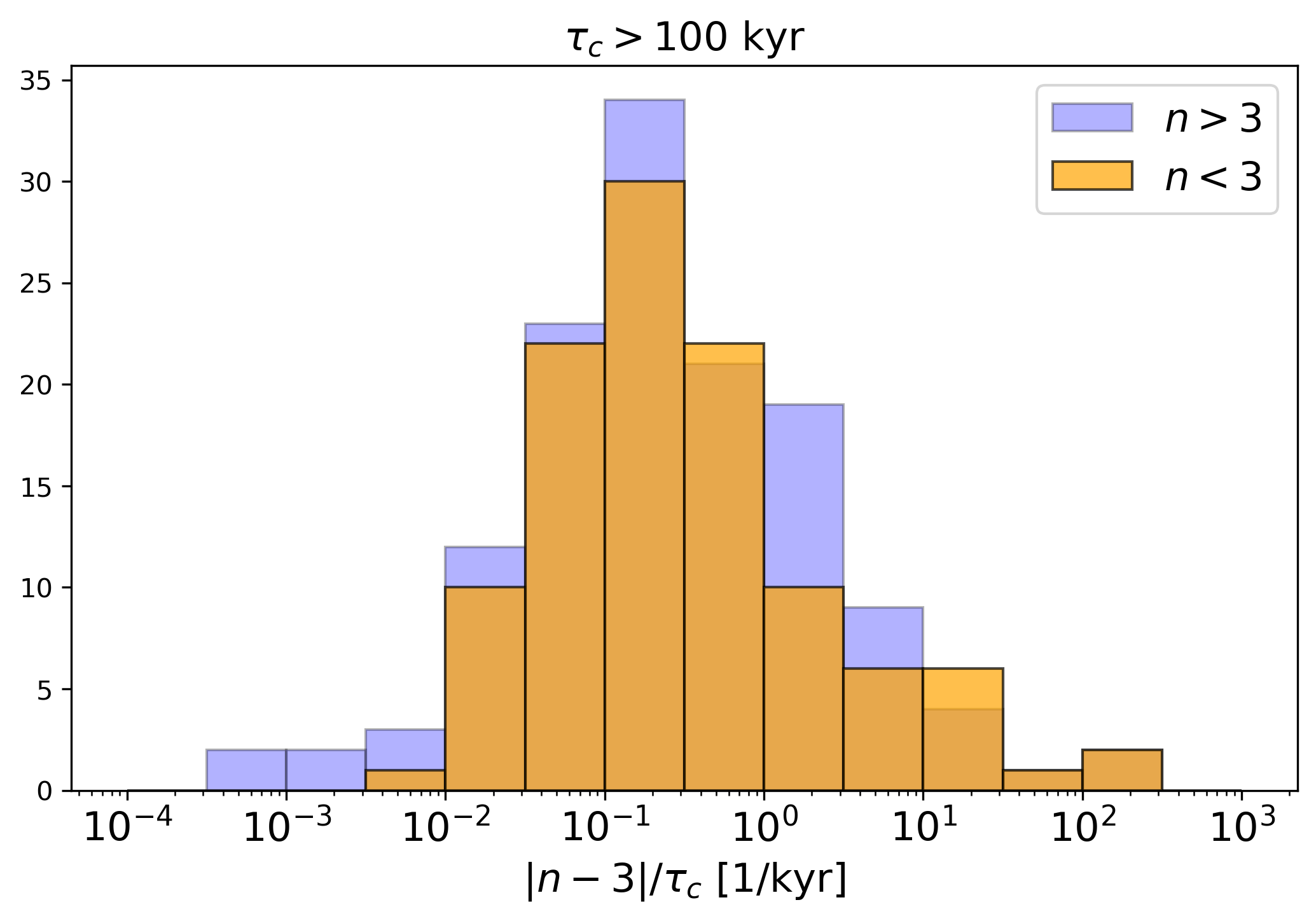}
\caption{Observed braking index positive (blue) and negative (orange) de-trended deviations from $n=3$. We show $|n-3|/\tau_c$ as a function of the characteristic age (left), and $|n-3|/\tau_c$ distributions for middle-age (centre) and older pulsars (right). Error bars are shown in the left panel, although mostly comparable to or smaller than the marker size, given our selection cut on the relative uncertainty of $n$; individual measurement errors are not shown in the binned distributions of the centre and right panels. }
\label{fig:obs}
\end{figure*}

The braking index $n$ is a commonly used diagnostic of pulsar spin-down evolution, offering a first-order probe of the physics governing the underlying magnetospheric torque (e.g. \citealt{espinoza17}). It is defined through $\dot\Omega \propto -\Omega^n$, i.e.
\begin{equation}
    n = \frac{\Omega \ddot{\Omega}}{\dot \Omega^2} = 2 -\frac{\ddot{P}P}{\dot P^2}, 
\end{equation}
where $\dot\Omega, \ddot\Omega$ and $\dot P, \ddot P$ are the first and second time derivatives of the spin angular frequency $\Omega$ and spin period $P=2\pi/\Omega$, respectively. For a standard magnetic dipole braking, with constant magnetic field strength and inclination angle, and no additional torque variations due to e.g. magnetospheric or internal superfluid dynamics (see below), one has $n=3$ (see Sect.~\ref{sec: braking index}). However, the measured braking index almost always deviates significantly from this standard value, with a clear correlation between the deviation $|n|$ and the characteristic age $\tau_c$ of a radio pulsar \citep{pons12,parthasarathy19,parthasarathy20}. Values of $n<3$ are of particular interest, since they imply a growing magnetospheric torque. Conversely, $n>3$ is naturally explained by a decrease of it, which can be consistent with a decay of the dipolar magnetic field component $B_d$, typically expected to occur on a timescale of $\sim10^{5}$~yr (see Sect.~\ref{sec: Bp variation} for a detailed discussion).

While $\Omega$ and $\dot \Omega$ are associated with tiny relative errors (short-term variations aside, as induced by glitches and magnetar outbursts), $\ddot \Omega$ is much harder to estimate reliably, both because it is usually intrinsically small and because its measurement is highly sensitive to the baseline of the timing solution \citep{cordes80,hobbs10,ou16,parthasarathy19} --- even for the cases (usually young, energetic pulsars) for which the timing solution and the estimate of $n$ are more reliable \citep{zhang12}. This is indicative of an underlying physical complexity in the torque dynamics, also often visible in structured timing residuals, and points to torque fluctuations on a variety of timescales that can be as short as months to years \citep{ou16}. As a consequence, in many cases the reported $n$ takes large positive or negative values regardless of the (often large) associated uncertainties, indicating that short-term torque variations dominate over the long-term secular evolution in shaping $\ddot \Omega$ and $n$. Dedicated studies exist on the determination of $n$, for which we refer the reader to the recent review by \cite{abolmasov24}, and to, e.g. \cite{hobbs04,biryukov07,hobbs10,parthasarathy19,parthasarathy20} for further observational work on pulsar timing irregularities and braking-index measurements.

With the caveats above in mind, we compiled braking-index measurements from the Australia Telescope National Facility (ATNF) pulsar catalogue \citep{manchester05} as of June 2026, applying the following selection criteria: we exclude the bulk of recycled millisecond pulsars with an accretion history (which alters the rotational evolution) by considering only pulsars with $P>15$~ms, and we retain only measurements with a relative error on $n$ --- propagated from the reported $\ddot\Omega$ uncertainty --- below 50\%.\footnote{\cite{pons12} considered a sample $\sim4$ times smaller than ours, filtering the by-then available data with a reported $\ddot\Omega$ uncertainty $<10\%$. Overall, changing the uncertainty filter leads to similar trends and histograms.}

Figure~\ref{fig:obs} shows the observed distribution of braking indices as a function of characteristic age, separating the sample (440 objects) into pulsars with $n>3$ (blue, decreasing torque) and $n<3$ (orange, growing torque). The left panel shows values of $n$ of order unity only for $\sim$kyr-old pulsars, with much larger positive and negative values that linearly correlate with the characteristic age $\tau_c$, in agreement with previous studies \citep{pons12,parthasarathy19,parthasarathy20} and with the definition of $n$ itself (see below). Notably, while previous studies have generally stressed a roughly even split between positive and negative braking indices, we find an overall slight excess ($58\%$) of $n>3$ cases. Such excess is significant and is clearly not uniform across ages: the central and right panels of Fig.~\ref{fig:obs} show the de-trended distribution of braking-index deviations, $|n-3|/\tau_c$, for middle-age (3 kyr $<\tau_c<100$ kyr) and older ($\tau_c > 100$~kyr) pulsars. This quantity factors out, by construction, the implicit linear dependence of $(n-3)$ on $\tau_c$ mentioned above; it therefore represents the inverse of the variation timescale associated with the underlying physical mechanism, allowing a more meaningful comparison between pulsars of different ages.

From the histograms, we find a significant excess ($78\%$) of positive values $n>3$ for the middle-age population, with values typically $|n-3|/\tau_c \sim 0.3-10$. This excess arises predominantly from middle-age pulsars ($\tau_c \sim 20$~kyr), with negative values appearing both at very young ages, and when approaching 100 kyr. This excess marks a sharp contrast with the handful of younger pulsars $\tau_c<3$ kyr, which all show $n<3$, as shown by previous dedicated studies with reliable $n$ estimates (\citealt{espinoza11} and references therein), indicating a growing magnetospheric torque. Older pulsars show another distribution yet, with more even distribution of large positive ($\sim55\%$) and negative ($\sim45\%$) values of $n-3$. Qualitatively, the picture that emerges is an increase of the magnetospheric torque at young ages, previously explained by magnetic-field fallback burial and re-emergence \citep{muslimov96,ho11,vigano12,pons12,igoshev16}, or by inverse cascade in simplified setups \citep{gourgouliatos15}. This is followed by a predominant, but not universal, torque decrease at middle age, which converges to a more random distribution of large $n$ values at late ages.  

Note also that the presence of a wide-orbit, unseen companion can contribute to large-amplitude oscillation in the values of $n$, and such companions could be expected in a minority of cases (\citealt{igoshev20} and references therein). 

Admittedly, the systematic errors associated with $\ddot\Omega$ could also be a fundamental factor in broadening the distribution shown here: the reported errors have not necessarily been inferred in the same way, and only a few dedicated studies exist with a homogeneous and more reliable timing-uncertainty assessment \citep{hobbs10}. For instance, a careful systematic analysis restricts the sample of measured values by more than one order of magnitude: it consists of only 19 young pulsars (two of which are negative, and all with $|n|\lesssim 100$) in \citet{parthasarathy19,parthasarathy20}. Although such observational caveats are not easily disentangled from physical effects, they are unlikely to be the main driver of the peculiar $\tau_c$-dependent deviations from $n=3$. These limitations prevent us from having a clear view of how the distribution of $n$ is shaped solely by the torque dynamics rather than observational uncertainty.

Keeping these caveats in mind, in the following we investigate, via 3D magneto-thermal simulations, how much the obliquity and $B_d$ evolution can contribute to the observed distribution of $n$ in middle-age pulsars, assuming the standard spin-down model presented in the following section. We restrict the study to ages $\lesssim 10^6$~yr (but characteristic ages $\tau_c$ which can be substantially larger depending on the model, see below), for which our evolutionary models are more numerically stable and physically reliable, and for which the $n$ distribution is less prone to dispersion from overlooked systematic errors on $\ddot\Omega$.

\section{Physical contributions to braking index values}
\label{sec: braking index}

The rotational evolution of an isolated neutron star is obtained by equating the rotational energy losses, $-\dot E = - I\Omega \dot\Omega$, where $I$ is the moment of inertia, to the spin-down power associated with the electromagnetic torque (see e.g. \citealt{deutsch55} for vacuum, \citealt{spitkovsky06} for force-free magnetospheres):
\begin{equation}\label{eq:spindown}
    {\cal P} =  \frac{R^6 f_\alpha B_d^2\Omega^4}{4 c^3}{\cal W}\, , 
\end{equation}
where $R$ is the stellar radius, $c$ is the speed of light, $f_\alpha$ is a function of the obliquity $\alpha$, and ${\cal W}$ is a generic, order-unity function accounting for short-term (indicatively, weeks to years) variations power output, hereafter referred to as a ``wobble'' term. Some possible physical origins are summarised in Sect.~\ref{sec: wobbling} below: in brief, it represents relatively fast phenomena contributing to the timing noise and unrelated to the long-term interior evolution. Hence, we can write
\begin{equation}
 P\dot P = \frac{\pi^2R^6}{Ic^3} B_d^2 f_\alpha {\cal W}.
\label{eq:ppdot}
\end{equation}
The function $f_\alpha$ would be $(2\sin^2\alpha)/3$ in the unrealistic vacuum case \citep{deutsch55}, while, for force-free magnetosphere we adopt in this and previous works, it is \citep{spitkovsky06}
\begin{equation}
    f_\alpha \simeq 1 + \sin^2\alpha\, .
\end{equation}
The braking index can then be written as
\begin{equation}
 n =  3 - \tau_c\left(\frac{1}{\tau_B} + \frac{1}{\tau_\alpha} + \frac{1}{\tau_{\cal W}} \right) \, , 
 \label{eq: braking index}
\end{equation}
where $\tau_c = P/2 \dot{P}$ is the characteristic age, $\tau_\alpha$ is the timescale associated with obliquity variations, $\tau_B$ is the timescale associated with the $B_d$ evolution, and $\tau_{\cal W}$ is the timescale associated with the short-term power variations:
\begin{eqnarray}
    && \frac{1}{\tau_\alpha} = 2\frac{\dot f_\alpha}{f_\alpha} = 4 \frac{d \sin\alpha}{dt} \frac{\sin\alpha}{\sin^2\alpha +1}~, \\
    && \frac{1}{\tau_B} = 4\frac{\dot B_d}{B_d}~, \\
    && \frac{1}{\tau_{\cal W}}=2\frac{\dot{\cal W}}{{\cal W}}~.
\end{eqnarray}
These timescales are signed: an increase (decrease) of $B_d$, $f_\alpha$, or ${\cal W}$ contributes towards values $n<3$ ($n>3$).

Here we assume standard electromagnetic spin-down. This mechanism dominates over, e.g. gravitational-wave losses for all observed pulsars, the latter becoming relevant only for large quadrupole deformations or ultra-fast rotation. The wind braking model \citep{tong13} is an alternative to the electromagnetic spin-down model and has been used to model the long-term evolution of the braking index \citep{tong17}. Since these models have mostly been applied to magnetars -- the class for which braking-index measurements are also the least reliable -- we focus on the electromagnetic spin-down model only. Nevertheless, note that wind models have been proposed to account for $n<3$ \citep{xu01}.

\subsection{Surface dipolar field evolution}
\label{sec: Bp variation}

The magnetic energy in the crust is expected to decrease resistively, since no active dynamo can replenish it, and mechanisms such as the CME \citep{dehmanpons25,dehman26} or thermoelectric effects can contribute only at very early ages. If the surface dipolar component $B_d$ is dominant and tracks this global magnetic-field decay on a timescale of $\sim10^4$--$10^6$~yr, a contribution towards $n>3$ is expected. A few studies have examined this evolution of $B_d$ and its imprint on $n$ in detail, in axisymmetric \citep{pons12,gourgouliatos15} and 3D simulations \citep{gourgouliatos18}, or from semi-analytical approaches \citep{igoshev20}. In general, three stages can be identified: (i) an early stage with several mechanisms leading to slight growth of the torque (see below), consistent with the above-mentioned values for young pulsars $n\lesssim3$; (ii) an intermediate stage of near-pure Ohmic decay, during which $n>3$; and (iii) a late stage, once the interior has cooled to very low temperatures, in which Hall-induced oscillations of the dipolar component \citep{pons12,gourgouliatos14} drive correspondingly signed oscillations of $n$, with amplitudes of several tens at $t\gtrsim 10^5$ yr in the most extreme cases (thus, unable to explain the much larger values and the trend with $\tau_c$).

Similarly, recent simulations accounting for the coupling between simplified core-field dynamics and crustal evolution show that giant Hall waves can produce oscillating values of $n$ up to $\sim\pm10$ \citep{bransgrove25}. The inclusion of more complex core dynamics (ambipolar diffusion, superfluid--superconducting components, etc.) is not expected to produce larger variations, since the core magnetic-evolution timescales are generally longer than those in the crust \citep{goldreich92}; even in the more optimistic models that consider initial fast rearrangement stage \citep{gusakov20}, these timescales are at most comparable to the crustal ones. What drives the very large positive or negative $n$ values often observed in middle-age and old pulsars therefore remains an open question, though the above-mentioned observational caveats need to be kept in mind.

In contrast, for the $\sim$ dozen young pulsars with reliable values $n\lesssim3$, convincing explanations exist: magnetic field burial via fallback accretion and subsequent re-emergence \citep{ho11,vigano12,pons12,ho15,eksi17}, or a moderate inverse cascade arising from an initially weak dipole combined with a strong toroidal field \citep{gourgouliatos15}. Extending these ideas, other explanations, not yet specifically studied so far in the context of effects on timing properties, include the early-time growth of the dipole driven by the CME through an inverse-cascade-like process \citep{dehmanpons25,dehman26}, or the evolution of an initially turbulent, small-scale-dominated field undergoing an inverse cascade \citep{WH09,WH10,Cho11,brandenburg2020,dehmanbrande25}. Such scenarios are assessed and compared with large-scale setups below.

\subsection{Obliquity evolution}
\label{sec: alpha variation}

The obliquity can change for two nearly independent reasons: internal dynamics, which determine the internal and surface magnetic field configuration, and long-term alignment driven by external mechanical torques (possibly accompanied by precession in the case of structural asymmetries; \citealt{igoshev20}). The two effects are largely decoupled, though the magnetospheric configuration should slightly affect the simulated internal dynamics through the surface boundary condition. Since, currently, all existing 3D magneto-thermal simulations prescribe a potential-field solution outside the star, we cannot include this effect, which is expected anyway to provide minor modifications to the outcome of this study. 

\subsubsection{Spontaneous magnetic axis drift}\label{sec:wandering}
Although neutron stars lack the convective dynamo responsible for the continuous, short-term variability seen in convective stars, brown dwarfs, and planets, their magnetic field configuration nonetheless evolves over time. As mentioned above, the initial field geometry plays an important role in this respect: configurations that start out with most of their energy at small scales tend to be more dynamically active, undergoing an inverse cascade due to the presence of initial magnetic helicity \citep{brandenburg2020,sarin2023,dehmanbrande25}, or due to early-time chiral anomalies \citep{dehmanpons25,dehman26}. These effects can partially transfer magnetic energy from small to large scales, enriching the dynamics which already consist of a Hall-like cascade and Ohmic dissipation. In this scenario the dipolar component is naturally more prone to change its axis. As a matter of fact, a dipole-dominated configuration is supported by large-scale current systems, which have greater ``inertia'' than the small-scale current systems supporting a small-scale-dominated configuration. In the latter case, the currents supporting the field fluctuate in space and time, so that their large-scale average, which determines the large-scale magnetic field component is also more variable.

\citet{gourgouliatos18} explored the resulting internal-dynamics-driven obliquity variation with 3D Hall-MHD simulations in the context of young pulsars, finding results roughly consistent with the observed gradual changes in the Crab pulsar's pulse profile \citep{lyne15}. They considered initial field configurations dominated by large-scale magnetic structures, evolved over a kiloyear timescale. We aim to fill this gap by systematically studying this effect over longer timescales and for a variety of initial configurations, including small-scale-dominated ones.

\subsubsection{Magnetospheric external feedback (alignment)}\label{sec:alignment}

The magnetospheric torque itself tends to act on the obliquity, minimising energy losses by driving the spin and dipole axes towards alignment. According to the force-free simulations of \citet{philippov14}, the obliquity angle evolves as
\begin{equation}
    \frac{d\alpha}{dt} = -\frac{\pi^2R^6}{Ic^3} \frac{B_d^2}{P^2} \sin\alpha\cos\alpha
    \label{eq: obliquity evol}
\end{equation}
consistent with observational evidence for a slow, long-term alignment of spin and magnetic dipolar axis \citep{tauris98}.
A caveat is that this torque, and the prefactor calibrating it, is obtained under the assumption of a purely dipolar surface field \citep[e.g.][]{spitkovsky06,philippov14,kalapotharakos09}. Cases with multipolar inclusions, e.g. quadrupoles and octupoles, and offset dipoles have been tested in a few cases, with minor modifications to the dipole-calibrated torque \citep{petri15,petri21}; similarly, force-free models of axisymmetric magnetospheres with arbitrary surface fields show that higher multipoles mainly reshape the polar caps, while the spin-down remains set by the dipolar component \citep{Gralla16}. However, to our knowledge, no study has assessed how this torque coefficient changes for a genuinely small-scale-dominated, oblique surface field, so that we will assume that the dipolar calibration is universal. This assumption does not apply to strongly twisted magnetospheres, as expected in magnetars, where currents along closed field lines and particle winds can substantially enhance the spin-down \citep[e.g.][]{thompson02}; such configurations are beyond the scope of this work.

Based on the expected long-term alignment due to the external torque, \citet{johnston17} proposed obliquity decay as a major driver of the observed timing properties of the pulsar population, while \citet{eksi16} used the long-term obliquity evolution to place constraints on a pulsar with $n>3$.

The two obliquity-variation mechanisms differ in a key aspect: mechanical alignment systematically reduces the torque ${\cal P}$, and so contributes towards $n>3$, whereas the internal mechanism is expected to be essentially random, since it is governed by Hall-driven internal dynamics alone. If the dipole is initially subdominant, we expect internal rearrangement of the field via Hall dynamics or the CME to compete with, or even dominate over, this external feedback --- a competition we investigate in this study.

\subsection{Short-term wobbling}
\label{sec: wobbling}
In most spin-down models, short-term variations of the spin-down power are implicitly assumed to be negligible compared to the secular, long-term evolution. While this assumption is reasonable for the inferred values of $P$ and $\dot P$ in the majority of pulsars, there are several known cases in which the spin-down rate changes abruptly (e.g.\ \citealt{wang20}). Moreover, timing noise (see Sect.~\ref{sec: Timing}) affects the inferred second- and higher-order frequency derivatives. \citet{lower25} studied such short-term variability in a sample of hundreds of radio pulsars, finding that its amplitude correlates with $\dot\Omega$.

Physically, such variations may originate from, e.g. stochastic fluctuations in plasma dynamics and the associated magnetospheric currents and configuration, from waves propagating through the magnetosphere, or from free precession \citep{igoshev20}. For instance, \citet{gao16} estimated the impact of particle winds on the highly variable, noisy braking-index measurements of magnetars.

Another contribution to short-term variations may come from superfluid vortex dynamics (e.g. \citealt{alpar06,melatos14,ou16,oliveira18,lower21,gugercinoglu23,abolmasov24,gugercinoglu26a,gugercinoglu26}), whereby crustal vortex (un)pinning dynamics and/or interactions with superconducting flux tubes can intermittently decouple and recouple the superfluid interior from the crust -- the component directly coupled to the external electromagnetic braking. Indeed, glitches produce a transient increase in the inferred braking index, up to $n\sim10$, which averages out over sufficiently long timing baselines and does not directly bias the long-term braking index; independently, young glitching pulsars are systematically observed to show $n<3$ \citep{espinoza17,lower21}.

These possible short-term effects are represented by the order-unity function ${\cal W}$ in Eq.~(\ref{eq:ppdot}), similar to the function $TN(t)$ introduced by \citet{johnston17}. If such variation timescale is much shorter than the others ($\tau_{\cal W}\ll \tau_B,\tau_\alpha$), the braking index will naturally attain large positive and negative values, superimposed on the long-term evolution. The importance of such short-term fluctuations was already proposed by \citet{ou16} to explain the roughly equal numbers of positive and negative measured values of $\ddot\Omega$.

Our magneto-thermal simulations cannot model such terms, so that we can only indirectly infer their relevance by checking whether the above-mentioned obliquity and $B_d$ variations can explain the observed $n$ distribution, or short-term variations are instead needed.

\section{Long-term 3D magneto-thermal simulations}
\label{sec: 3D simulations}

\subsection{Numerical setup} 
\label{sec: numerical setup}

We aim to assess the change in obliquity angle in young and middle-aged pulsars ($\lesssim1$~Myr) driven by the long-term internal dynamics of the magnetic field in the neutron-star crust. Our simulations directly compute the long-term variation of $B_d$, driven by internal dynamics (Sect.~\ref{sec: Bp variation}), and of the obliquity, driven by both internal dynamics and the alignment induced by the magnetospheric torque (Sect.~\ref{sec: alpha variation}). Comparing these simulated contributions with the observed braking index distribution then allows us to infer indirectly whether or not they dominate over the short-term wobbling discussed in Sect.~\ref{sec: wobbling}.

To this end, we perform realistic 3D simulations using the coupled magneto-thermal evolution code \MATINS \citep{dehman23matins,ascenzi24}. Specifically, we solve the induction equation accounting for Ohmic dissipation and the Hall drift term in the crust \citep{dehman23matins,dehman23ccsn,dehmanbrande25}. We neglect the CME \citep{dehmanpons25,dehman26}, since (i) it is expected to be relevant only during the first centuries of neutron-star evolution, whereas we focus here on the long-term evolution over $\lesssim 1$~Myr, and (ii) resolving it is computationally expensive, requiring finer spatial resolution and correspondingly short timesteps of order days.

We model the entire neutron-star volume but restrict the magnetic-field evolution to the crust, as in most quantitative magneto-thermal evolution studies. Simulations that include core dynamics do not show faster magnetic-field variations (e.g.\ \citealt{castillo17,bransgrove18,vigano21,igoshev23,castillo25,moraga25}), although none of them incorporates all relevant physical ingredients simultaneously, i.e. the coupled effects of normal, superfluid, and superconducting components together with ambipolar and Ohmic diffusion. Given these results, and the timescales at play \citep{gusakov20}, core long-term dynamics are very unlikely to contribute to timing noise or to large braking-index values, as commented above.

We impose potential-field (current-free) boundary conditions at a mass density of $\rho=10^{10}\,\mathrm{g\,cm^{-3}}$, which defines the numerical surface of the star, $R_b$, located at the base of the liquid envelope where it transitions to the solid crust. Since the envelope is thin (typically $\sim0.01\,R$; see below for our case) and more resistive than the crust, it is not expected to host significant electrical currents (unless there are important magnetospheric current loops, here neglected). Currents are then mostly confined to the inner, highly conductive crust (e.g.\ \citealt{vigano12code,vigano13,pons13,wood15}), so that the magnetic field can safely be assumed to be potential in the envelope. Perfect-conductor boundary conditions are imposed at the crust--core interface, i.e.\ the radial magnetic field and the tangential electric field vanish there, thereby confining the magnetic field to the crust. A field threading the core would generally evolve on longer timescales and, if it had an appreciable dipolar component, would tend to stabilise the orientation of the dipolar axis, since it is sustained mostly by slowly evolving currents circulating in the core. The simplification of considering only crustal dynamics therefore mostly affects the results of large-scale dominated configurations, whereas small-scale-dominated ones, for which the sustaining currents mostly circulate in the crust, are mainly governed by the crustal dynamics here modelled. Therefore, the inclusion of a core-threading field will enhance further the differences, shown below, between the negligible dipolar axis shift in large-scale dominated configurations and the relevant obliquity changes seen in the small-scale ones.
We refer the reader to \citet{dehman23matins} for details on the magnetic boundary conditions, grid, numerical methods, and the spherical-harmonic decomposition of the magnetic field.

The thermal evolution is computed by solving the heat-diffusion equation as described in \citet{ascenzi24}. For the blanketing envelope, which provides the thermal boundary condition \citep{pons26}, we adopt the model of \citet{potekhin2015}, which assumes a heavy-element composition (e.g. iron) and accounts for magnetic-field effects. For a discussion of the impact of different envelope compositions and magnetisation on the magneto-thermal evolution, we refer the reader to \citet{dehman2023c}. Note that not all models presented here solve the heat-diffusion equation; most instead use simplified temperature evolution, as detailed in Sect.~\ref{sec: Tevol}, as a computational cost compromise.

\MATINS incorporates state-of-the-art microphysical ingredients, including the specific heat capacity, conductivities, and neutrino emissivities (see also \citealt{vigano13,vigano21,dehman23ccsn,ascenzi24}). In particular, the temperature-dependent anisotropic conductivities are computed using the IOFFE conductivity code\footnote{\url{http://www.ioffe.ru/astro/conduct/}} \citep{potekhin2015}. The stellar background in \MATINS is constructed using zero-temperature equations of state (EOSs) from the CompOSE database.\footnote{\url{https://compose.obspm.fr/}} In this work, we adopt the BSk24 EOS \citep{pearson2018}, which has been shown to be compatible with the observed thermal luminosities of a wide range of weakly magnetised neutron stars \citep{marino24}. We consider a canonical $1.4\,M_\odot$ neutron star, yielding a stellar radius of $R=12.59$~km, a numerical-domain radius (envelope bottom) $R_b=12.41$~km, a crust thickness (solid crust only, excluding the envelope) of $0.83$~km, and a moment of inertia $I=1.97\times10^{45}$~g\,cm$^2$. \MATINS also allows for the exploration of alternative EOSs and stellar masses, but for the aims of this work (comparing different initial magnetic-field configurations for a fixed structure) this is not required. We use the reported values of $I$ and $R$ throughout for self-consistency, noting that the relevant combination $I/R^6$ typically varies by a factor $\lesssim3$ across different EOSs and masses, which would not qualitatively change our results.

We follow the evolution over the first 1~Myr of the neutron star's life (unless instabilities kick in before), with magnetic-evolution timesteps of a fraction of a year. Not all runs reach this age: once the star enters the photon-cooling era ($t\gtrsim50$--$100$~kyr, see below), the resistivity drops, the magnetic-field dynamics become increasingly Hall-dominated, and the code correspondingly less stable \citep{pons26}. Moreover, the microphysical ingredients and the thermal evolution implemented here are calibrated for the temperatures typical of young pulsars ($\sim10^8$~K) and become much less reliable for internal temperatures $\lesssim10^7$~K typical of $\sim$ Myr stars.

We discretise the crust with 40 radial grid points, resolving structures down to scales of tens of metres. The angular directions are discretised using a cubed-sphere grid \citep{dehman23matins} with $42^2$ points in each of the six patches, corresponding to 168 points along the equator, resolving angular scales of several hundred metres, more than sufficient to capture the main physical effects of interest.

\subsection[]{Dipolar field $B_d$ and obliquity $\alpha$}\label{sec: bd alpha}

\label{sec: evo params}
For each simulation, we compute the spherical-harmonic weights $b_{\ell,m}$ of the three $m$-components ($m=-1,0,1$) of the dipolar ($\ell=1$) term. Specifically, we compute them at the numerical surface (see \citealt{dehman23matins} for details), multiplying these values by $(R_b/R)^3\sim0.958$ (as in the neutron star model used here, see above) to account for the small correction from the cubic decrease of the dipolar components with distance across the envelope. The dipolar component intensity at the surface, which regulates the rotational evolution, is then 
\begin{equation}
 B_d = \sqrt{b_{1,-1}^2+b_{1,0}^2+b_{1,1}^2}\, . 
\end{equation}
As in all existing 3D magneto-thermal simulations, we assume that, in the absence of differential rotation, the spin and the location of the spin axis do not affect the internal dynamics. Assuming that internal dynamics and obliquity are decoupled means that, over a large pulsar population, spontaneous variations of $\alpha$ are expected to contribute to either $n>3$ or $n<3$ with roughly equal likelihood at a given age. The main caveat of this approach is that it neglects any feedback of the magnetospheric return currents on the internal field, particularly through the open-field-line region. However, except for very fast-rotating pulsars, this region is small, and the assumption of no relevant spin feedback on the magneto-thermal evolution is reasonable. Current loops are expected to circulate in closed field lines only in magnetars, which are not the object of this study.

In practice, for a given simulation, the relative initial orientation between the dipolar axis and the spin axis, i.e. the initial obliquity $\alpha_0$, is chosen a posteriori. For each simulation, we consider three representative geometries, labelled by $i$. For simplicity, we consider the initial dipolar field (which in our simulations corresponds to the $m=0$ mode, see above) to lie along three orthogonal directions, labelled $x,y,z$, in a Cartesian frame where $z$ is the spin-axis direction. The corresponding obliquity $\alpha_i$ between spin axis $i$ and the surface dipolar magnetic axis is then simply defined by
\begin{equation}
\cos\alpha_i = \frac{b_{1,i}}{B_d} \qquad i=x,y,z\,.
\end{equation}
This allows us to obtain, for each simulation, three representative calculations for the spontaneous obliquity evolution. No $R_b/R$ correction is needed here because, as discussed above, there are no currents in the envelope, so the dipolar axis direction is insensitive to any distance $r>R_b$. Note that the evolution of the timing properties will also change due to the $f_\alpha$ factor.

Separately, we consider the expected alignment driven by the external magnetospheric torque, via Eq.~(\ref{eq: obliquity evol}). By default, we use an initial obliquity $\alpha_{e,0}=\pi/4$, a value which minimises the initial alignment timescale, thus representing an upper limit on the contribution to the braking index; we will also briefly discuss the cases $\alpha_{e,0}=0.1\pi$ and $0.4\pi$. We label this obliquity calculation $\alpha_e$. Notice that a more self-consistent obliquity evolution could be obtained by simultaneously accounting for the internal dynamics and the external feedback, redefining at each time step the externally driven change of the spin axis. However, in this first work we treat the two effects separately, in order to assess clearly their relative importance.

For each of these cases ($\alpha_x,\alpha_y,\alpha_z,\alpha_e$), we integrate Eq.~(\ref{eq:ppdot}), neglecting the wobbling term (${\cal W}=1$), to obtain the corresponding long-term period evolution $P(t)$. We consider two different initial periods $P_0$: a fast one, $P_0^f=10$~ms, and a slower one, $P_0^s=50$~ms -- the latter being more representative of the average initial period, $P_0\sim40$--$200$~ms with a large dispersion, inferred from population synthesis studies (\citealt{pardo25}, and the compilation of previous estimates in their Table~2). For large enough fields and small enough initial periods (two conditions not always met in our simulations, see below), the evolutionary tracks converge quickly to the same one, and the memory of the initial period is lost, since the spin-down power scales with $\Omega^4$ and is therefore very effective at early times. Finally, for each $P(t)$, we compute the corresponding $\dot P$ via Eq.~(\ref{eq:ppdot}), and the characteristic age $\tau_c=P/(2\dot P)$, which shows deviations from the real age $t$. See Appendix~\ref{app: rotational} for more details on the rotational evolution of our models.

\subsection{Initial configurations}
\label{sec: Initial cond}

The configurations considered here fall into three categories, according to the range of scales over which the magnetic energy is distributed (see App.~\ref{app: magnetic evolution} for the detailed initial spectra):
\begin{enumerate}
\renewcommand{\labelenumi}{\textit{(\roman{enumi})}}
\item \CCSN: a small-scale configuration inspired by the magnetic field spectrum obtained at saturation (i.e.\ after many MHD timescales) in shell dynamo setups representing the typical density and angular-velocity background found hundreds of milliseconds after core collapse in dedicated supernova simulations \citep{reboul2021}. We adopt $B_d=10^{12}$~G and a volume-averaged intensity $B_{\rm avg}\approx10^{14}$~G. The magnetic energy is shared between a large-scale, axisymmetric toroidal quadrupole ($\ell=2$) and a smooth, non-axisymmetric spectrum of poloidal and toroidal components peaking again at small scales, around $\ell=30$, as adopted by \citet{dehman23ccsn} to study the long-term magneto-thermal evolution.
\item \texttt{SS1}, \texttt{SS2}, \texttt{SS3}: three configurations in which the magnetic energy is almost entirely toroidal ($\approx90\%$) and concentrated at small scales, with $\gtrsim80\%$ contained in multipoles of degree $18\lesssim\ell\lesssim33$. The dipolar component is weak in all three and is the main difference between them, with $B_d=10^{11}$~G, $5\times10^{10}$~G and $8\times10^{9}$~G, respectively; all three share $B_{\rm avg}\approx1.2\times10^{14}$~G. They consequently span a range of
dipolar evolutions: the dipole mostly decays in \texttt{SS1}, alternately grows and decays in \texttt{SS2}, and can only grow in \texttt{SS3}, whose initial dipolar energy is so small that any leakage towards $\ell=1$ amplifies it, whether or not a systematic inverse cascade operates. These values of $B_d$ are too low to let the resulting $P$--$\dot P$ tracks cross the bulk of the observed population (App.~\ref{sec: PPdot}), but they are numerically tractable and allow the inverse cascade to be followed, as shown below.
\item \texttt{D12} and \texttt{D14}: two dipole-dominated configurations with polar surface field strengths $B_d=10^{12}$~G and $B_d=10^{14}$~G, respectively. In both, $\gtrsim95\%$ of the magnetic energy is poloidal and concentrated in the dipolar ($\ell=1$) component, with $B_{\rm avg}\approx B_d$. They represent the configurations commonly adopted for simplicity in previous studies, and serve both as a reference case and as the counterpart to the complex, small-scale configurations described above. Their results are presented in App.~\ref{app: large scales}.
\end{enumerate}
In all models, the weights of the poloidal and toroidal scalar functions defining the initial field ($\Phi^0_{\ell m}$ and $\Psi^0_{\ell m}$, see App.~\ref{app: magnetic evolution} and App.~B of \citealt{dehman23matins}) are close to zero for ($\ell=1$, $m=\pm1$),\footnote{In other words, the dipolar axis initially points close to the centre of the northern patch of the cubed-sphere grid described in \citet{dehman23matins}. The numerical reconstruction of the surface multipoles, obtained by integrating the radial field component weighted by spherical harmonics, introduces a small error in the recovered dipolar axis, of at most a few degrees for the cases with the weakest dipolar component.} so that all models start with $\sin\alpha_z\simeq0$ and $\sin\alpha_x\simeq\sin\alpha_y\simeq1$. For each multipole $\ell\geq2$, the $2\ell + 1$ weights corresponding to $m\in[-\ell,\ell]$ are set by random values, and rescaled together so that their joint contributions to magnetic energy give the prescribed $E_M^{\rm in}(\ell)$. The $z$-axis thus identifies the initial dipolar direction, while the spin axis is chosen a posteriori.

We do not include the effects of a possible post-collapse burial of the magnetic field \citep{ho11,vigano12}. All simulations are run for $1$~Myr, except \CCSN and \texttt{CCSN-3DMT}, which develop numerical instabilities after $200$~kyr -- a limitation common to all existing simulations of relatively cold, strongly magnetised neutron stars.

\subsection{Temperature treatment}
\label{sec: Tevol}

Over the long-term evolution considered here ($\lesssim10^6$~yr), the internal temperature of a neutron star is neither constant nor homogeneous \citep[see][for a review]{potekhin2015}. After an early thermal relaxation phase lasting $\sim10$--$100$~yr \citep[e.g.][]{gnedin01}, during which crust and core reach a quasi-isothermal state, the star continues to cool through copious neutrino emission for the first $\mathcal{O}(100~{\rm kyr})$. The exact duration of this stage depends on the neutrino cooling channels, the envelope composition, the equation of state and the mass \citep[e.g.][]{dehman2023c,ascenzi24}, during which the internal temperature is typically $1$--$3\times10^{8}$~K. Once neutrino losses drop below the surface photon luminosity, cooling is dominated by photon emission.

Inhomogeneities in the internal and surface temperature arise from the anisotropic conductivity induced by the strong magnetic field, and from the non-uniform Ohmic dissipation that heats the crust locally. As the star cools the magnetic diffusivity decreases, so that the field dissipates more slowly; this coupling between the thermal and the magnetic evolution can in turn affect the rotational evolution.

A fully coupled 3D magneto-thermal evolution in \MATINS is computationally expensive. In most of the models presented here we therefore evolve only the magnetic field, adopting the analytical approximation of \citet{yakovlev2011} for the spatially homogeneous, time-evolving redshifted temperature $e^{\nu}T$:\footnote{Here $e^{\nu}$ is the Schwarzschild metric factor, which increases from a minimum value at the centre to $\sim0.8$ at the surface, and approaches unity far from the star \citep[see, e.g.][]{pons26}.}
\begin{equation}
     e^{\nu}T(t) = 3.45 \times 10^8~\text{K} \left(1-\frac{2GM}{c^2R}\right) \left[1+0.12 \left(\frac{R}{10~\text{km}} \right)^2\right] \left(\frac{t_c}{t} \right)^{1/6}\, ,
    \label{eq: fit cooling}
\end{equation}
where $G$ is the gravitational constant and $t_c$ is a fiducial normalisation timescale, set to the age of the Cas~A supernova remnant ($330$~yr). This prescription was calibrated on unmagnetised cooling models during the neutrino-cooling era \citep{yakovlev2011}, and gives $\sim10^{8}$~K at $\sim50$~kyr.

Two effects limit its accuracy. First, fast cooling \citep{marino24} or a high rate of Ohmic dissipation \citep{vigano13} can drive the temperature respectively lower or higher by a factor of a few. Second, the temperature can vary by a comparable factor in radius and angle, owing to localised Ohmic heating and anisotropic conduction, which in turn modify the local resistivity. These limitations are most severe at late ages ($\gtrsim10^{5}$~yr), when photon emission accelerates the cooling; in young, highly magnetised stars, where Ohmic heating is strongest; and in massive stars, where fast neutrino processes operate. For this reason we also run the \CCSN initial configuration with the fully coupled 3D magneto-thermal evolution (model \texttt{CCSN-3DMT}), in order to quantify the differences with respect to the simplified treatment adopted for the remaining models.

\begin{figure*}
    \centering
   \includegraphics[width=0.45\textwidth]{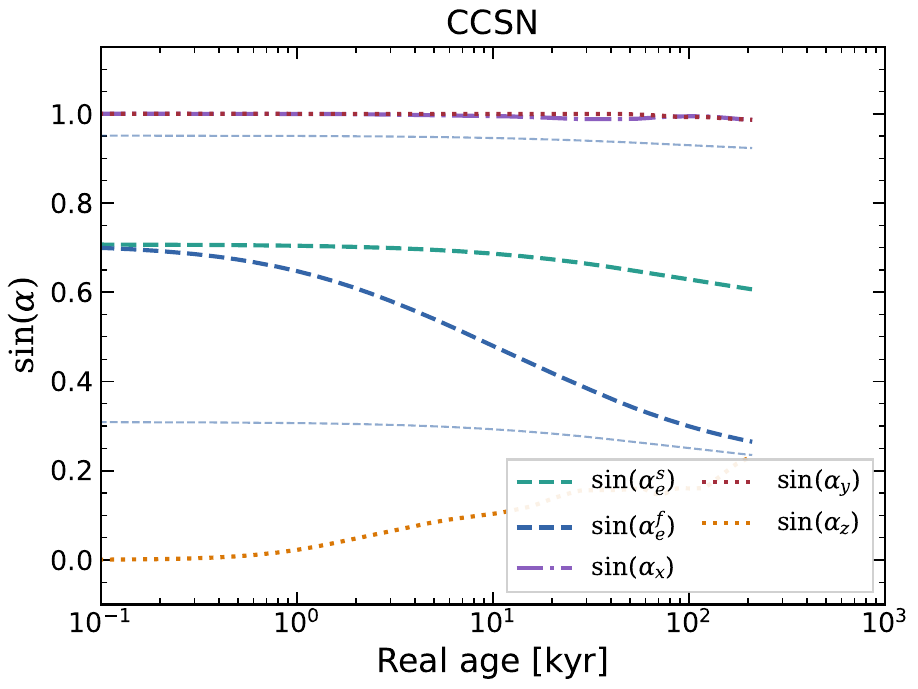}  
   \hspace{1em}
   \includegraphics[width=0.45\textwidth]{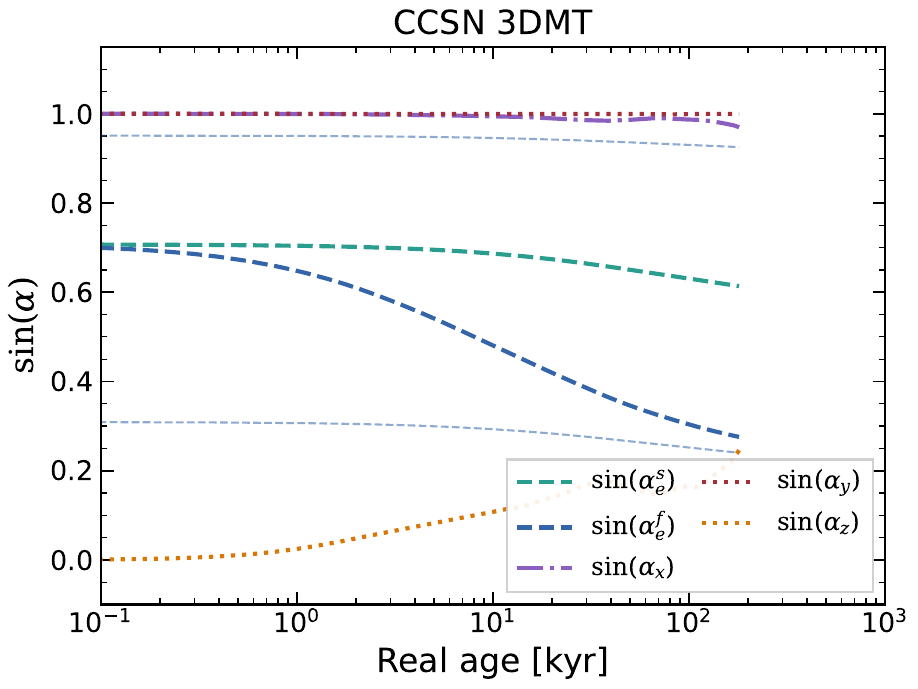}
   \vspace{5pt}
     \includegraphics[width=0.45\textwidth,height=0.33\textwidth]{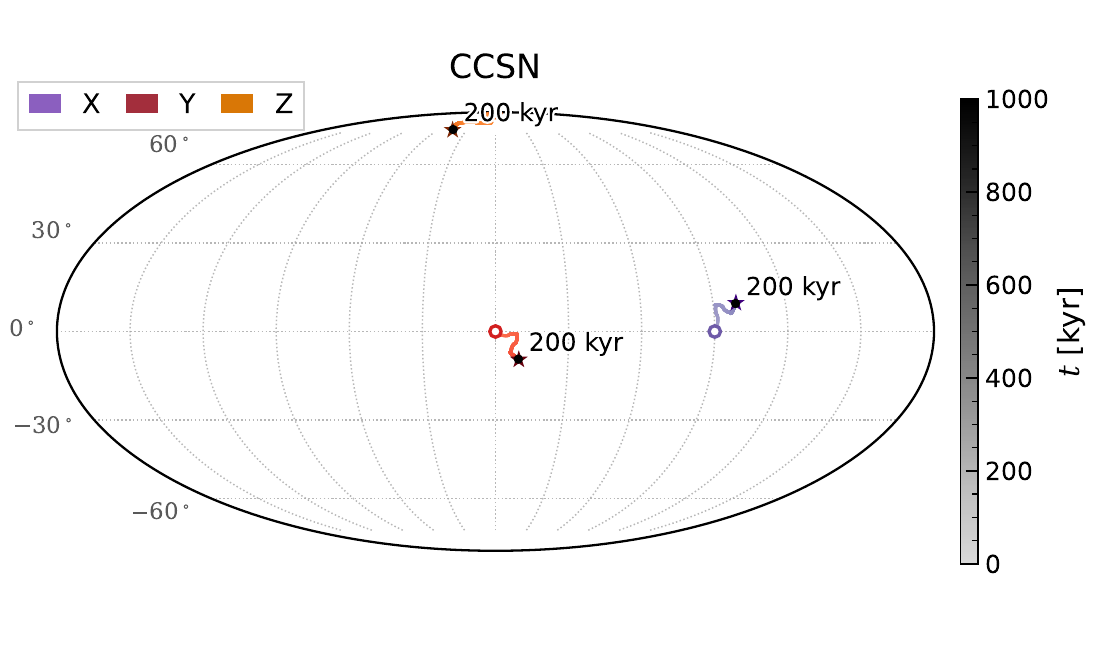}  
   \hspace{1em}
   \includegraphics[width=0.45\textwidth,height=0.33\textwidth]{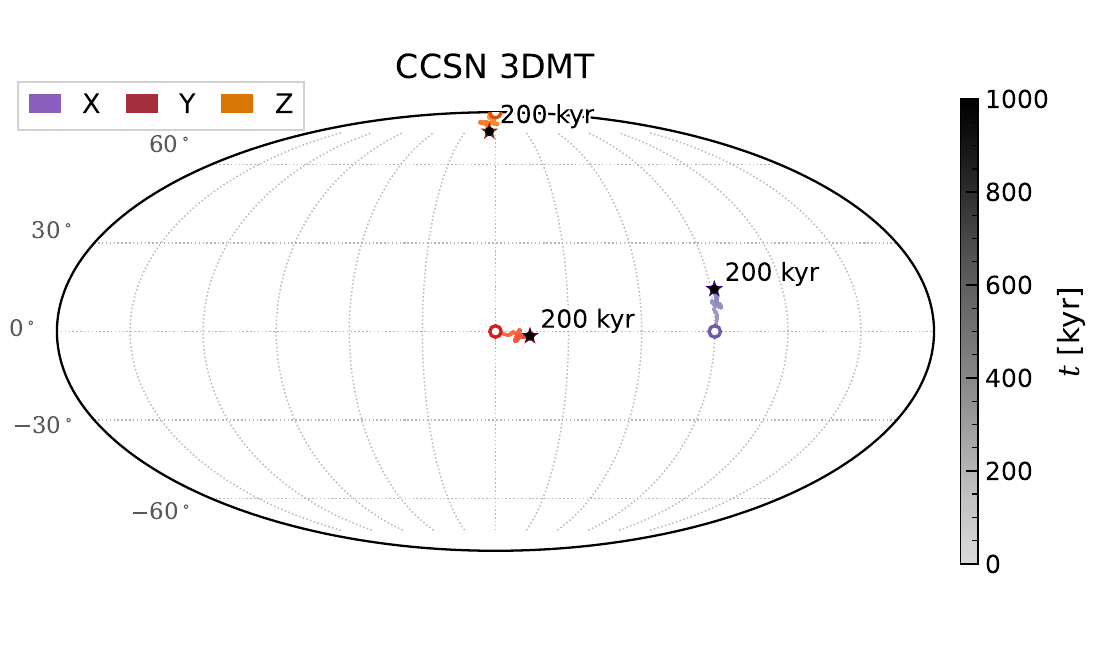} 
\caption{Obliquity evolution in the \CCSN (\emph{left}) and \texttt{CCSN-3DMT} (\emph{right}) models. \emph{Top}: evolution of $\sin\alpha$ driven either by alignment (blue dashes for $P_0^f=10$~ms, thick for $\alpha_{e,0}=\pi/4$, thin for $\alpha_{e,0}=0.1\pi$ and $0.4\pi$; cyan dashes for $P_0^s=50$~ms) or by spontaneous drift, for the three choices of initial dipolar-axis direction along $x$, $y$ and $z$ (purple dash-dotted, red dotted and orange dotted). \emph{Bottom}: Mollweide projection of the dipolar-axis trajectory for the same three choices. The spin axis is directed north--south ($z$-axis in this frame), and the original simulation frame is rotated so that the initial dipolar moment points towards $x$ (equator, central longitude, purple), $y$ (equator, centre-east, red) or $z$ (north, orange). The latitudinal distance of the dipolar axis from the north pole is $\alpha_i$, whose sine is plotted in the top panels. Within each track, colour darkens with age, spanning the range indicated by the grey colour bar shared by the three tracks. The circle marks $t=0$ and the star the end of the simulation.}
\label{fig: CC evol}
\end{figure*}

\section{Results}\label{sec: results}

This section focuses on the contributions of the magnetic and obliquity evolution to the braking index. Further details of the magnetic and rotational evolution (magnetic spectra, $B_d(t)$, period, period derivative, and characteristic versus real age), obtained with standard techniques \citep{pons26}, are left to App.~\ref{app: magrot evolution}. We discuss here the configurations that display a non-trivial obliquity evolution, leaving the benchmark dipole-dominated cases \texttt{D12} and \texttt{D14} to App.~\ref{app: large scales}. In brief, with $\gtrsim95\%$ of their magnetic energy in the dipolar component, those models show no appreciable spontaneous drift: the dipolar axis stays essentially fixed, so that the long-term contributions to the braking index are set by $B_d$ and by the alignment torque alone, as in previous dedicated studies \citep{pons12,gourgouliatos15,igoshev20}.

\subsection{\texttt{CCSN} and \texttt{CCSN-3DMT}}
\label{sec: reference run}

We begin with \CCSN and \texttt{CCSN-3DMT}, which illustrate our results for a fiducial configuration while comparing the two temperature treatments.

\begin{figure*}
    \centering   
\includegraphics[width=0.45\textwidth]{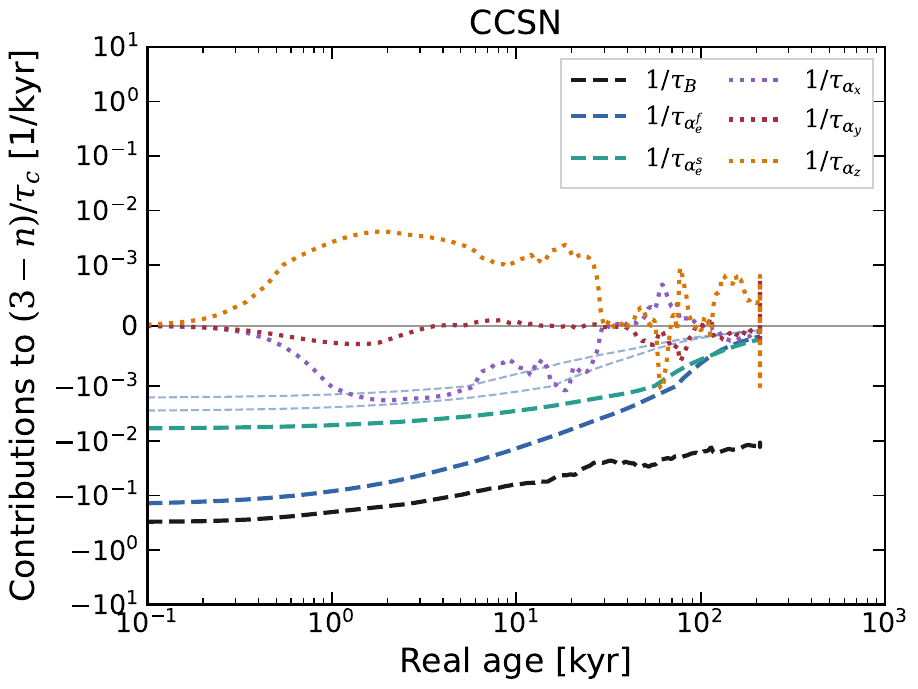}
\hspace{1em}
\includegraphics[width=0.45\textwidth]{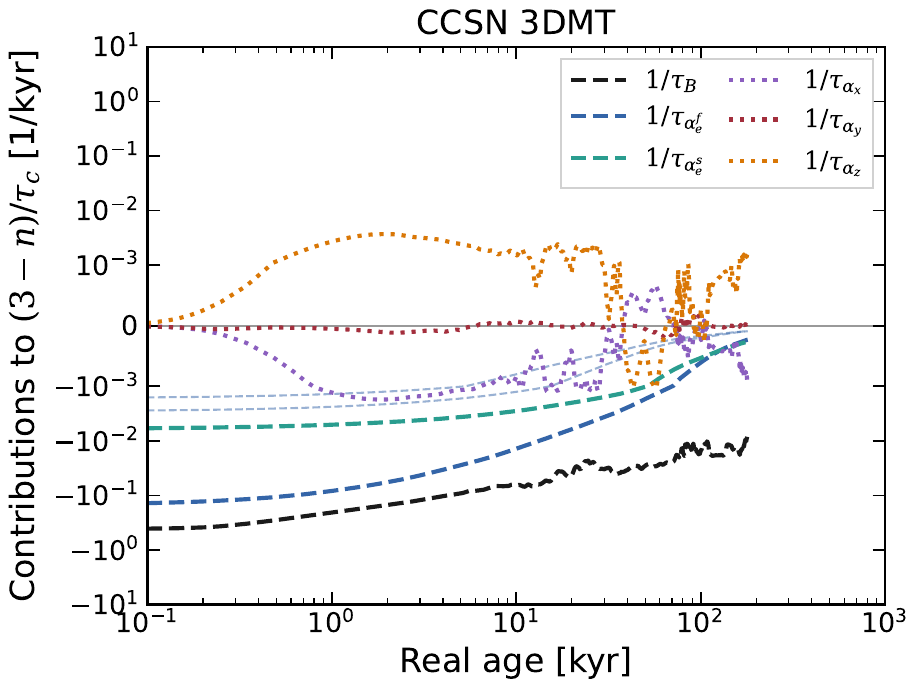}
\caption{Evolution of the contributions to the de-trended braking index for the \CCSN (\emph{left}) and \texttt{CCSN-3DMT} (\emph{right}) models: $1/\tau_B$ (black dashed), $1/\tau_{\alpha_i}$ (purple, red and orange dotted for $i=x,y,z$), and $1/\tau_{\alpha_e}$ for a fast ($P_0^f=10$~ms, blue dashes, thick and thin as in Fig.~\ref{fig: CC evol}) and a slow ($P_0^s=50$~ms, cyan dashes) rotator. The $y$-scale is linear within $[-10^{-3},10^{-3}]$ and logarithmic outside.}
\label{fig: braking index CC}
\end{figure*} 

\subsubsection{Obliquity evolution}

The top panels of Fig.~\ref{fig: CC evol} show $\sin\alpha_k$ ($k=x,y,z,e$) over the $200$~kyr simulated, for \CCSN (left) and \texttt{CCSN-3DMT} (right). The dashed lines follow the alignment dynamics $\sin\alpha_e(t)$ (Sect.~\ref{sec:alignment}) for the fast ($P_0^f=10$~ms, blue) and slow ($P_0^s=50$~ms, cyan) initial periods; both move appreciably towards alignment, more so for the fast rotator because of the $1/P^2$ factor in Eq.~(\ref{eq: obliquity evol}). For the fast case we also show two alternative initial obliquities, $\alpha_{e,0}=0.1\pi$ and $0.4\pi$ instead of $0.25\pi$, which as expected align substantially more slowly.

The purple dash-dotted, red dotted and orange dotted lines follow the spontaneous drift $\sin\alpha_i(t)$ for the three geometric choices (Sect.~\ref{sec:wandering}). The $\alpha_x$ and $\alpha_y$ geometries, both initially orthogonal rotators, vary negligibly. The variation of $\sin\alpha_z$, in contrast, exceeds that of every alignment track except the fast rotator with $\alpha_{e,0}=\pi/4$, which it approaches without quite matching.

The bottom panels of Fig.~\ref{fig: CC evol} display the same drift directly on the sphere. The spin axis lies along the north--south direction, and each track follows the magnetic dipolar axis for one of the three initial geometries ($x$, $y$ and $z$: purple, red and orange, respectively), with colour darkening with age. Each track starts at the position set by its geometry: aligned with the spin axis, at the north pole, in the $z$ case, and orthogonal to it, on the equator, in the $x$ and $y$ cases. As the system evolves, the three trajectories drift away from these initial configurations, with a total net displacement of $\mathcal{O}(10)$ degrees over the $200$~kyr simulated.

The direction of this drift matters: only latitudinal motion, measured with respect to the spin axis, changes the obliquity, whereas longitudinal motion does not. The weight of a given track therefore depends on the choice of spin axis, and this is why $\alpha_z$ produces by far the largest variation in $\sin\alpha$. The three geometries drift by a comparable $\sim10$ degrees, but the same angular displacement maps onto very different changes in the sine: $\sin\alpha_z\to\alpha_z+\mathcal{O}(\alpha_z^3)$ as $\alpha_z\to0^+$, which is linear, whereas $\sin\alpha_x\to1-(\pi/2-\alpha_x)^2/2+ \mathcal{O}\big((\pi/2-\alpha_x)^4\big)$ as $\alpha_x\to\pi/2^-$, which is only quadratic. At early times $\alpha_y$ varies much less than $\alpha_x$, because most of the drift in the $y$ geometry is longitudinal.

These differences between the three representative geometries underline the importance of the orientation of the drift track, which is set by the relative orientation of the spin and dipolar axes. Other initial geometries -- for instance intermediate values of $\alpha$, with tracks starting at mid latitudes -- would produce different tracks, but their variations of $\sin\alpha$ would generally fall within the range spanned by these three representative cases.

The two thermal treatments give a very similar dipolar drift up to late ages, where the tracks start to differ in shape, although the variations remain of the same order of magnitude. The origin of the difference is that the analytical fit of Eq.~(\ref{eq: fit cooling}) is reliable during the neutrino-cooling era, whereas the self-consistent 3D evolution of \texttt{CCSN-3DMT} also captures the temperature inhomogeneities and the photon-cooling era; during the latter the star cools further, the conductivity increases, and the dynamics become more Hall-dominated. These differences are nonetheless minor compared with those produced by the initial magnetic-field
configuration (Sect.~\ref{sec: SS} and App.~\ref{app: large scales}), by the relative orientation of the spin and dipolar axes, and by the initial period that sets the alignment.

\subsubsection{Contributions to the braking index}

Figure~\ref{fig: braking index CC} shows the individual terms contributing to $(3-n)/\tau_c$: $1/\tau_B$ and $1/\tau_{\alpha_k}$ with $k=e,x,y,z$, where $\alpha_e$ is evaluated for both initial periods and for the different initial obliquities discussed above.

In both models $1/\tau_B$ (black dashed) remains strongly negative throughout, decreasing in magnitude from $\sim-0.4$~kyr$^{-1}$ at $t\sim10^2$~yr to $\sim-0.05$~kyr$^{-1}$ by $t\sim2\times10^5$~yr. The two thermal prescriptions give nearly identical curves, apart from slightly stronger fluctuations in the 3D magneto-thermal case, caused by its richer Hall dynamics at late times. The $1/\tau_B$  term exceeds every obliquity contribution by one to two orders of magnitude. The only exception is $\alpha_e^f$ with $\alpha_{e,0}=\pi/4$ (thick blue dashes), which has the same sign and a comparable magnitude until $t\sim1$~kyr; the spontaneous drift (purple, red and orange dotted), the slow-rotator external torque (cyan dashes), and the fast-rotator torque for $\alpha_{e,0}=0.1\pi$ and $0.4\pi$ (thin blue dashes) are much smaller at all times. The decay of the surface dipolar field is therefore the dominant contribution in these models: it shows no early-time growth of $B_d$, and systematically drives $n>3$ over the whole timespan considered.

The evolution of $1/\tau_B$ and $1/\tau_{\alpha_e}$ is smooth and monotonic, whereas the spontaneous-drift contributions $1/\tau_{\alpha_i}$ oscillate rapidly between negative and positive values, slightly more abruptly in \texttt{CCSN-3DMT}. Their magnitude is $\lesssim5\times10^{-3}$~kyr$^{-1}$, and they become comparable to, or larger than, the externally driven contribution for middle-aged stars ($t\gtrsim50$~kyr). This stochastic behaviour is a promising candidate to explain the fluctuating braking indices observed in neutron stars older than $\sim50$~kyr (Sect.~\ref{sec: Timing}).

Overall, the two thermal treatments differ too little to change the order of magnitude of any contribution to the braking index. The remaining configurations therefore adopt the computationally inexpensive analytical prescription for the temperature, bearing in mind that a fully coupled treatment would mainly yield a less smooth, more dynamical obliquity evolution at late times, without qualitative consequences for our conclusions.

\subsection{The small-scale dominated configurations}
\label{sec: SS}

\begin{figure*}
\centering
\includegraphics[width=0.33\linewidth]{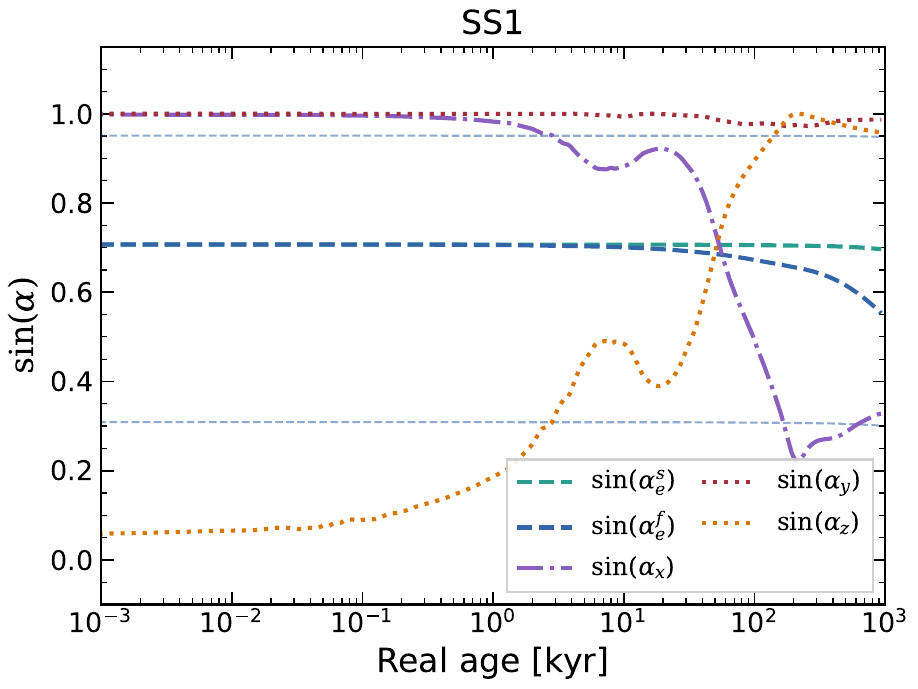} 
\hfill
\includegraphics[width=0.33\linewidth]{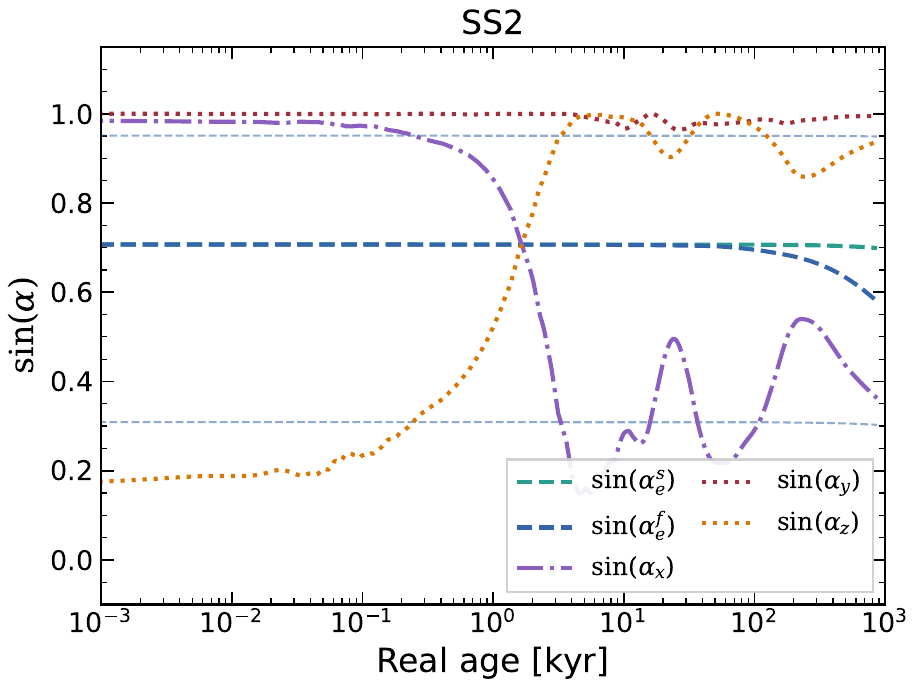} 
\hfill
\includegraphics[width=0.33\linewidth]{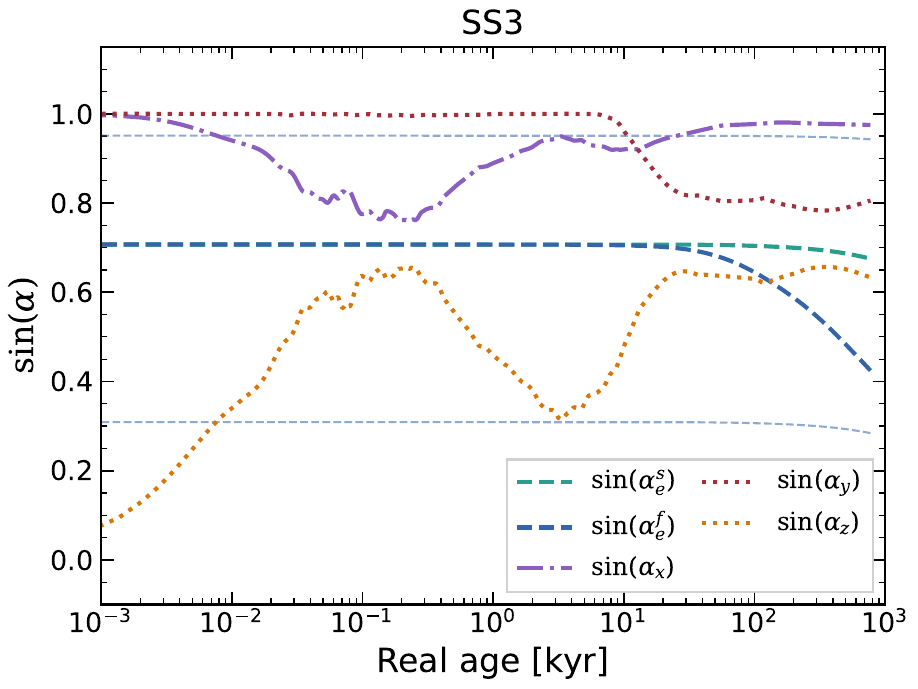} 
\vspace{5pt}
\includegraphics[width=0.33\linewidth,height=0.26\textwidth]{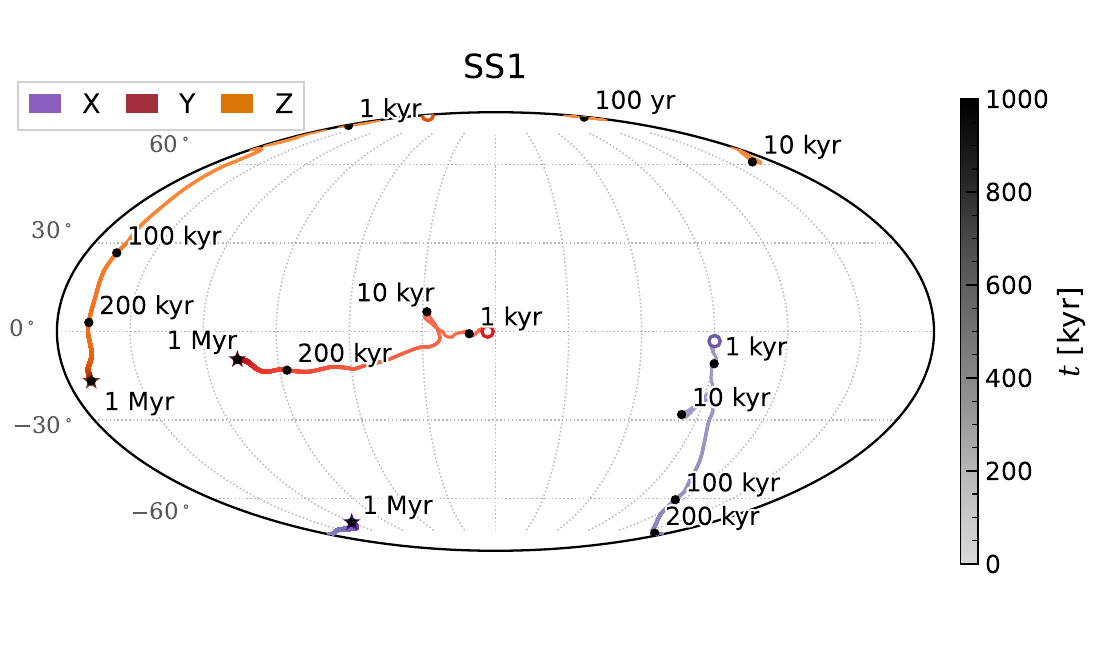} 
\hfill
\includegraphics[width=0.33\linewidth,height=0.26\textwidth]{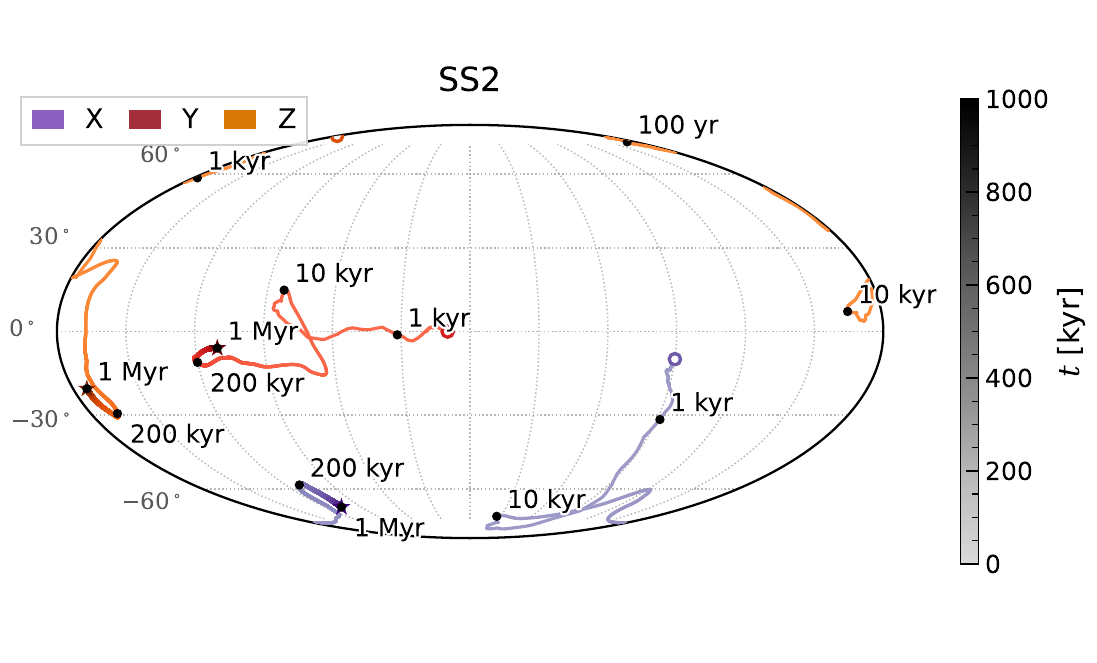} 
\hfill
\includegraphics[width=0.33\linewidth,height=0.26\textwidth]{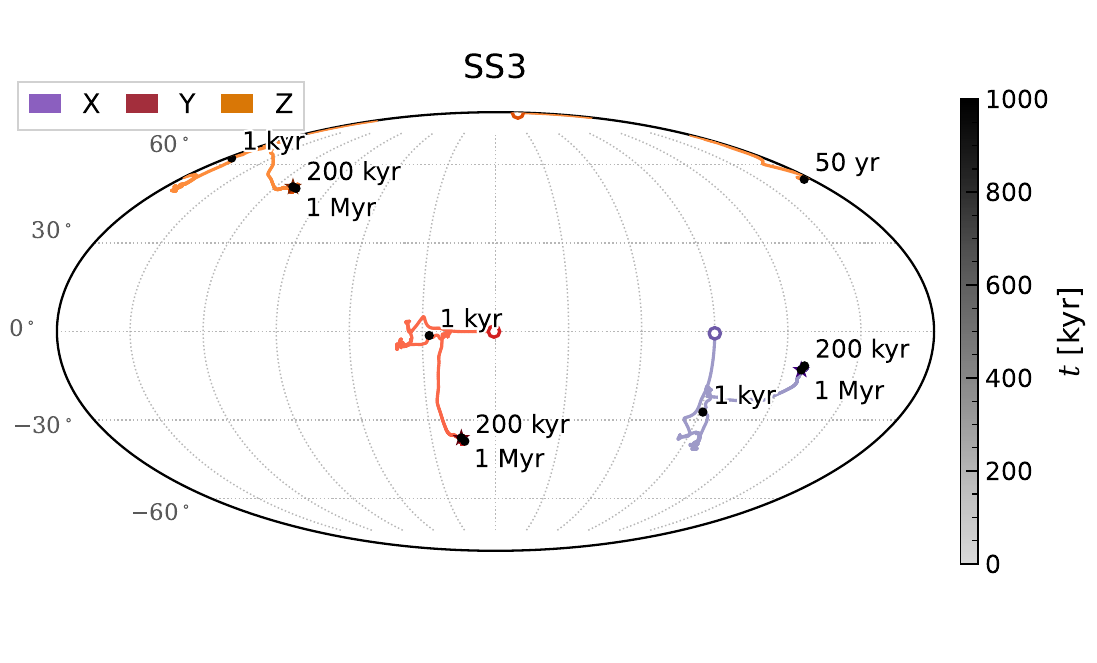} 
\vspace{5pt}
\includegraphics[width=0.33\linewidth]{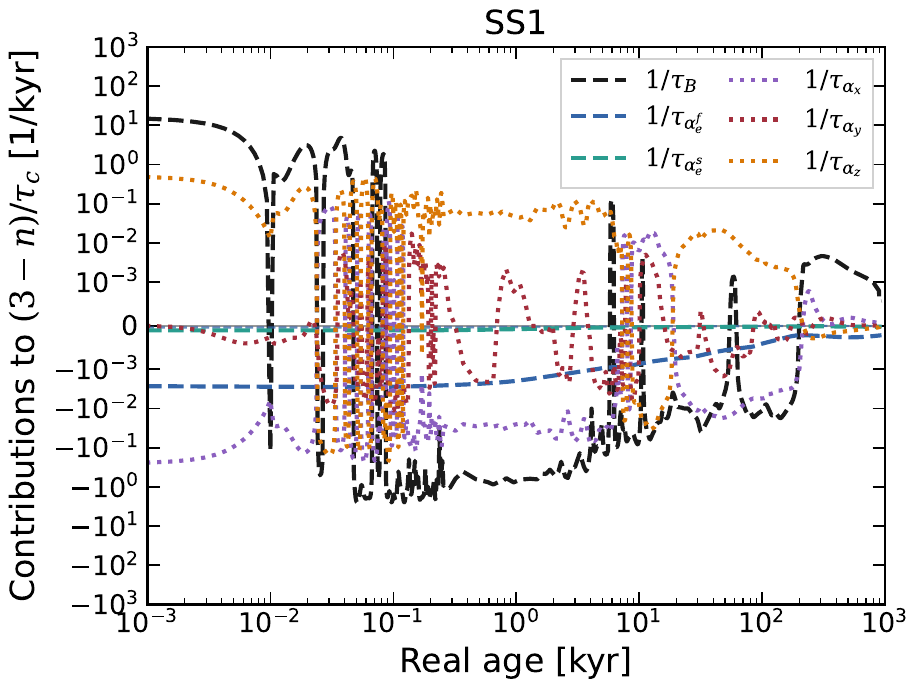} 
\hfill
\includegraphics[width=0.33\linewidth]{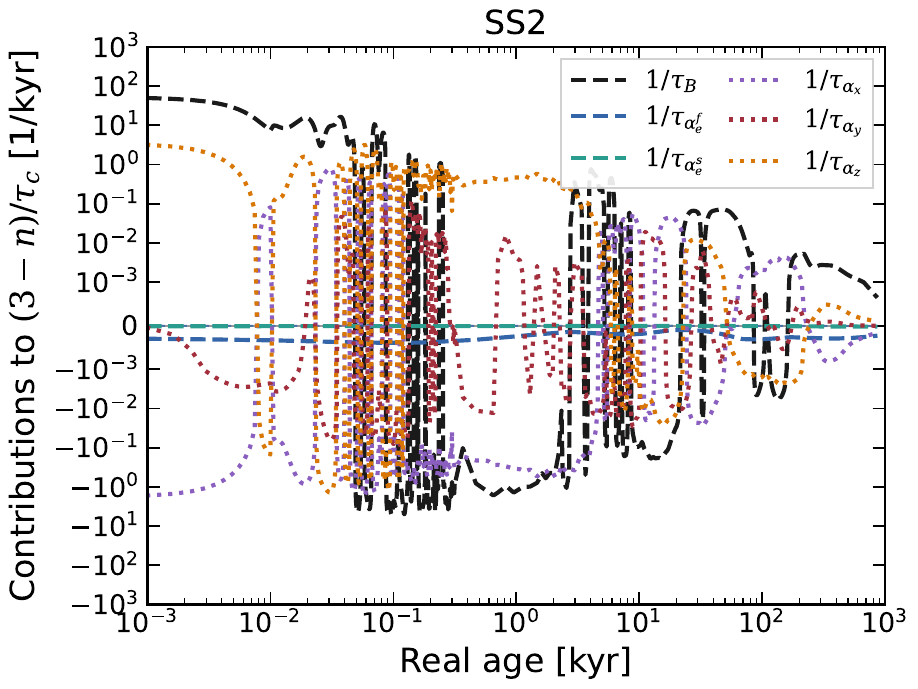} 
\hfill
\includegraphics[width=0.33\linewidth]{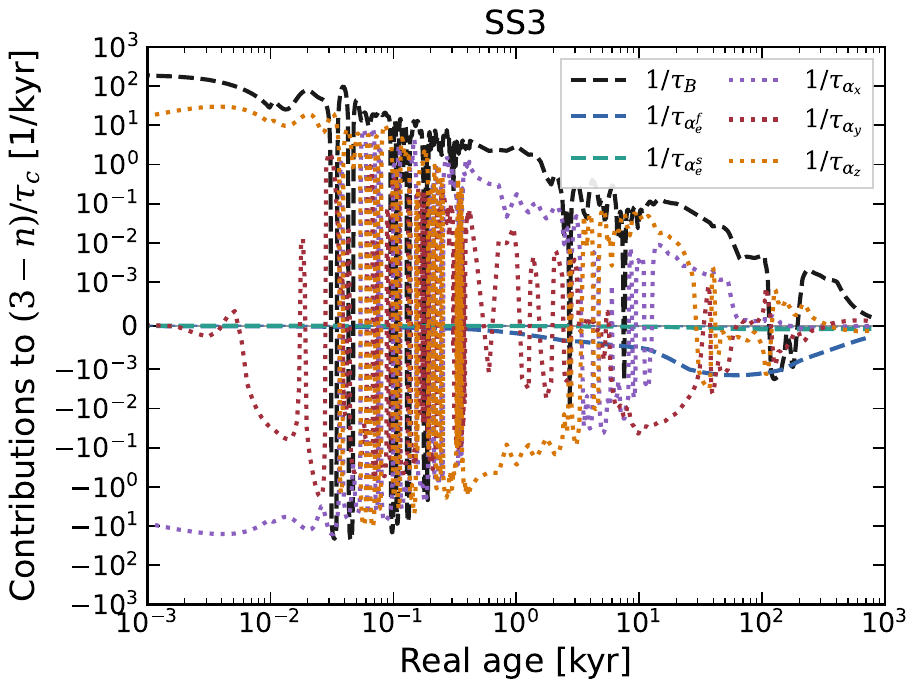} 
\caption{Time evolution of $\sin\alpha$ (\emph{top}), map of the dipolar-axis drift (\emph{middle}), and contributions to $(3-n)/\tau_c$ (\emph{bottom}), for the \texttt{SS1} (\emph{left}), \texttt{SS2} (\emph{centre}) and \texttt{SS3} (\emph{right}) runs. Colours and line styles are as in Figs.~\ref{fig: CC evol} and~\ref{fig: braking index CC}. Selected ages are marked along the tracks in the middle panels.}
\label{fig: SS run}
\end{figure*}

Figure~\ref{fig: SS run} shows the three small-scale runs. They share the same qualitative behaviour: a large, erratic dipolar-axis drift accompanied by a strongly fluctuating contribution to the braking index. They differ, however, in the amplitude and timing of the drift and in the evolution of the dipolar intensity, so that their $n(t)$ are qualitatively different.

In all three the dipolar axis drifts much farther than in \CCSN, by several tens of degrees rather than $\mathcal{O}(10)$ degrees. Part of this is simply a longer baseline -- \texttt{SS1}--\texttt{SS3} are followed to $\sim1$~Myr against the $200$~kyr of \CCSN -- but the comparison holds at fixed age as well: restricted to the first $200$~kyr, the three runs have already accumulated most of their total excursion. The details of that drift are nonetheless specific to each realisation. In \texttt{SS3} the axis moves by $\sim40^\circ$ within the first $\sim50$~yr, when the dipole is very weak and therefore most prone to drift, and thereafter wanders back and forth within that cone for the rest of the evolution. In \texttt{SS1} it moves by only a few degrees during the first kyr and then performs a single broad sweep of $\sim80^\circ$, completed by $\sim200$~kyr. In \texttt{SS2} it reorients by a comparable amount already at $\sim1$~kyr and then keeps oscillating, with swings of a few tens of degrees, until the end of the run. These runs are therefore a more extreme version of \CCSN: with almost all the magnetic energy at small scales, the weak dipolar component is entirely controlled by the small-scale dynamics, and the externally induced alignment is negligible except for the fast rotator at late ages.

The contributions to the braking index reflect this. The spontaneous-drift terms $1/\tau_{\alpha_i}$ are orders of magnitude larger than in \CCSN, with an amplitude that varies from run to run, reaching a few tens of kyr$^{-1}$ at early ages in \texttt{SS3} against $\sim0.3$~kyr$^{-1}$ in \texttt{SS1}. They rarely dominate the budget -- $1/\tau_B$ remains the largest single term throughout most of the evolution -- but they are now comparable to it, and fluctuate frequently, rather than being negligible as in \CCSN. More importantly, $1/\tau_B$ is itself no longer smooth: it reverses sign several times, as do all the other internal terms, while the external-torque contributions remain orders of magnitude smaller at all times. These reversals follow directly from the non-monotonic evolution of $B_d$ shown in App.~\ref{app: magrot evolution}, and make the predicted braking index oscillate about its secular trend instead of following it smoothly.

Beyond these rapid reversals, the secular sign of $1/\tau_B$ is the clearest systematic difference between the runs, and simply follows the evolution of the dipolar component: it is predominantly negative in \texttt{SS1}, whose dipole mostly decays, driving $n>3$; it alternates in \texttt{SS2}, whose dipole grows and decays in turn; and it is positive over most of the evolution in \texttt{SS3}, whose dipole is amplified by the transfer of magnetic energy from small to large scales, driving $n<3$. In small-scale dominated configurations
the obliquity contribution is therefore robustly present and sign-alternating, whereas the sign of the $B_d$ term is not a generic property of the class but depends on the particular initial spectrum, and in particular on how much energy the dipole starts with.

Note that the rotational evolution of these models is marked by very large values of $\tau_c/t$ at early ages, particularly for slow initial periods, which decrease only slowly and still exceed unity at the end of the run (App.~\ref{app: magrot evolution}). Where the dipole grows, the corresponding tracks in the $P$--$\dot P$ diagram first rise towards larger $\dot P$ and only later turn over, once $B_d$ begins to decay -- a behaviour compatible with the timing properties of young pulsars showing $n<3$ \citep{espinoza11}.

\subsection{Synthetic comparison of models with data}
\label{sec: comp with obs}

\begin{figure*}
\centering
\includegraphics[width=0.33\linewidth]{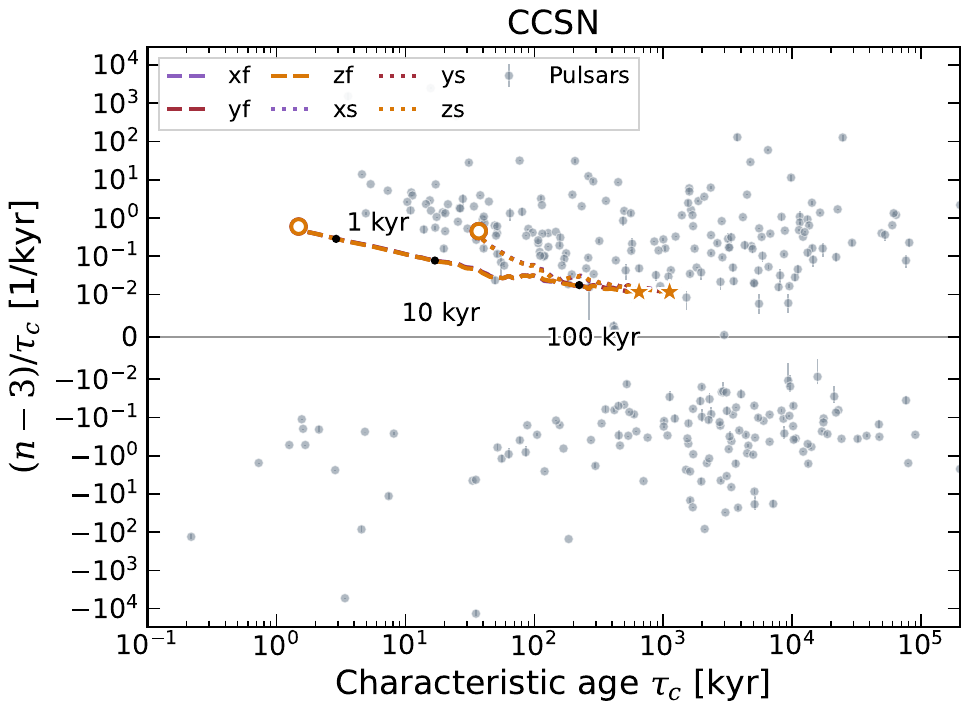}
\includegraphics[width=0.33\linewidth]{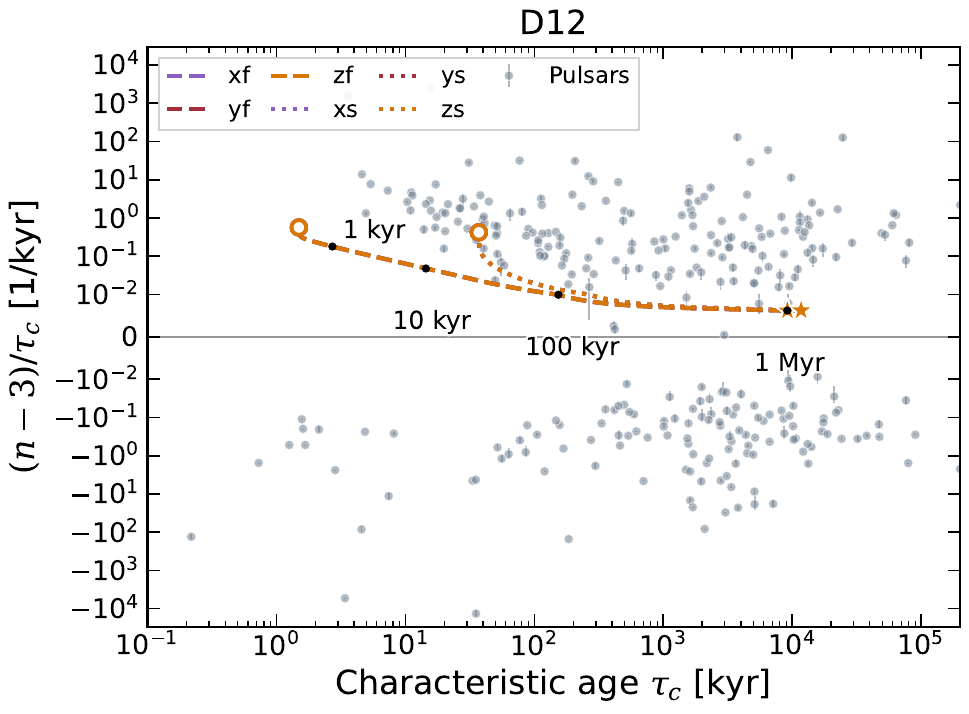}
\includegraphics[width=0.33\linewidth]{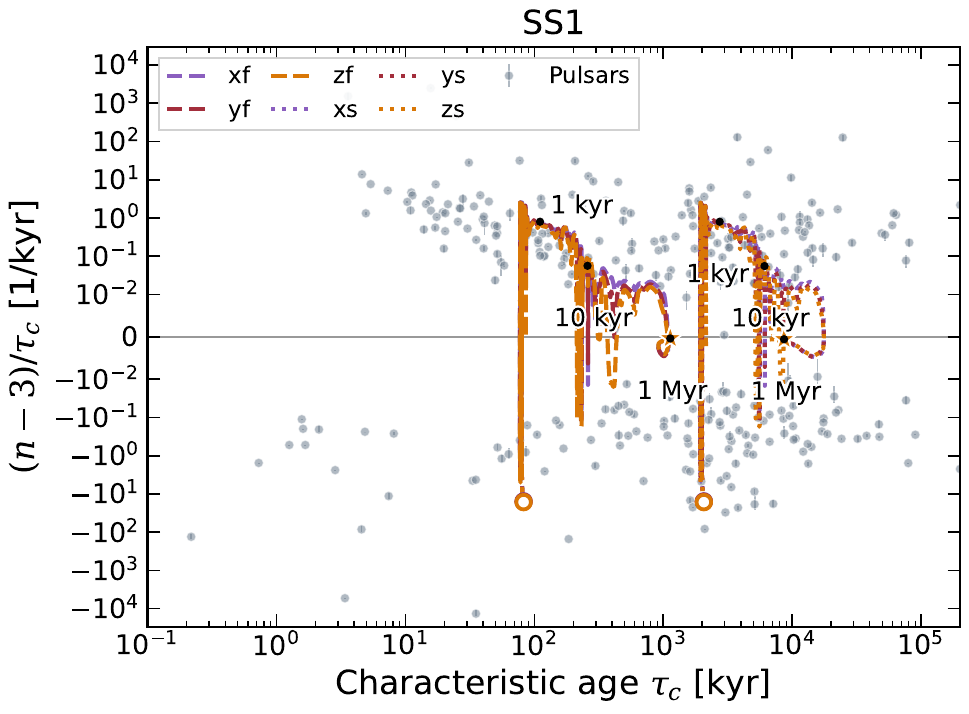}
\includegraphics[width=0.33\linewidth]{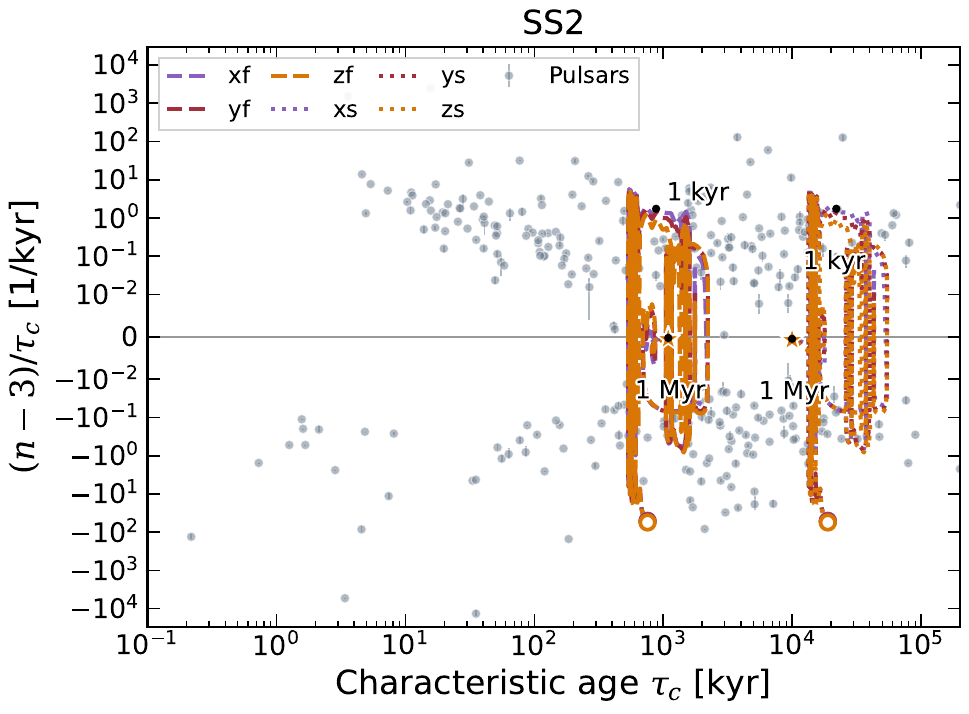}
\includegraphics[width=0.33\linewidth]{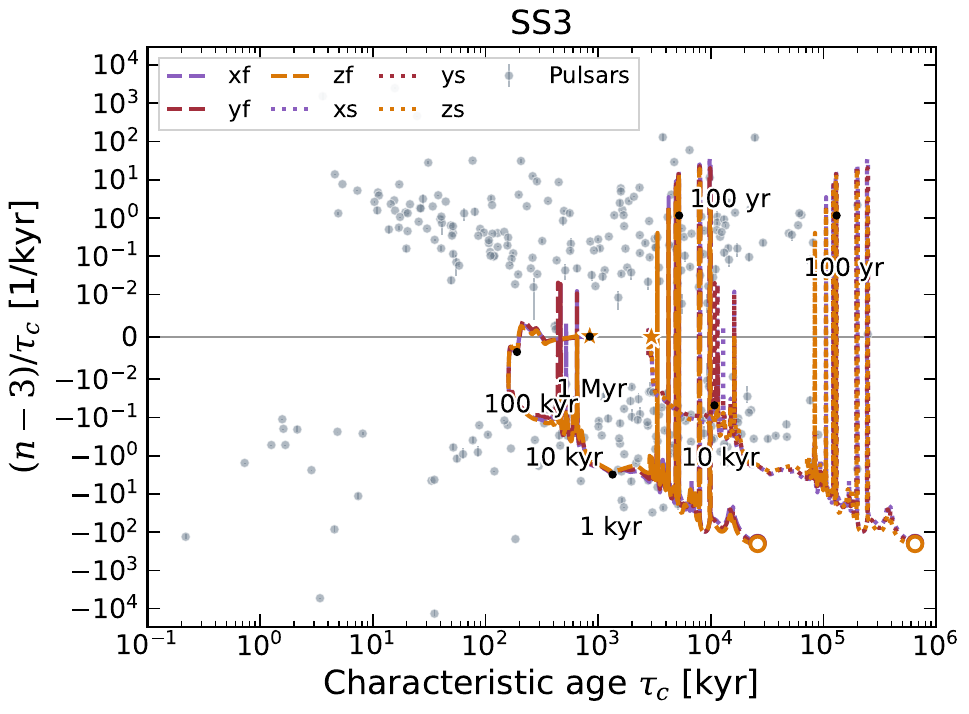}
\includegraphics[width=0.33\linewidth]{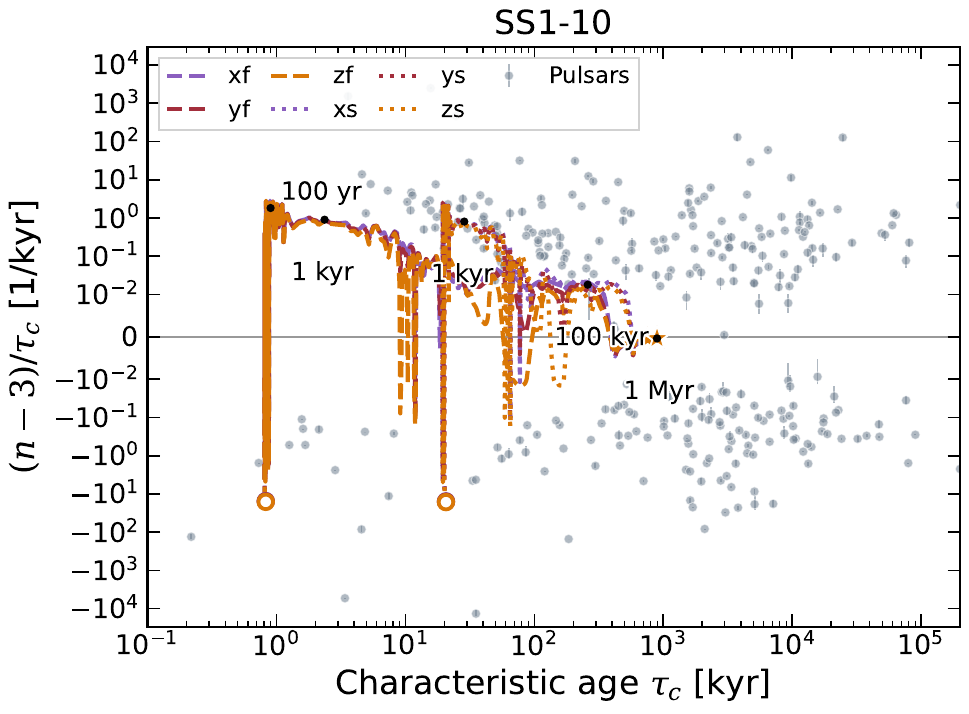}
\caption{Synthetic de-trended braking index, $(n-3)/\tau_c=-(1/\tau_{\alpha_i}+1/\tau_{\alpha_e}+1/\tau_B)$, as a function of the characteristic age, for the \CCSN, \texttt{D12}, \texttt{SS1}, \texttt{SS2}, \texttt{SS3} and \texttt{SS1-10} models (left to right, top to bottom), computed for the different geometries $i$ (colours as in the previous figures) and initial periods $P_0^f$ (dashes) and $P_0^s$ (dots). Open circles mark $t=0$ and stars the end of each track, along which selected ages are labelled. To allow all tracks to be compared in the same plot, the characteristic age is computed from the $P(t)$ evolution that includes $\alpha_e(t)$; using $\alpha_i(t)$ instead produces only minor differences. The scale is linear in the range $[-10^{-2},10^{-2}]$ and logarithmic outside it. Grey points with error bars are the observed pulsars of our sample (Sect.~\ref{sec: Timing}).}
\label{fig: n tau models}
\end{figure*}

Figure~\ref{fig: n tau models} summarises the combined contributions from $B_d(t)$ and $\alpha(t)$ to the de-trended braking index,\footnote{As discussed in Sect.~\ref{sec: bd alpha}, a more self-consistent approach would couple the obliquity changes due to alignment and drift, and consider a single timescale $\tau_\alpha$. Given the order-of-magnitude differences between models, a simple sum of the inverse estimated timescales, $1/\tau_{\alpha_i}+1/\tau_{\alpha_e}$, suffices for our purposes.} $(n-3)/\tau_c=-(1/\tau_{\alpha_i}+1/\tau_{\alpha_e}+1/\tau_B)$, as a function of $\tau_c$, for the different models, initial obliquities (colours) and initial periods (dashes for the fast rotator, dots for the slow one). The tracks are directly comparable with the observed data of Sect.~\ref{sec: Timing}, shown as grey points.

The characteristic age depends crucially on the assumed initial period -- both the fast and the slow case are shown -- and on the evolution of $B_d$ (App.~\ref{app: rotational}). For the relatively low values of $B_d$ studied here, characteristic ages are large at both early and late times, the most extreme case being \texttt{SS3} with a slow initial period. Although our simulations span $\lesssim1$~Myr, the characteristic age can therefore exceed that value by orders of magnitude, covering a substantial fraction of the observed sample of Fig.~\ref{fig:obs}. Conversely, the most extreme cases never reach the observed pulsars with small characteristic ages. A combination of different initial periods and configurations, run over longer timescales than considered here, would arguably cover the full observed range of $\tau_c$, but this calls for a costly population synthesis approach that lies beyond the scope of this work. We aim instead at sampling representative points in a vast parameter space.

The models studied here separate roughly into two classes. In \CCSN and \texttt{D12} (and \texttt{D14}, not shown here), the de-trended braking index varies smoothly and monotonically with $\tau_c$. Since the dipolar-axis drift is negligible in the \texttt{D} runs and subdominant in \CCSN, the predicted braking index is set by the steady decay of $B_d$ and by the alignment, both of which drive $n>3$. As noted in Sect.~\ref{sec: intro}, oscillations can also arise in dipole-dominated cases when a large-scale (e.g.\ quadrupolar) toroidal field is present, allowing a continuous exchange with the component associated with $B_d$ \citep{pons12}. Such oscillations reach at most $|n|\sim10$, however, so that the observed braking indices would then clearly indicate that a fluctuating wobbling term, magnetospheric or superfluid in origin, dominates over the $B_d$ and $\alpha$ evolution.

In the small-scale dominated configurations (\texttt{SS1}--\texttt{SS3}) the tracks instead oscillate rapidly, crossing the $n=3$ line repeatedly and spanning the bulk of the observed range, $|(n-3)/\tau_c|\lesssim10$~kyr$^{-1}$: the dipolar axis drifts by several tens of degrees, the spontaneous obliquity contribution becomes comparable to or larger than that of $B_d$, and the dipole itself may grow rather than decay, fed by the transfer of magnetic energy from the small scales. This dichotomy, rather than the behaviour of any individual model, is the central result of this work.

The three runs differ, however, in the trend about which they oscillate. The sign-alternating character is common to all of them, varies on timescales far shorter than the stellar age, and shows no tendency to weaken as the star gets older; the trend it is superimposed on follows the alignment, which drives $n>3$ at middle ages, and the evolution of $B_d$ described in Sect.~\ref{sec: SS}, giving $n>3$ in \texttt{SS1}, both signs in \texttt{SS2}, and $n<3$ in \texttt{SS3}. The amplitude and rate of the fluctuations differ from run to run as well. What is generic to the class is therefore the fluctuating behaviour, not its sign.

An important limitation of these models is that their synthetic $P$--$\dot P$ tracks remain well away from the bulk of the observed sample (see Fig.~\ref{fig: ppdot} and App.~\ref{sec: PPdot}). Reaching it would require initially larger values, $B_d\sim10^{12}$--$10^{13}$~G, but more magnetised cases are numerically challenging for such tangled configurations, in which the volume-averaged field is orders of magnitude larger than the dipolar surface value. As a simplified alternative we consider a further case, \texttt{SS1-10} (bottom right panel of Fig.~\ref{fig: n tau models}), obtained by multiplying the $B_d(t)$ track of \texttt{SS1} by a factor of ten, so that the resulting $P$--$\dot P$ track lies closer to the observed population. Since the Hall term is non-linear, a genuine simulation with a ten times stronger field would not be a simple rescaling of the weaker one, and would in fact be even more dynamic; this post-processing therefore provides a conservative estimate of what a more realistic $B_d$ implies for the rotational evolution and the braking index of a small-scale field. The stronger dipole brakes the star far more efficiently and displaces the tracks by about two orders of magnitude towards lower $\tau_c$, closer to the real age, as expected from $\tau_c\propto P^2/B_d^2$. In this particular case the large fluctuations end up at values of $\tau_c$ that are too small, but the same exercise can be extrapolated to the other models: varying only the intensity of $B_d$ and the initial period, at fixed initial spectral shape, is likely enough to cover the bulk of the observed values. Small-scale dominance in the initial spectrum thus predicts the fluctuations themselves, while the range of $\tau_c$ over which
they are observed, and the trend they are superimposed on, are set by the field intensity and the initial period.

Overall, the tracks of these models reach the bulk of the observed $n(\tau_c)$ values (grey points), although a substantial fraction of the sample shows even larger variations. A wobbling term is therefore likely to be needed even in small-scale models: in other words, we would expect $\tau_{\cal W}\lesssim\tau_\alpha,\tau_B$ also in this case.

\section{Discussion}
\label{sec: discussion}

Braking index measurements are affected by several caveats related to the complexity of the timing residuals, but they display a clear correlation with $\tau_c$, expected by construction, and three distinct regimes (Fig.~\ref{fig:obs}): at early ages, $\tau_c\lesssim3$~kyr, the measured values are $n<3$, requiring an increasing torque; at middle ages, $\tau_c\sim10$--$50$~kyr, they are systematically $n>3$, with $n\sim10$--$100$; at later ages, large positive and negative values become almost equally likely. Besides possible observational biases, this behaviour must reflect non-trivial variations in the torque governing the rotational evolution.

We have considered a variety of initial models, going well beyond the usual assumption of almost purely large-scale, intense fields, and exploring instead a more realistic distribution of the initial magnetic energy over a broad range of scales. This extends the few existing studies in this direction \citep{pons12,gourgouliatos15,gourgouliatos18}, which considered almost purely large-scale initial configurations -- arguably hard to produce through the turbulent, all-scale dynamics of a core collapse -- dominated by an almost purely dipolar poloidal and/or quadrupolar toroidal field. The two components are tightly Hall-coupled \citep[e.g.][]{pons07}, so that a sufficiently strong $\ell=2$ toroidal field can make $B_d$ grow at early ages, yielding $n<3$ \citep{gourgouliatos15}. In such configurations, however, the large-scale modes necessarily remain dominant throughout the evolution, with no dipolar-axis drift and only a modest fraction of the energy flowing into a Hall direct cascade, $\EM\propto\ell^{-2}$. It remains unclear how such specific configurations could be systematically produced by the turbulent dynamics of core collapse, which tend to spread the magnetic energy across all spatial scales \citep{reboul2021}.

Small scales are relevant to the torque evolution not through any direct contribution, which is negligible \citep[e.g.][]{igoshev20}, but because they drive the dipolar component to tilt, grow, or decay faster. In particular, the spontaneous dipolar-axis drift can dominate over the externally triggered alignment, which complicates the simple prescriptions for rotational evolution commonly adopted in population synthesis studies \citep[][and references therein]{pardo25}, making them strongly dependent on the initial configuration. The large-scale configurations usually assumed raise no such difficulty.

A simple decay of $B_d$ cannot account for the large and fluctuating values of  $n$ observed at large characteristic ages, whereas small-scale dominated fields produce them naturally. The systematic $n>3$ at middle ages follows directly from the decay of $B_d$, which dominates the braking index in the large-scale dominated models. The large, sign-alternating values at later ages instead require small-scale dominated configurations, whose stochastic obliquity contributions persist without decaying as the star ages, and which reach $|n-3|$ comparable to the observed values. Their variability timescale is also of the right order: inverting the peak of the distribution of de-trended values at late ages (right panel of Fig.~\ref{fig:obs}) yields $\langle\tau_c/|n-3|\rangle\sim1$--$20$~kyr, and the internal terms in our small-scale runs fluctuate on comparable timescales, although the detailed comparison depends on the initial configuration, the field intensity and the initial period. Our models do not produce such fast fluctuations at middle ages, where the required $\sim0.3$--$5$~kyr is shorter than they typically provide; this is consistent with the absence of any observed fluctuating behaviour there, since a faster and systematic process -- the decay of $B_d$ combined with the externally induced alignment -- is already dominant.

The trends $n(\tau_c)$ thus involve the long-term evolution of three intertwined ingredients: $B_d$, the long-term alignment -- both frequently considered in the past -- and the dipolar-axis drift, barely studied before and never in full 3D, long-term simulations. The models presented here sample only a small region of this parameter space and have not been fine-tuned, so that they probably under-represent the variety of possible cases. Several ingredients could in principle be adjusted to reproduce the data more closely. The shape of the initial spectrum, and in particular the fraction of energy in the dipole, controls the evolution of $B_d(t)$ and hence the general trend of $n(t)$, about which $n$ fluctuates because of the dipolar drift. Conversely, the overall field strength and the initial period set where the tracks fall in $\tau_c$ and in the $P$--$\dot P$ diagram. A systematic exploration of these degrees of freedom should allow a more quantitative assessment of the relative importance of the spontaneous obliquity drift and of the other magnetospheric or superfluid factors -- referred to here as wobbling contributions -- responsible for the stochastic oscillations of $n$ at large $\tau_c$.

We conclude that the observed trends at middle and late ages can be reproduced by the spontaneous dipolar drift, with predicted values of $n$ compatible with the bulk of the data, while the $n<3$ measured at early ages can be explained by the inverse cascade \citep{dehmanpons25,dehman26}. A more systematic assessment is nonetheless needed. On the one hand, since our models are not fine-tuned, we cannot exclude more extreme variations of $n$ caused by the obliquity drift and the field decay. On the other, we suspect that the largest values of $|n|$ among pulsars with $\tau_c\gtrsim100$~kyr -- the tails in the right histogram of Fig.~\ref{fig:obs}, which our models do not reproduce -- will still require a short-term wobbling contribution from magnetospheric processes or superfluid dynamics. Such mechanisms are probably also needed to explain the specific shapes of the timing residuals. Throughout, one must keep in mind the observational uncertainties on $n$, their dependence on the baseline, and more generally the fact that timing residuals show a complexity that goes well beyond a single braking index value.

\begin{acknowledgements}
CD is supported by the Ministerio de Ciencia, Innovaci\'on y Universidades (JDC2023-052227-I), co-funded by AEI (MCIN/AEI/10.13039/501100011033), the European Social Fund Plus (ESF+), and the Universidad de Alicante. CD acknowledges support from the Conselleria d'Educaci\'o, Cultura, Universitats i Ocupaci\'o de la Generalitat Valenciana (grant CIPROM/2022/13), and the computing resources allocated by the Swedish National Allocations Committee at the Center for Parallel Computers at the Royal Institute of Technology in Stockholm (Sweden). DV is supported by the European Research Council (ERC) under the European Union's Horizon 2020 research and innovation programme (ERC Starting Grant ``IMAGINE'', No.~948582), and by the project PID2025-172309OB-I00, funded by MCIU/AEI/10.13039/501100011033 and by the ESF+. We acknowledge the ``Mar\'ia de Maeztu'' award from the Spanish Ministry of Science and Innovation to the Institut de Ci\`encies de l'Espai (CEX2020-001058-M), the MaX-CSIC Excellence Award MaX4-SOMMA-ICE, and the support of the Generalitat de Catalunya/CERCA programme. The publishing costs were covered by the IEEC award for the best PhD thesis of 2023, granted to CD and her supervisors (DV, Nanda Rea, and Jos\'e A. Pons).
\end{acknowledgements}

\bibliographystyle{aa} 
\bibliography{aa63071-26} 

\begin{appendix}
\nolinenumbers

\section{The dipole-dominated configurations}
\label{app: large scales}

We discuss here the \texttt{D12} and \texttt{D14} models, in which the magnetic energy is almost entirely in the dipolar component. Such initial conditions are directly comparable to those adopted in several previous studies, which considered only large-scale configurations: the spherical-harmonic degree was mostly fixed to $\ell=1$ or, in the few cases where a range of configurations was explored, did not exceed $\ell=4$ \citep{pons12,gourgouliatos18}. They serve as the counterpart to the complex, small-scale configurations studied in the main text.

These configurations develop some small-scale structure through a direct Hall cascade \citep{dehman23matins}, but the dipole itself does not drift, since the underlying large-scale currents dissipate smoothly and without substantial reorganisation. An intense large-scale component at birth therefore leaves a lasting imprint on the subsequent evolution, keeping the dipolar axis far more stable than in more complex initial configurations.

\begin{figure*}
    \centering  
\includegraphics[width=0.45\textwidth]{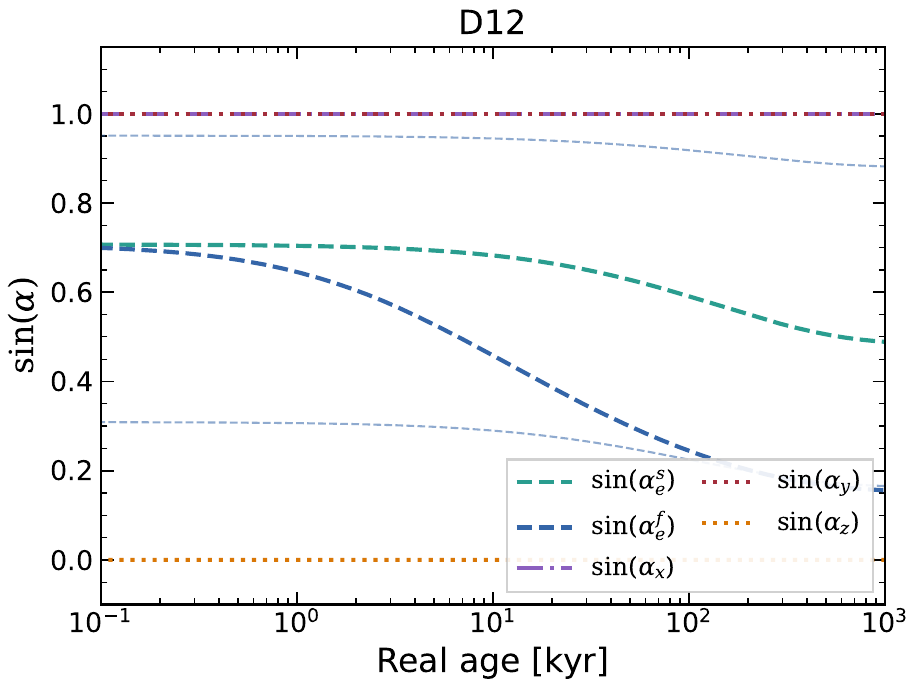}
  \hspace{1em}
\includegraphics[width=0.45\textwidth]{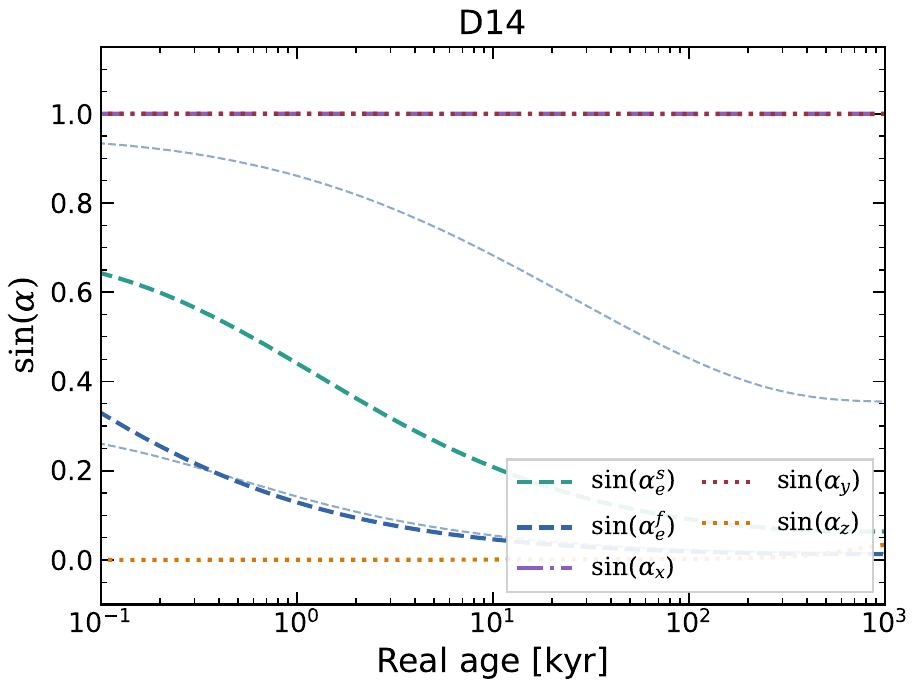}
    \vspace{0.5em}
\includegraphics[width=0.45\textwidth]{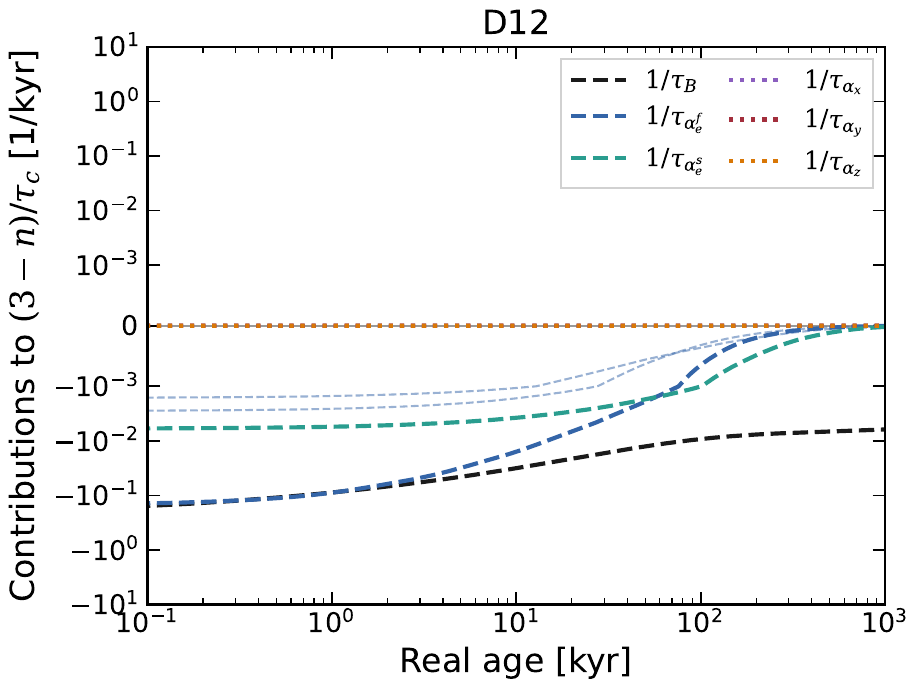}
\hspace{1em}
\includegraphics[width=0.45\textwidth]{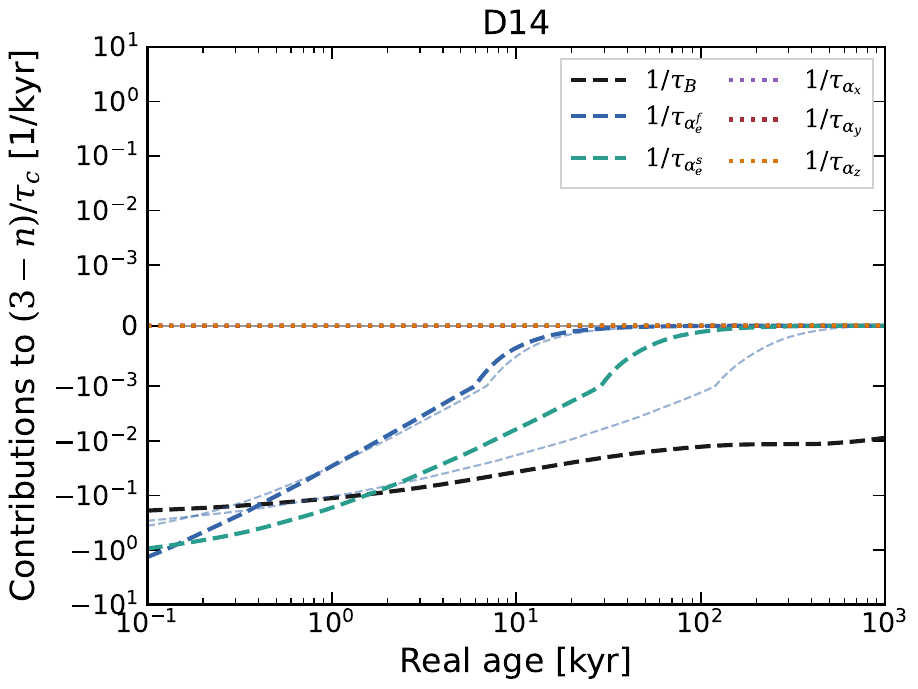}
\caption{Evolution of $\sin\alpha$ (\emph{top}) and contributions to $(3-n)/\tau_c$ (\emph{bottom}), for \texttt{D12} (\emph{left}) and \texttt{D14} (\emph{right}). Colours and line styles are as in Figs.~\ref{fig: CC evol} and~\ref{fig: braking index CC}. The maps of the dipolar-axis drift are omitted, as the drift is negligible on the scale of the map.}
\label{fig: D runs}
\end{figure*}

The top panels of Fig.~\ref{fig: D runs} show the evolution of $\sin\alpha$. The internal obliquities $\alpha_i$ remain strictly constant over the whole $1$~Myr covered by these runs, in contrast with \CCSN (Fig.~\ref{fig: CC evol}), so that any obliquity evolution driven by internal dynamics can safely be neglected. By contrast, the external torque is fully at play: in \texttt{D12} the evolution of $\sin\alpha_e$ is qualitatively similar to \CCSN for a given $P_0$. \texttt{D14} differs from both, since its stronger initial $B_d$ accelerates the alignment, as expected from the $B_d^2$ factor in Eq.~(\ref{eq: obliquity evol}): $\sin\alpha_e$ drops by a factor of $5$--$20$ depending on $P_0$ and $\alpha_{e,0}$, and the star is essentially aligned by $t\sim10$~kyr for the fast rotator and by $t\sim100$~kyr for the slow one.

The bottom panels show the corresponding contributions to $(3-n)/\tau_c$. The $1/\tau_B$ term is of the same order of magnitude in \texttt{D12}, \texttt{D14} and \CCSN: always negative, and dominant beyond a few tens of kyr, driving $n>3$. The internal contribution $1/\tau_{\alpha_i}$ vanishes identically, since there is no dipolar-axis drift. The externally driven term dominates at young ages -- overwhelmingly so in \texttt{D14}, where for the fast rotator it exceeds $1/\tau_B$ by two orders of magnitude -- and also drives $n>3$; it then dies away as the star approaches alignment, on a timescale set by $B_d$, $P_0$ and $\alpha_{e,0}$. No contribution driving $n<3$ appears at any time.

\citet{pons12} showed that Hall-driven oscillations of $B_d$ can arise at $t\gtrsim10^5$~yr, during the Hall-attractor phase typical of these configurations, and produce oscillations in the braking index. We do not find them here, probably because of the strong dominance of the poloidal dipolar component in these particular models, and because the analytical temperature prescription partly suppresses the Hall dynamics with respect to a fully coupled magneto-thermal calculation, which would give colder old stars. A toroidal-quadrupole-dominated configuration could produce oscillations with $|n|\sim10$, but should lead to the same conclusions regarding the obliquity, since that component is tightly Hall-coupled to the poloidal dipole. Given the dedicated studies already available for such configurations, we do not pursue this further.

\section{Magnetic field and rotational evolution}
\label{app: magrot evolution}

\subsection{Initial magnetic field and its evolution}
\label{app: magnetic evolution}

The initial field is constructed as in \citet{dehman23matins}. It is decomposed into poloidal and toroidal parts,
\begin{equation}
\mathbf{B} = - \boldsymbol{r} \times \Bnabla \Psi - \boldsymbol{r}\, \Delta \Phi + \Bnabla \frac{\partial}{\partial r} \left(r \Phi \right),
\label{eq: field pol tor}
\end{equation}
where the scalar functions $\Phi(t,\rr)$ and $\Psi(t,\rr)$ uniquely define the poloidal and toroidal components, respectively. They are expanded in spherical harmonics \citep{KR80}:
\begin{eqnarray}
    \Phi(t,r,\theta,\phi) &=& \frac{1}{r}\sum_{\ell m} \Phi_{\ell m}(r,t)\, Y_{\ell m}(\theta,\phi), \nonumber\\
    \Psi(t,r,\theta,\phi) &=& \frac{1}{r}\sum_{\ell m} \Psi_{\ell m}(r,t)\, Y_{\ell m}(\theta,\phi),
       \label{eq: phi and Psi scalar functions}
\end{eqnarray}
where $\ell = 1, 2, \ldots$ is the multipole degree and $m = -\ell, \ldots, \ell$ the azimuthal order.

At $t=0$, the radial dependence of the scalar functions is
\begin{eqnarray}
\Phi_{\ell m}(r) &=&
\begin{cases}
\Phi^0_{1m}\, k_r r \, \bigl[a(k_r r) + \tan(k_r R_b)\, b(k_r r)\bigr], & \ell = 1,\\[4pt]
-\,\Phi^0_{\ell m}\, \dfrac{(R_b-r)^2\,(r-R_c)^2}{(R_b-R_c)^4}, & \ell \geq 2,
\end{cases} \nonumber\\[4pt]
\Psi_{\ell m}(r) &=& -\,\Psi^0_{\ell m}\, \frac{(R_b-r)^2\,(r-R_c)^2}{(R_b-R_c)^4}, \qquad \ell \geq 1.
\label{eq: funa function}
\end{eqnarray}
Here $R_c$ is the radius of the crust--core interface and $R_b$ the numerical surface defined in Sect.~\ref{sec: numerical setup}. The dipolar poloidal profile follows \citet{aguilera2008}, with $a$ and $b$ the spherical Bessel functions of the first and second kind of order~1, and the radial wavenumber $k_r$ fixed by requiring $\Phi_{1m}(R_c)=0$; it vanishes at the crust--core interface and matches an external dipole at the surface. All other poloidal and toroidal components are confined to the crust, vanishing together with their radial derivative at $R_c$ and $R_b$. Such radial profiles are arbitrary, as in virtually all studies in the literature.

The coefficients $\Phi^0_{\ell m}$ and $\Psi^0_{\ell m}$ set the amplitude and orientation of each $(\ell,m)$ mode, and are chosen for each configuration as described in Sect.~\ref{sec: Initial cond}. The magnetic field is then computed from Eq.~(\ref{eq: field pol tor}) using the finite-volume curl operators of \MATINS on the cubed-sphere grid \citep{dehman23matins}, which guarantees an initial field that is divergence-free to machine precision and free of coordinate singularities. Finally, the poloidal field is rescaled so that the maximum of the dipolar component at the surface equals $B_d$, and the toroidal field is rescaled independently to the prescribed volume-averaged strength.

\begin{figure*}
    \centering
    \includegraphics[width=0.32\linewidth]{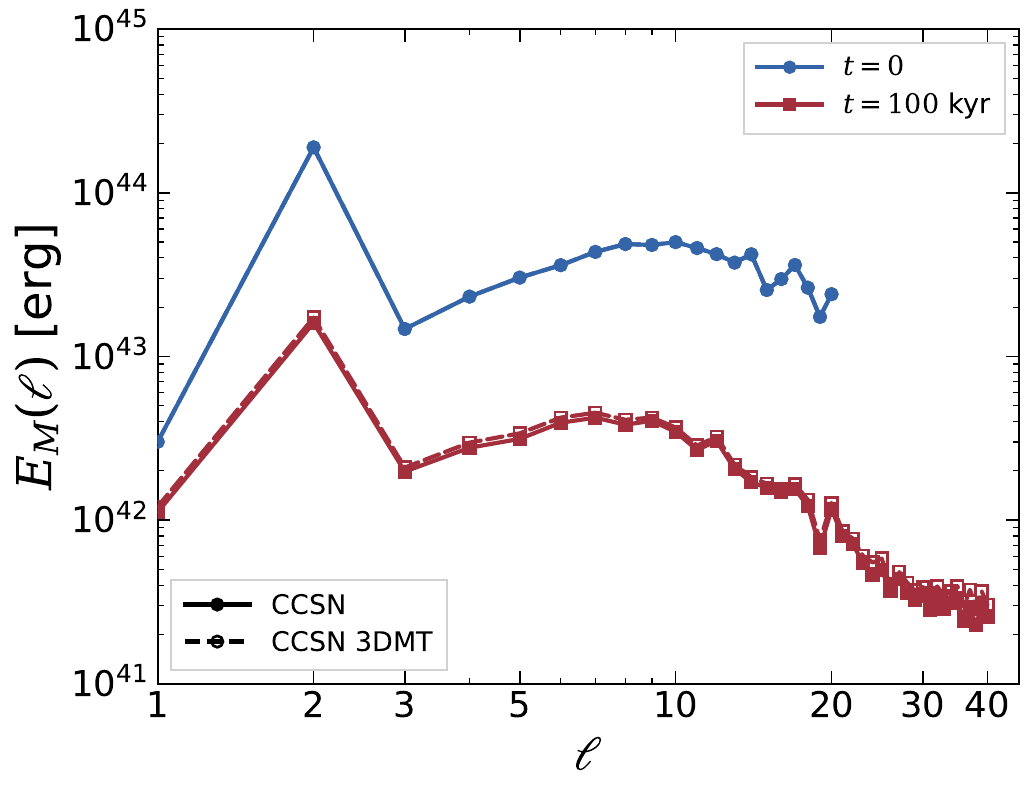}\hfill
    \includegraphics[width=0.32\linewidth]{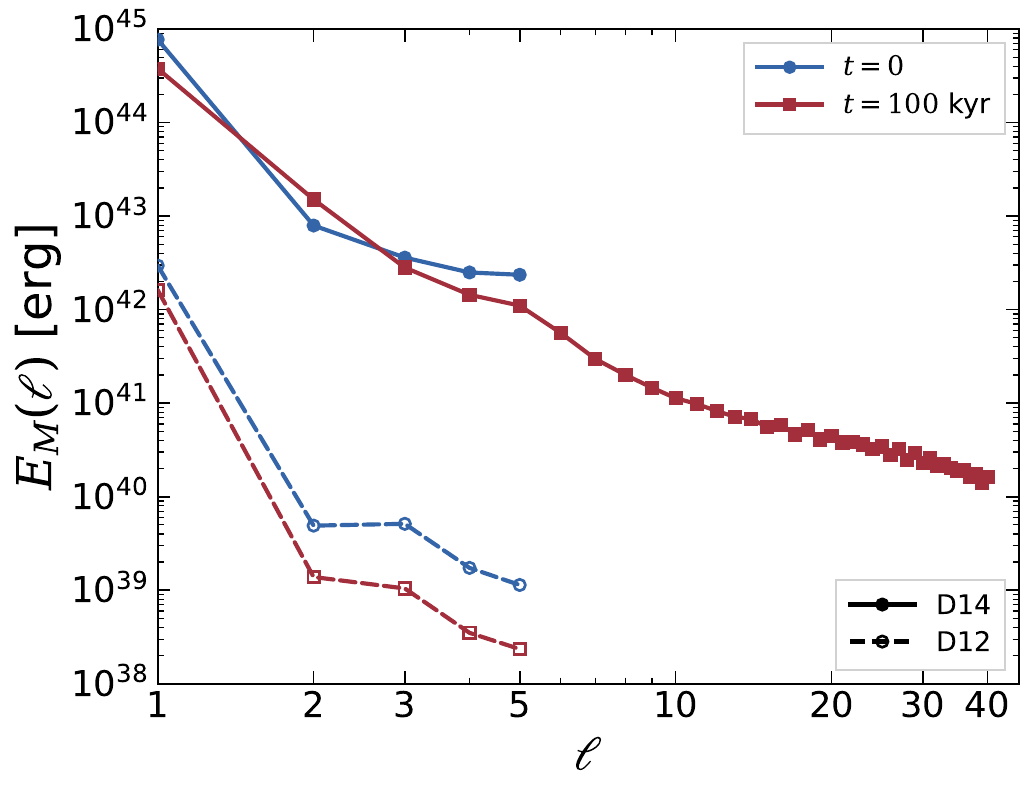}\hfill
    \includegraphics[width=0.32\linewidth]{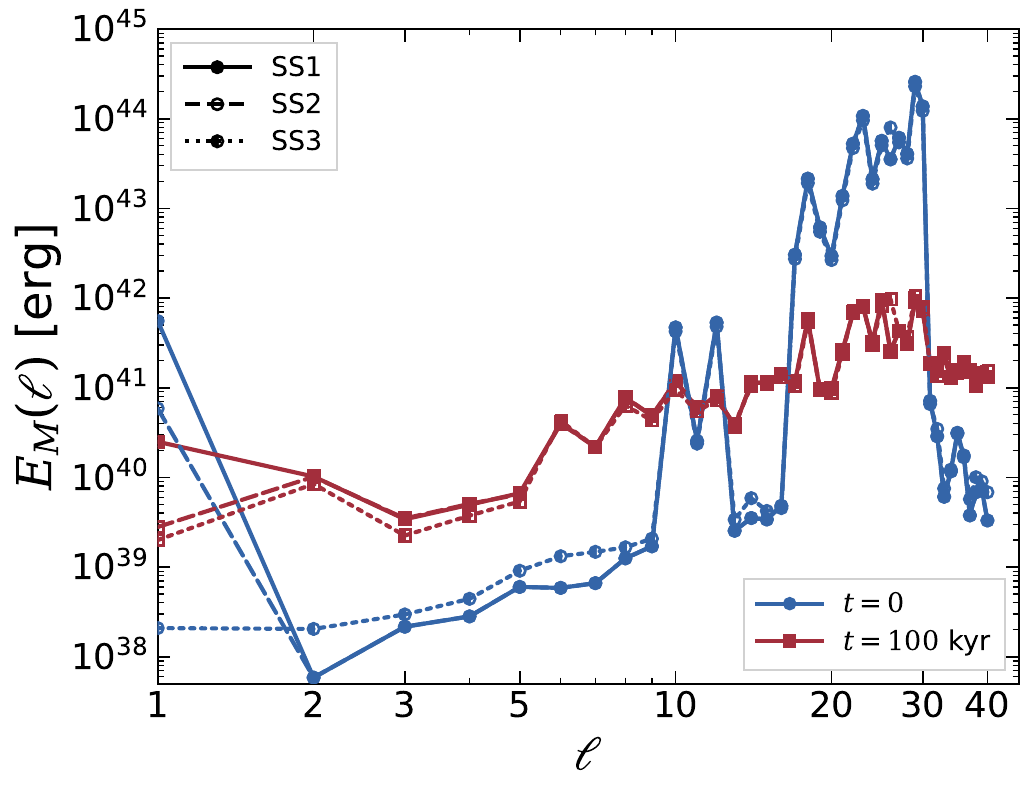}
\caption{Magnetic energy spectrum $E_M(\ell)$ at $t=0$ (blue circles) and $t=10^{5}$~yr (red squares) for different models. \emph{Left:} \CCSN (solid) and \texttt{CCSN-3DMT} (dashed). \emph{Centre:} \texttt{D14} (solid) and \texttt{D12} (dashed). \emph{Right:} \texttt{SS1} (solid), \texttt{SS2} (dashed) and \texttt{SS3} (dotted). Modes with $E_M<10^{38}$~erg are not shown. Note the different $y$-scales.}
\label{fig:energy_spectrum}
\end{figure*}

Figure~\ref{fig:energy_spectrum} shows the volume-integrated magnetic energy spectra of all the runs, at $t=0$ and after $100$~kyr.
The initial spectra differ markedly. The \CCSN one is dominated by the quadrupole, but already carries a broad, nearly flat plateau out to $\ell\simeq20$. The \texttt{D12} and \texttt{D14} models have a steeply decreasing spectrum, in which the dipole exceeds every other multipole by one to two orders of magnitude. The \texttt{SS1}--\texttt{SS3} configurations are the opposite limit: their energy is concentrated around $\ell\simeq18$--$30$, with essentially no power at $2\lesssim\ell\lesssim9$. The three differ only in the dipole, whose initial energy spans more than three orders of magnitude across the set; the right panel is therefore a controlled comparison at fixed small-scale content. By $100$~kyr the total energy has decreased in all runs, while the Hall cascade has flattened the spectra and redistributed the energy over all degrees permitted by the numerical setup, up to $\ell\simeq40$. The exception is \texttt{D12}, for which the Hall cascade is irrelevant and the spectrum remains purely large-scale.
 
The dipole, which controls the braking index, does not evolve in a single way. Where it is initially strong (\CCSN, \texttt{D12} and \texttt{D14}) it decays by only a factor of a few, appreciably more slowly than the energy stored at intermediate and high degrees, so that in relative terms it is more important at $10^{5}$~yr than at the beginning. In the small-scale runs the outcome depends on how much energy the dipole starts with: after $100$~kyr it lies more than an order of magnitude below its initial value in \texttt{SS1} and \texttt{SS2}, whereas in \texttt{SS3}, which starts essentially empty, it is generated during the evolution and grows by about an order of magnitude, showing that the cascade also proceeds towards the largest scales. The net effect is a convergence: the three initial dipole energies, spread over more than three orders of magnitude, differ by little more than one order of magnitude after $100$~kyr, as the dipole relaxes towards a level set by the small-scale reservoir that feeds it. Finally, \CCSN and \texttt{CCSN-3DMT} are nearly indistinguishable after $100$~kyr, indicating that the results are insensitive to the particular prescription adopted for the temperature evolution over this timespan.

We fix the initial spectrum $E_M^{\rm in}$ without attempting to enforce an MHD equilibrium, for two reasons that make such fine-tuning essentially irrelevant here, in contrast to the purely large-scale, static configurations often assumed. First, the Hall-MHD and MHD static equilibria differ and neither is a unique solution, so an initial transient from an MHD spectrum to a Hall cascade is expected in any case, with details depending on the initial configuration. Secondly, and possibly more importantly, the multiple scales spread across orders and degrees after the collapse imply that such static, purely large-scale solutions are unlikely to be reached. As in dynamo simulations reaching saturation and fully developed MHD turbulence, a steady state is probably attained, but it is not static: it is instead characterised by a continuous energy exchange between multipoles, preserving the same average spectrum up to minor fluctuations.

\subsection{Rotational evolution}
\label{app: rotational}

For each obliquity model ($\alpha_x,\alpha_y,\alpha_z,\alpha_e$), we integrate Eq.~(\ref{eq:ppdot}) neglecting the wobbling term (${\cal W}=1$), and obtain the long-term period evolution, identified below by the same subscripts:
\begin{equation}
    P(t)^2 = P_0^2 + \frac{2\pi^2 R^6}{Ic^3}\int_0^t f_\alpha(t') B_d(t')^2\,dt'.
    \label{eq: P evol}
\end{equation}
This equation is coupled to $\alpha_e(t)$ through Eq.~(\ref{eq: obliquity evol}), but not to the spontaneous wandering $\alpha_i(t)$, since the magnetosphere does not feed back on the internal dynamics.
 
As already mentioned, we consider two initial periods: a fast rotator, $P_0^f=10$~ms, and a slow one, $P_0^s=50$~ms. Since the spin-down power scales as $\Omega^4$, a sufficiently strong field combined with a sufficiently short initial period erases the memory of $P_0$ within a few kyr; neither condition is always met in our models, as shown below.

From each $P(t)$ we compute $\dot P$ from Eq.~(\ref{eq:ppdot}) and the characteristic age $\tau_c=P/(2\dot P)$, which deviates from the real age $t$ as
\begin{equation}
\frac{\tau_c}{t} = \frac{Ic^3}{2\pi^2 R^6} \frac{P_0^2}{f_\alpha(t)\,B_d(t)^2\,t} + \frac{\int_0^t f_\alpha(t') B_d(t')^2\,dt'}{f_\alpha\,B_d(t)^2\,t}~.
    \label{eq: tauc evol}
\end{equation}
The first term is always positive and carries the memory of $P_0$: it gives $\tau_c\gg t$ as long as $P$ remains comparable to $P_0$, which can persist for a long time when $P_0$ is large and/or $B_d$ is weak. The second term measures the accumulated torque relative to the instantaneous one; it equals unity if neither $B_d$ nor $\alpha$ evolve, and exceeds unity when the field decays \citep[typically at $t\gtrsim10^5$~yr; e.g.][]{vigano13,pons13}. We use $\tau_c$ to compare our models with the observed $n(\tau_c)$ trends.

\begin{figure*}
\centering
\includegraphics[width=0.45\linewidth]{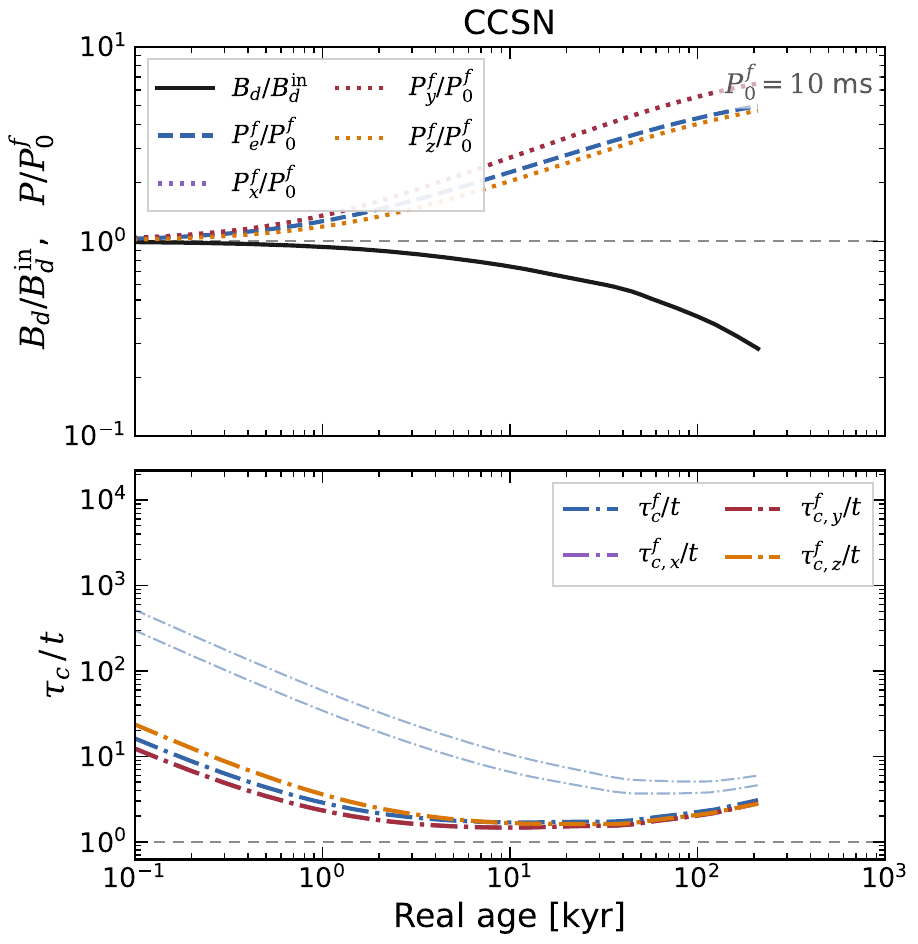} 
\includegraphics[width=0.45\linewidth]{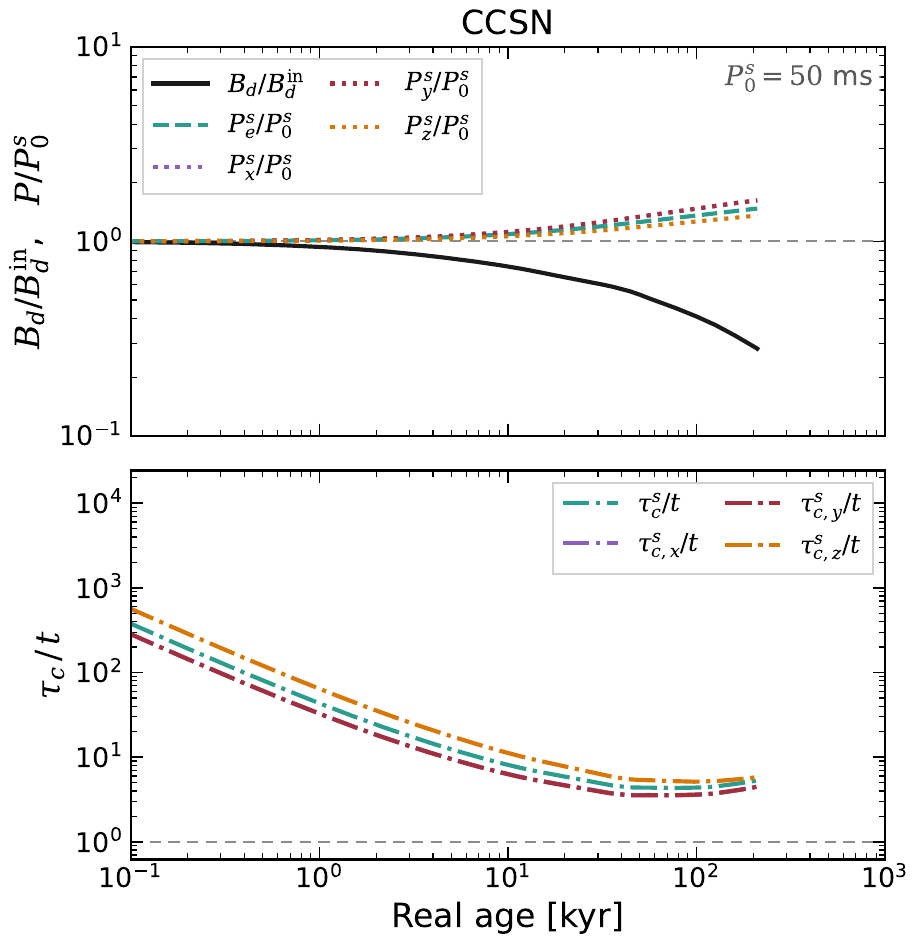}
\caption{Magneto-rotational evolution of the \CCSN model for an initial period $P_0^f=10$~ms (\emph{left}) and $P_0^s=50$~ms (\emph{right}). \emph{Top:} dipolar field normalised to its initial value, $B_d/B_d^{\rm in}$ (solid black), and spin period normalised to its initial value, $P/P_0$. \emph{Bottom:} characteristic-to-real age ratio $\tau_c/t$; the dashed grey line marks $\tau_c=t$. Colours identify the obliquity model of Fig.~\ref{fig: CC evol}: $\alpha_e$ in blue (left) or cyan (right), with thin lines for $\alpha_{e,0}=0.1\pi$ and $0.4\pi$, and $\alpha_i$ ($i=x,y,z$) in purple, red and orange. The $\alpha_x$ and $\alpha_y$ tracks, both initially orthogonal rotators, largely overlap.}
\label{fig: Bdip tau P CC}
\end{figure*}

Figure~\ref{fig: Bdip tau P CC} shows how the obliquity models propagate into the observable timing quantities for the \CCSN run. The dipolar field decays by a factor of a few over $\sim100$~kyr. As discussed in \citet{dehman23ccsn}, this reflects the decay of the total magnetic energy, combined with a transfer from small to large scales that is too weak to amplify the dipole. The \texttt{CCSN-3DMT} run (not shown) gives nearly identical results. The period evolution differs markedly between the two initial periods. For $P_0^s=50$~ms the period stays close to its initial value for several kyr and only reaches $P\simeq100$~ms by the end of the simulation, whereas for $P_0^f=10$~ms it grows from the very beginning and increases by nearly an order of magnitude. By $200$~kyr the two cases have converged to comparable periods ($P\simeq100$~ms), i.e.\ the memory of $P_0$ has been lost.

This difference is reflected in $\tau_c/t$. For the slow rotator the characteristic age always overestimates the real age, by at least a factor of a few: at early times because of the large $P_0^s$ (first term of Eq.~\ref{eq: tauc evol}), and later because the accumulated torque exceeds the instantaneous one as $B_d$ decays. For the fast rotator $\tau_c$ approaches $t$ between a few and a few tens of kyr -- after the memory of $P_0$ is lost and before field decay becomes important -- and rises again afterwards, for the same reason as in the slow case.

Compared with the choice of $P_0$, the obliquity evolution (different colours in each panel of Fig.~\ref{fig: Bdip tau P CC}) has a minor effect: different obliquities change the spin-down torque by at most a factor of two, through $f_\alpha=(1+\sin^2\alpha)$ (see Sect.~\ref{sec: braking index}).

\begin{figure*}
\centering
\includegraphics[width=0.45\linewidth]{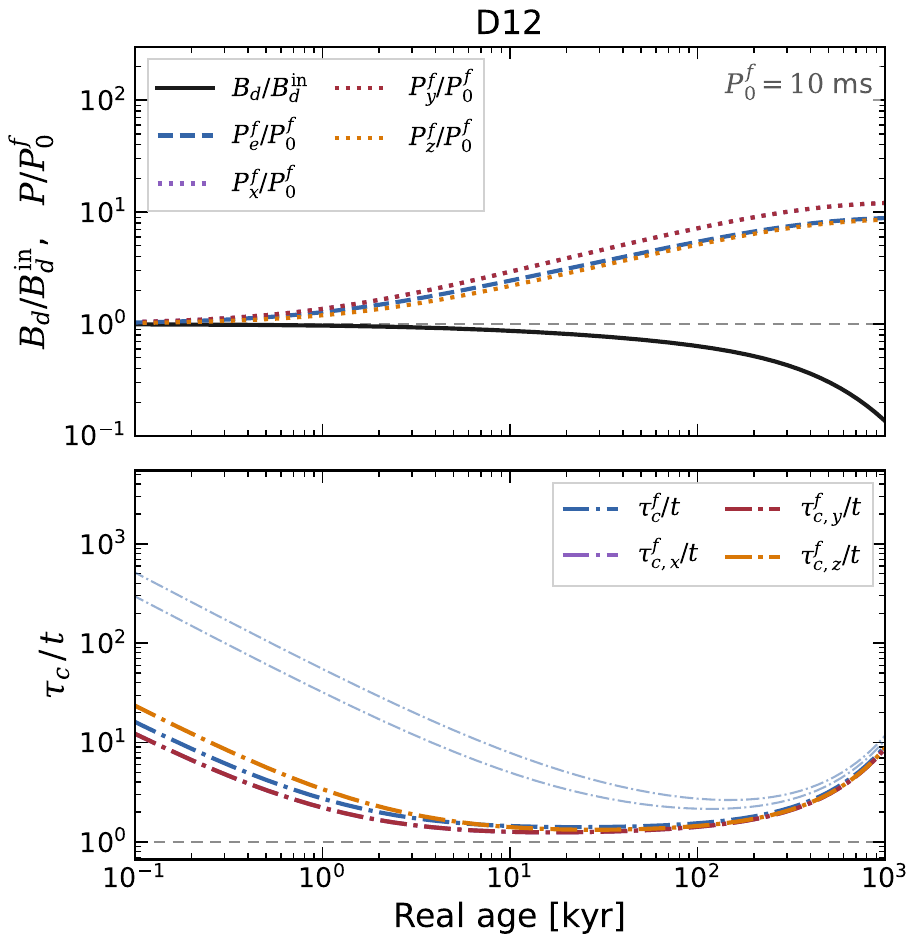}
\includegraphics[width=0.45\linewidth]{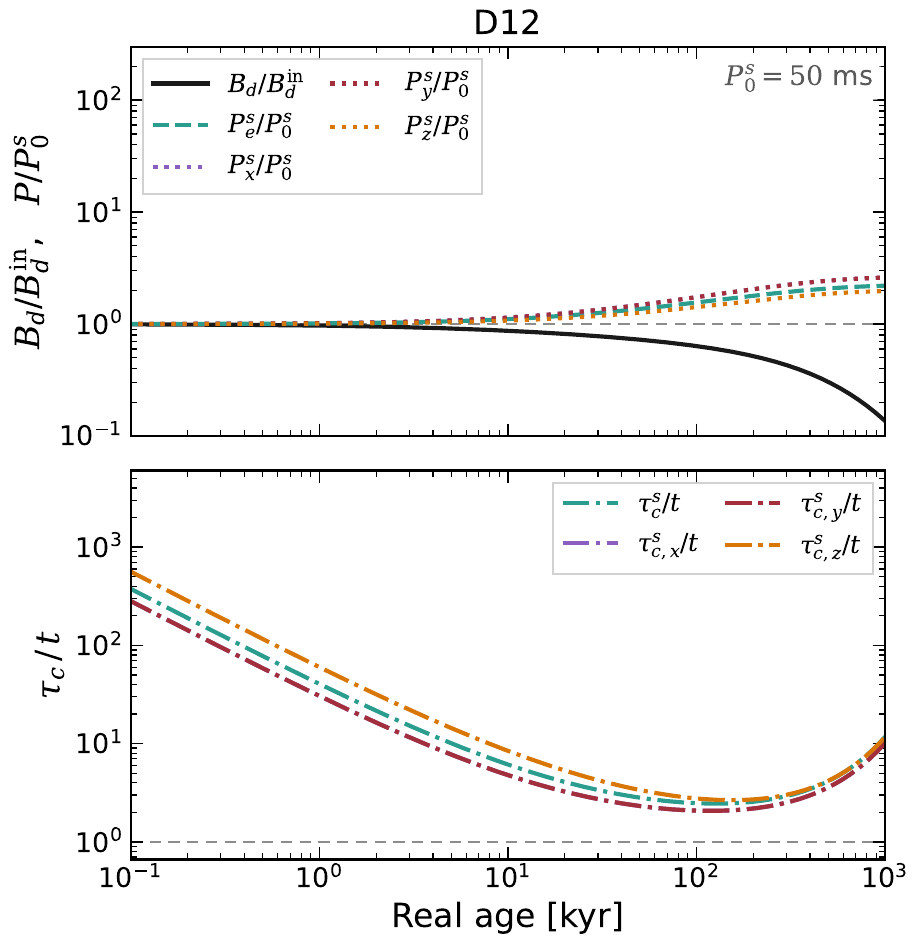}
\includegraphics[width=0.45\linewidth]{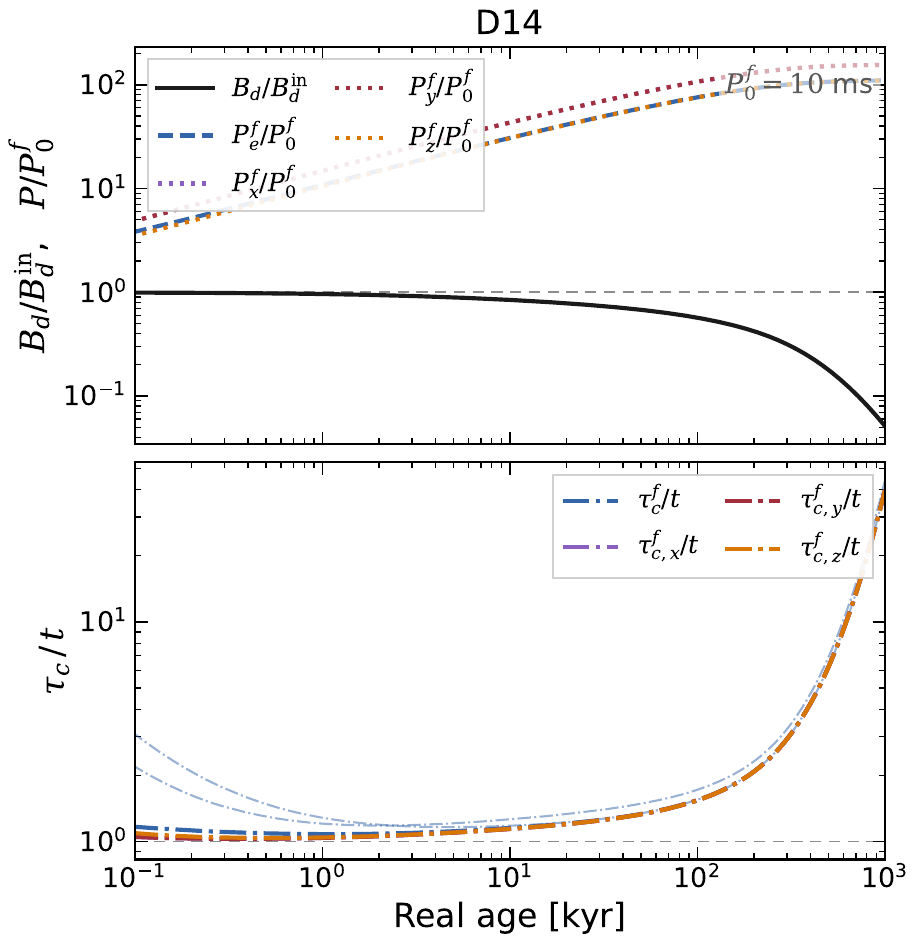}   \includegraphics[width=0.45\linewidth]{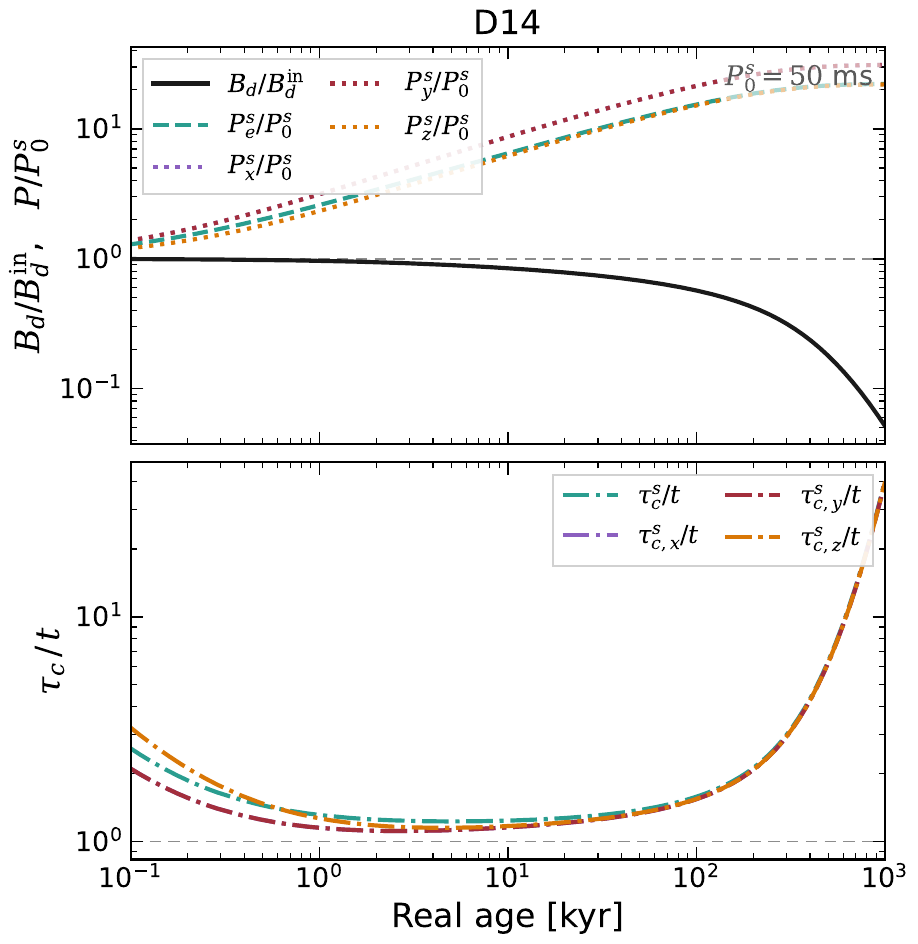}
\caption{Same as Fig.~\ref{fig: Bdip tau P CC}, for the \texttt{D12} (\emph{top}) and \texttt{D14} (\emph{bottom}) models, with $P_0^f=10$~ms (\emph{left}) and $P_0^s=50$~ms (\emph{right}).}
\label{fig: Bdip tau P D}
\end{figure*}

\begin{figure*}
\centering
\includegraphics[width=0.45\linewidth]{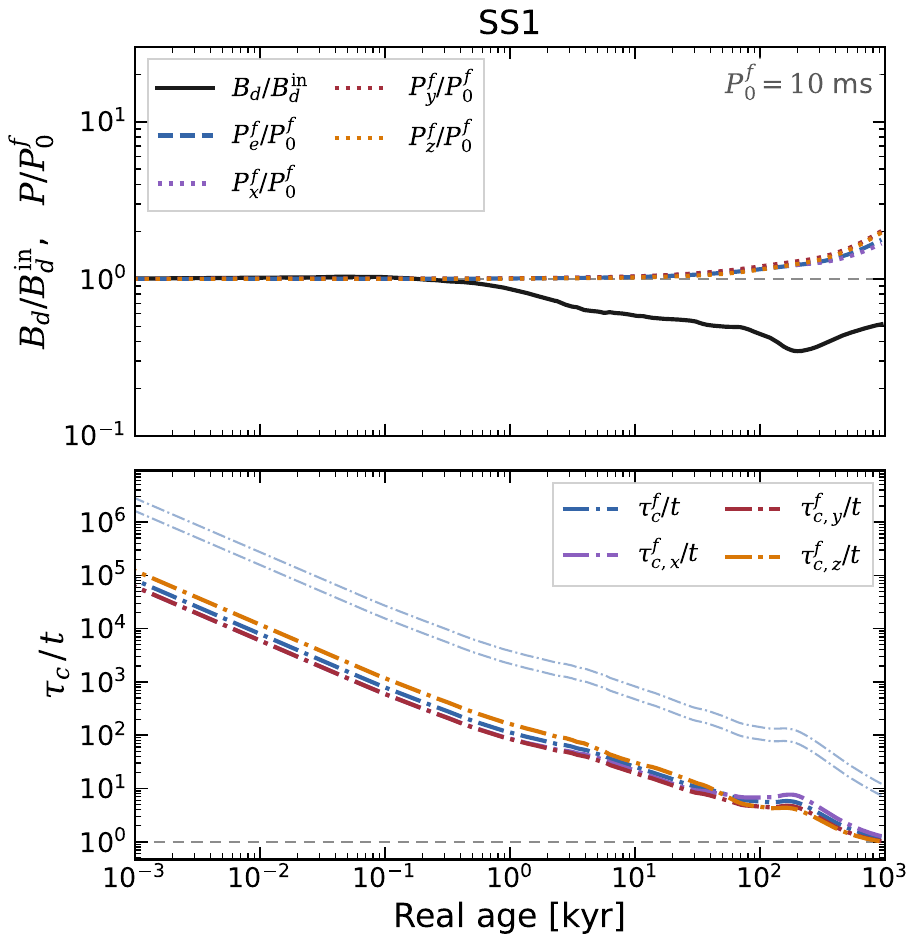}
\includegraphics[width=0.45\linewidth]{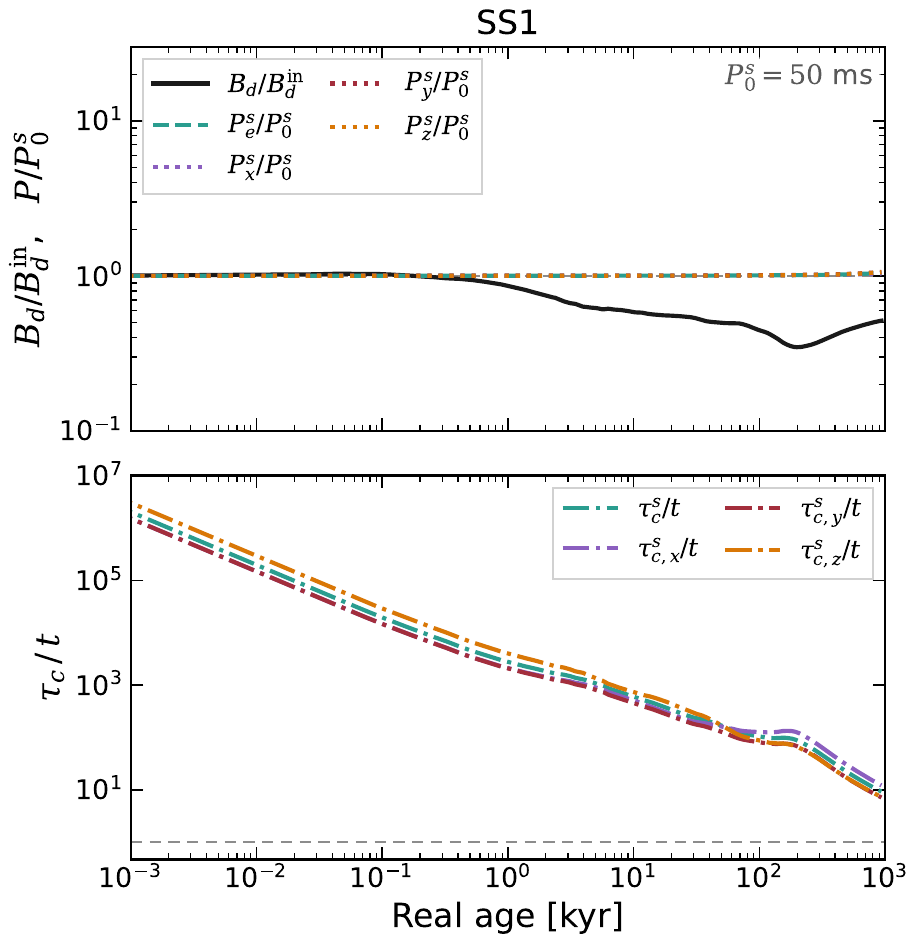}
\includegraphics[width=0.45\linewidth]{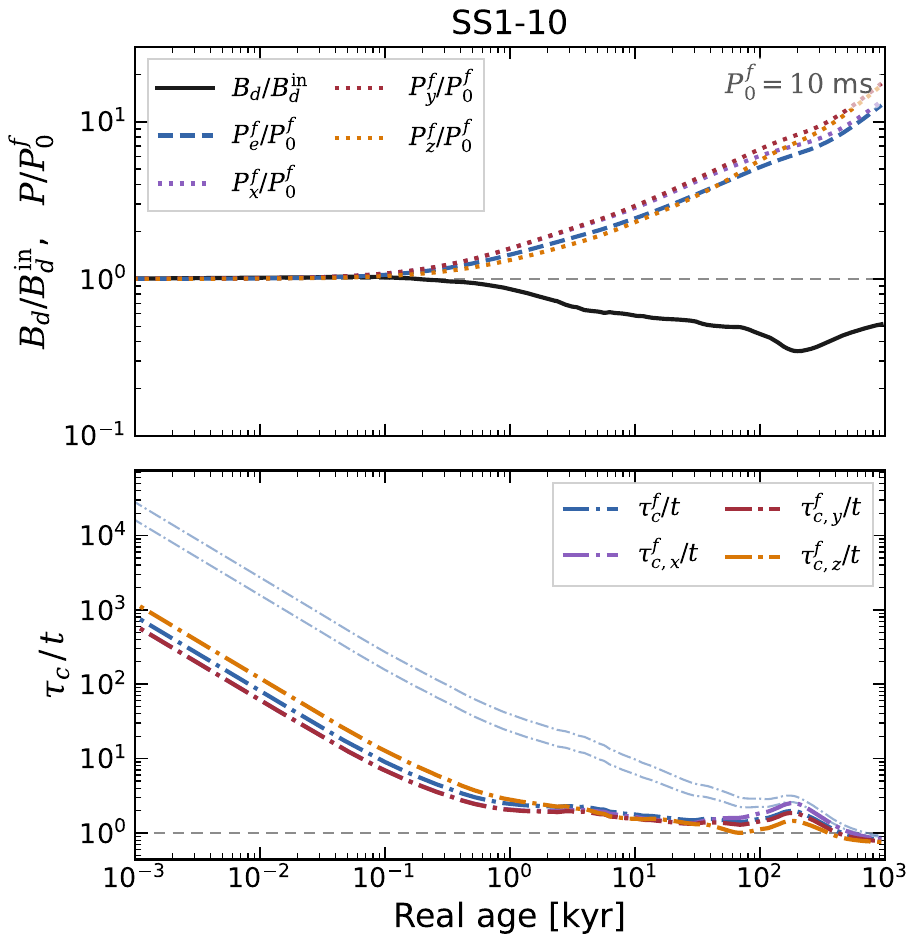}
\includegraphics[width=0.45\linewidth]{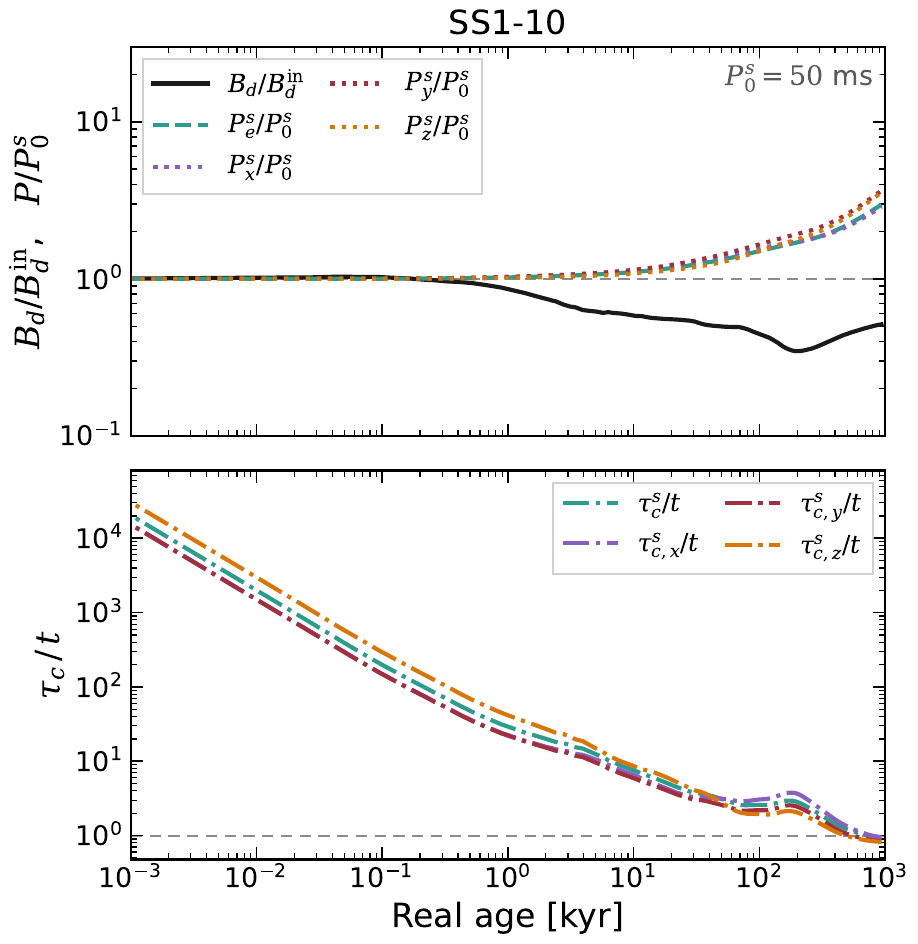}
\caption{Same as Fig.~\ref{fig: Bdip tau P CC}, for the \texttt{SS1} (\emph{top}) and \texttt{SS1-10} (\emph{bottom}) models, with $P_0^f=10$~ms (\emph{left}) and $P_0^s=50$~ms (\emph{right}).}
\label{fig: Bdip tau P SS1}
\end{figure*}

\begin{figure*}
\centering
\includegraphics[width=0.45\linewidth]{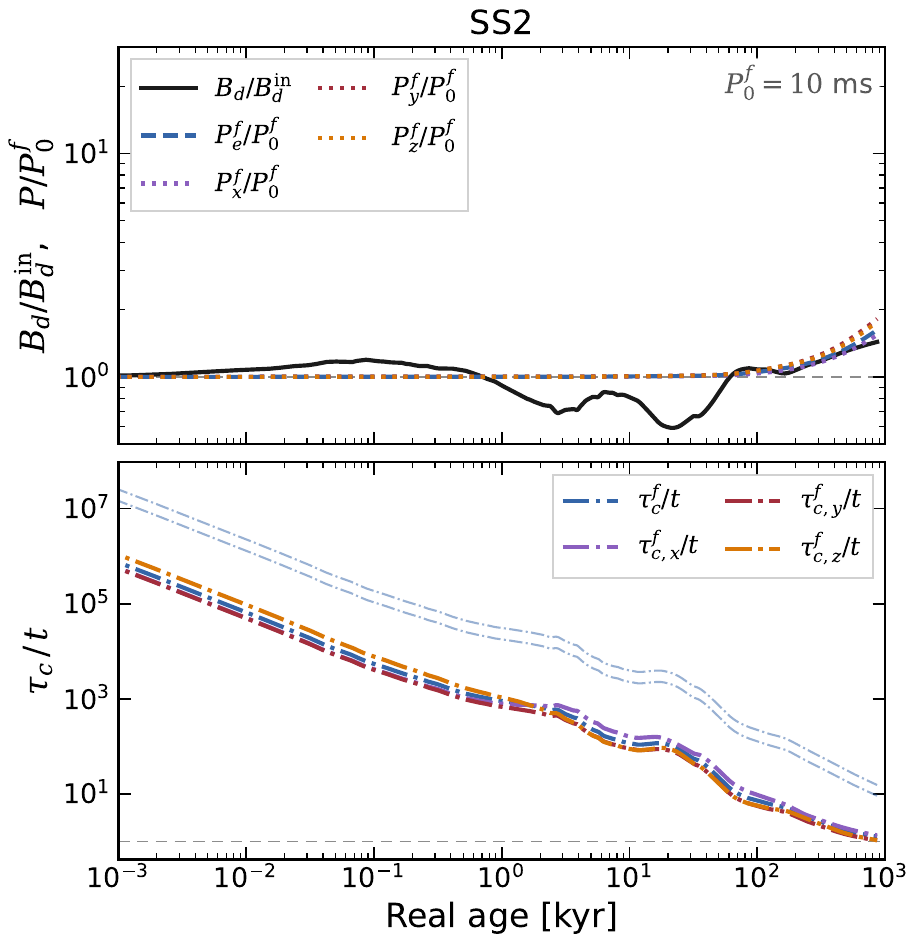}
\includegraphics[width=0.45\linewidth]{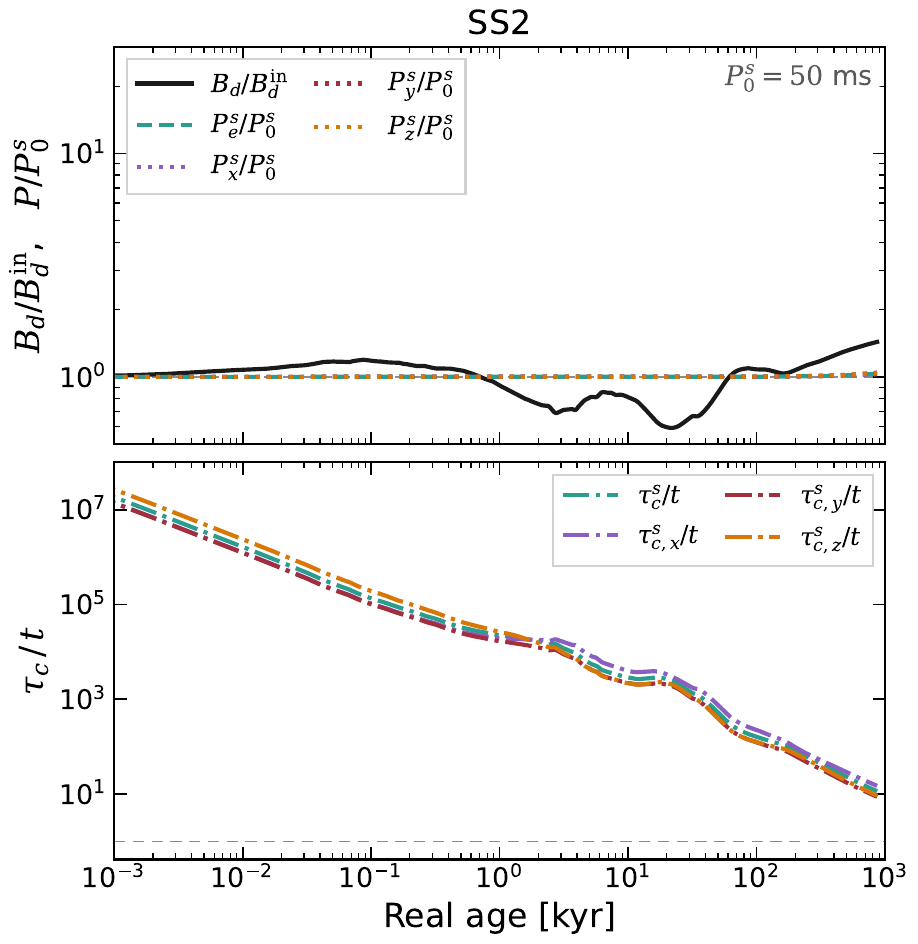} \\ 
\includegraphics[width=0.45\linewidth]{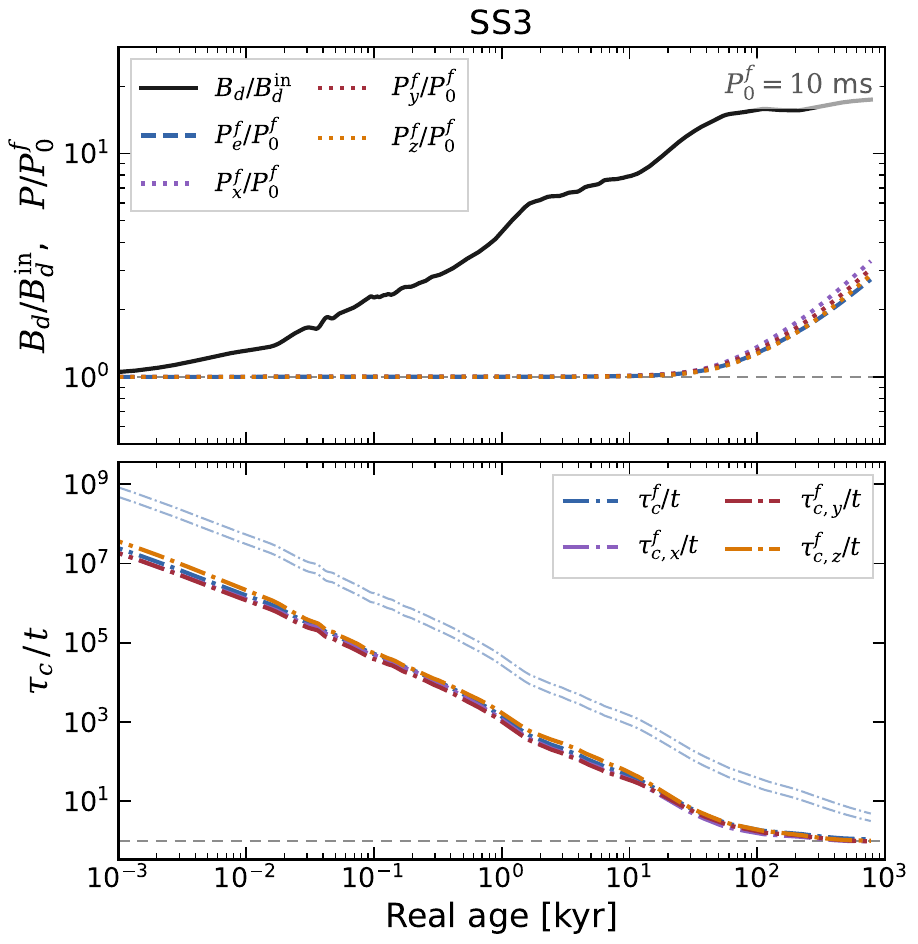}
\includegraphics[width=0.45\linewidth]{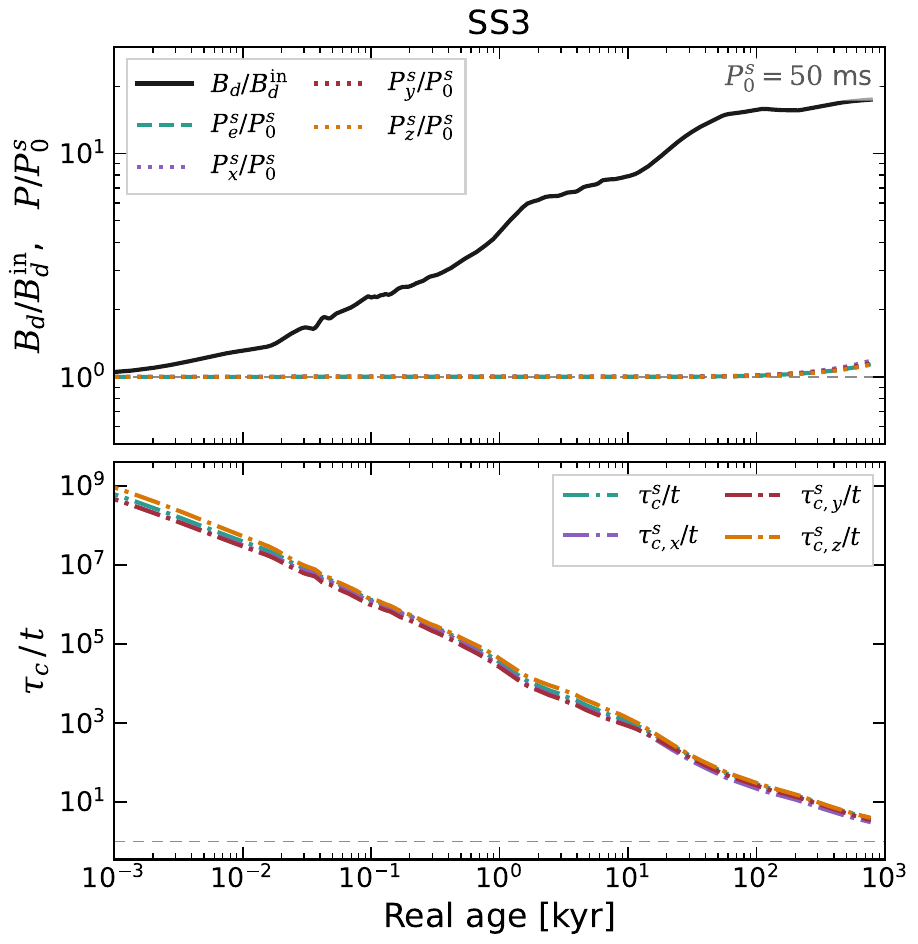}
\caption{Same as Fig.~\ref{fig: Bdip tau P CC}, for the \texttt{SS2} (\emph{top}) and \texttt{SS3} (\emph{bottom}) models, with $P_0^f=10$~ms (\emph{left}) and $P_0^s=50$~ms (\emph{right}).}
    \label{fig: Bdip tau P SS23}
\end{figure*}

Figures~\ref{fig: Bdip tau P D}--\ref{fig: Bdip tau P SS23} show the same quantities for the remaining models. The qualitative behaviour of each obliquity model is the same as described above for \CCSN, so we only highlight the differences, which are driven almost entirely by the strength of the initial dipolar component.

In \texttt{D14} the dipole is strong enough that the spin-down is efficient from the outset: the memory of $P_0$ is lost within $\lesssim0.1$~kyr, the two initial periods give indistinguishable tracks after that, and the star reaches $P\sim1$~s by $200$~kyr. Consequently $\tau_c\simeq t$ over almost the entire evolution, and departs from it only at $t\gtrsim100$~kyr, when the field decay
makes the second term of Eq.~(\ref{eq: tauc evol}) grow. This decay sets in earlier than in the other runs, owing to the faster Hall dynamics, which transfer energy efficiently towards the small scales where dissipation is most effective. This is the regime in which the characteristic age is a reliable age estimator. \texttt{D12} is intermediate: the period evolution resembles that of \CCSN, and $\tau_c/t$ approaches unity for the fast rotator over a broad interval, $t\sim3$--$100$~kyr, while for the slow rotator it never drops below $\simeq2$.

The small-scale models are the opposite limit, and the three differ according to how their dipole evolves. In \texttt{SS1} it decays by a factor of about three and grows back only at late times, whereas in \texttt{SS2} it grows and decays repeatedly throughout the evolution, ending slightly above its initial value; in \texttt{SS3}, which starts essentially empty, it is amplified by more than an order of magnitude, as energy is transferred from the small scales at which the field is initially concentrated. In all three it nevertheless remains too weak to brake the star appreciably: the period increases by at most a factor of a few for $P_0^f$, and is essentially unchanged for $P_0^s$. The first term of Eq.~(\ref{eq: tauc evol}) therefore dominates throughout, and the weaker the initial dipole, the more extreme the effect: at early times $\tau_c$ overestimates the real age by four to five orders of magnitude in \texttt{SS1} and by seven to nine in \texttt{SS3}, as expected from the $B_d^{-2}$ scaling. The discrepancy decreases only slowly, and except for the fast rotator at late times the characteristic age is a very poor estimator of the true age in these configurations.

\texttt{SS1-10} isolates the effect of the field strength alone. Its normalised dipolar evolution $B_d/B_d^{\rm in}$ is identical to that of \texttt{SS1} by construction, but the ten times stronger field brakes the star far more efficiently: the period grows by a factor of $\sim20$ instead of $\sim2$, and $\tau_c/t$ drops to unity within $\sim1$~kyr for the fast rotator and by $\sim100$~kyr for the slow one. It is the only small-scale case in which the characteristic age is a reliable age estimator over most of the evolution.

In all models the spread among the obliquity prescriptions is bounded by $f_\alpha=1+\sin^2\alpha$, which varies by at most a factor of two between an aligned and an orthogonal rotator, and hence by at most $\sqrt{2}$ in $P$. The bound is nearly saturated in the \texttt{D12}, \texttt{D14} and \CCSN models, and far from it in the \texttt{SS} runs, where the period barely evolves. In every case it is negligible compared with the differences produced by the evolution of the dipolar field, which is tied to the initial magnetic field configuration.

\subsection{$P$--$\dot P$ tracks}
\label{sec: PPdot}

\begin{figure*}
\centering
\includegraphics[width=0.37\linewidth]{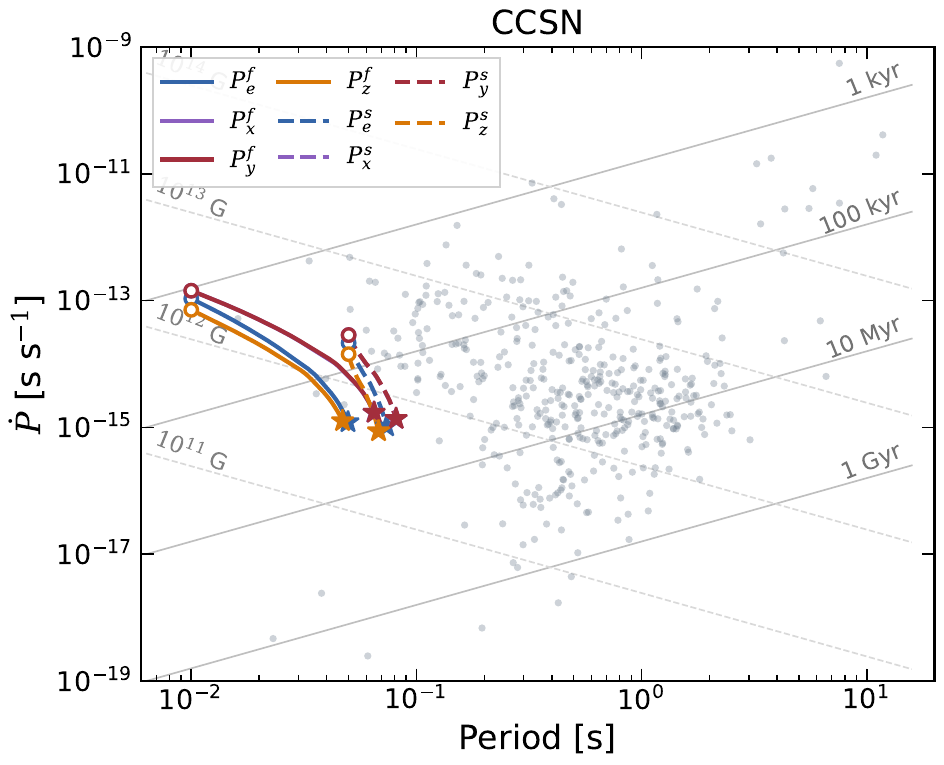}
\includegraphics[width=0.37\linewidth]{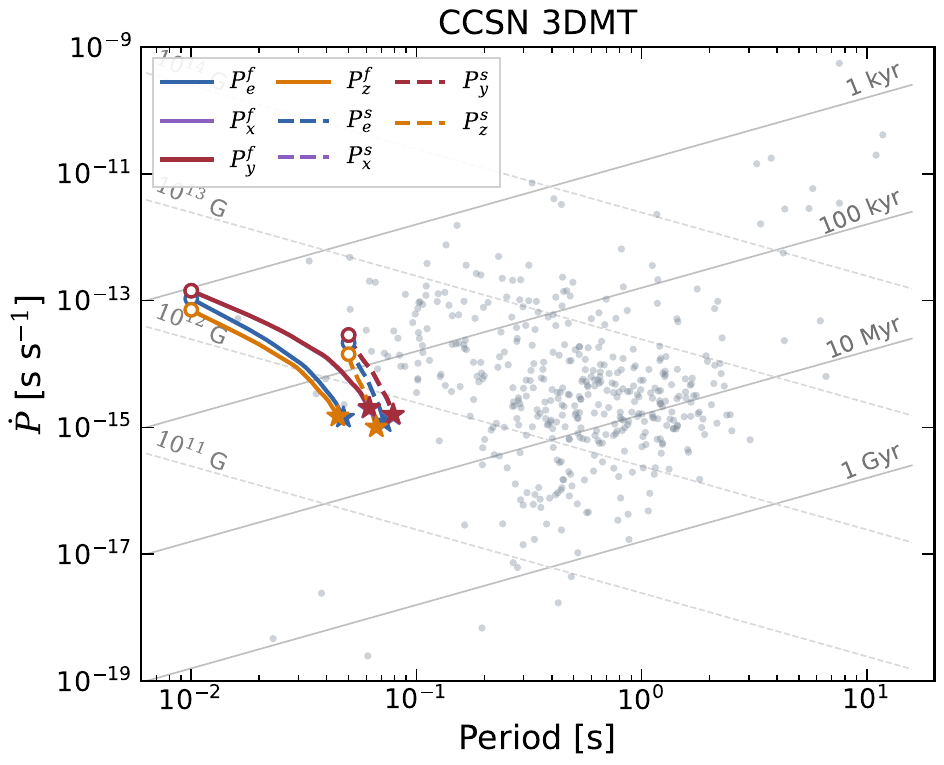}
\includegraphics[width=0.37\linewidth]{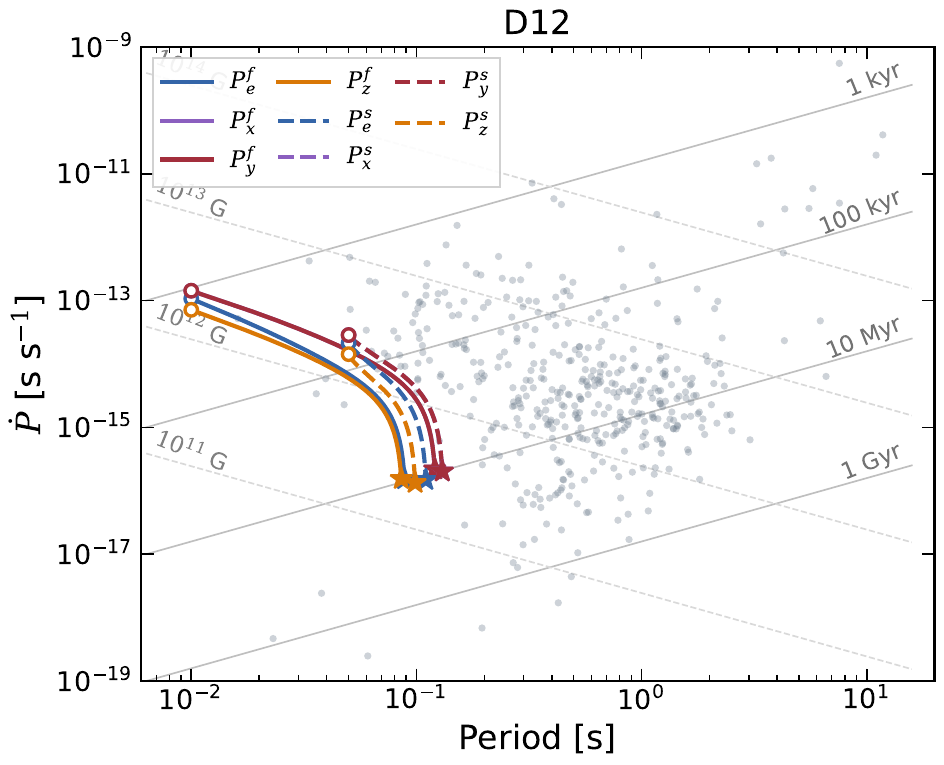} 
\includegraphics[width=0.37\linewidth]{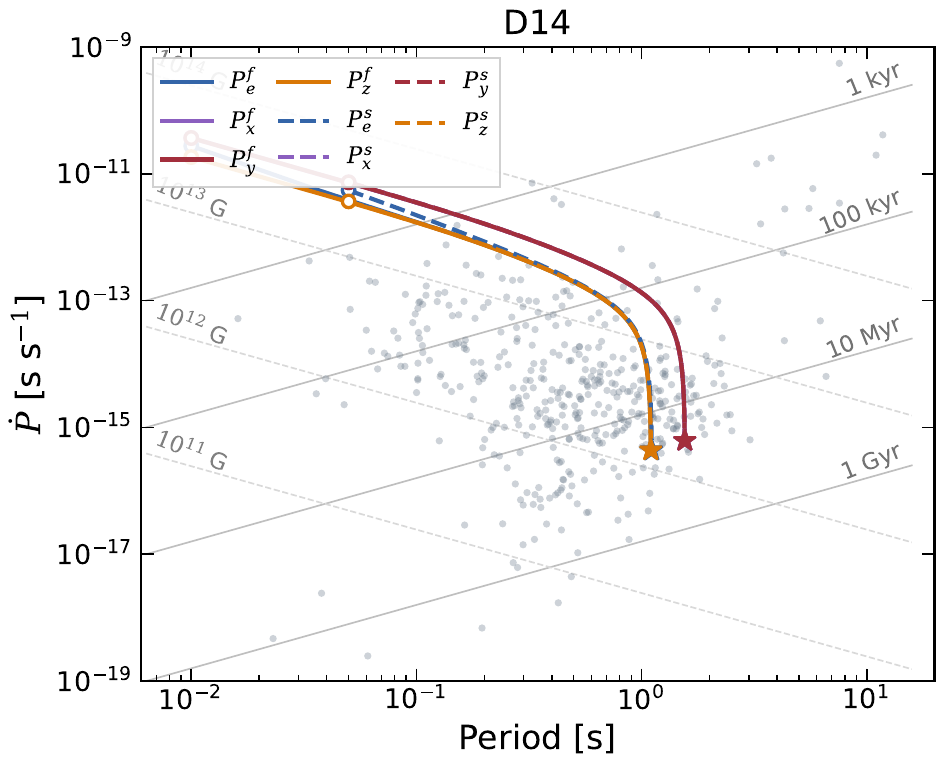}
\includegraphics[width=0.37\linewidth]{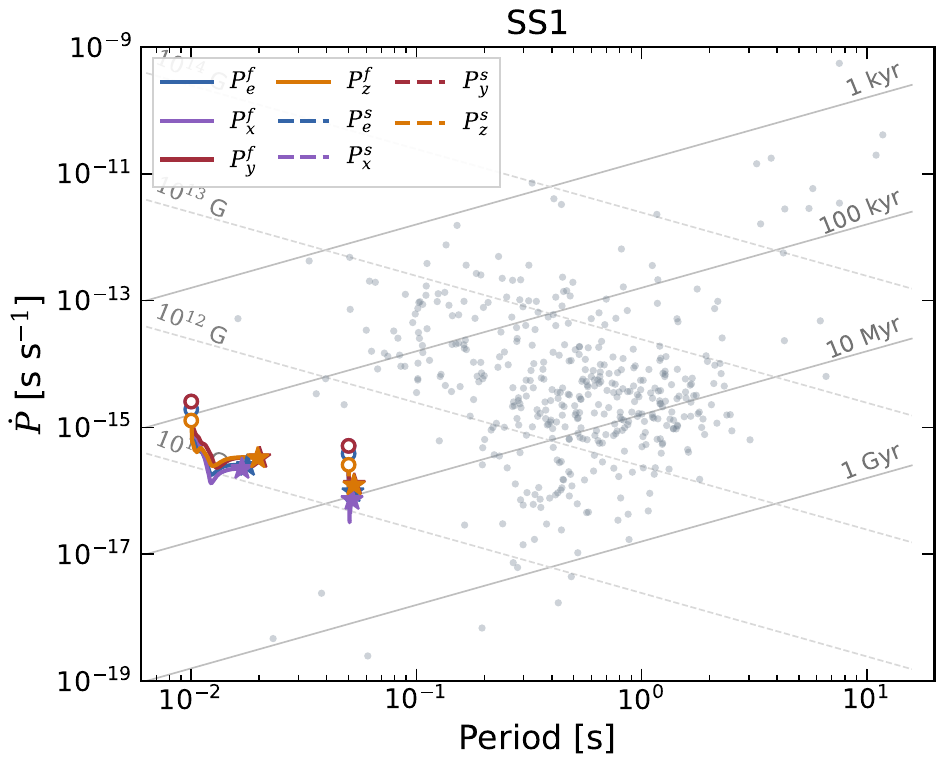} 
\includegraphics[width=0.37\linewidth]{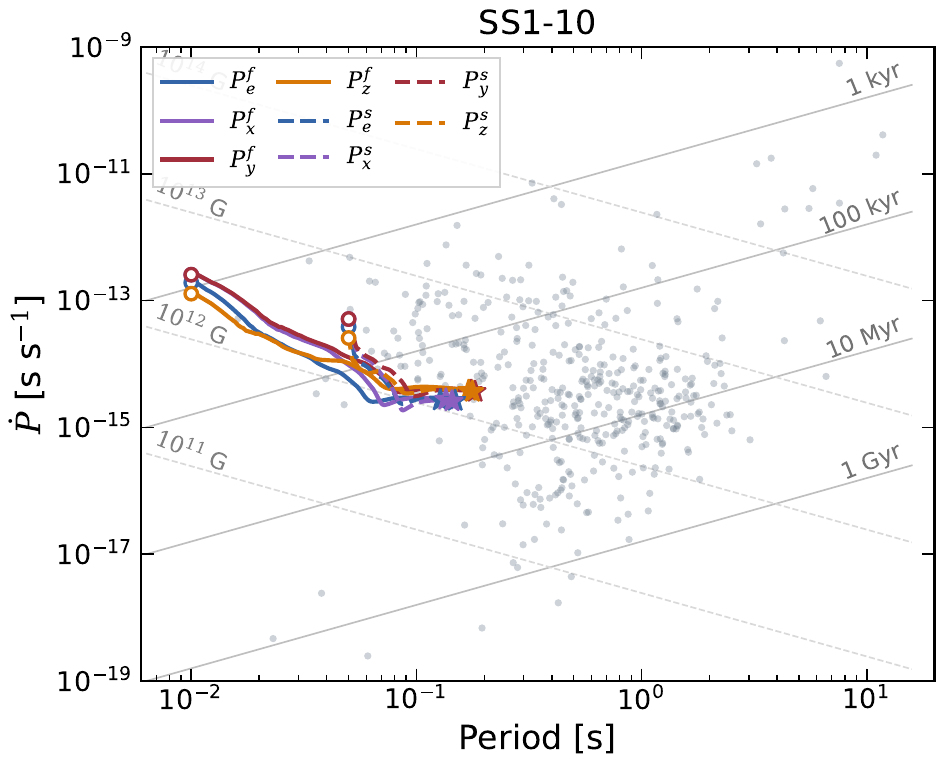} 
\includegraphics[width=0.37\linewidth]{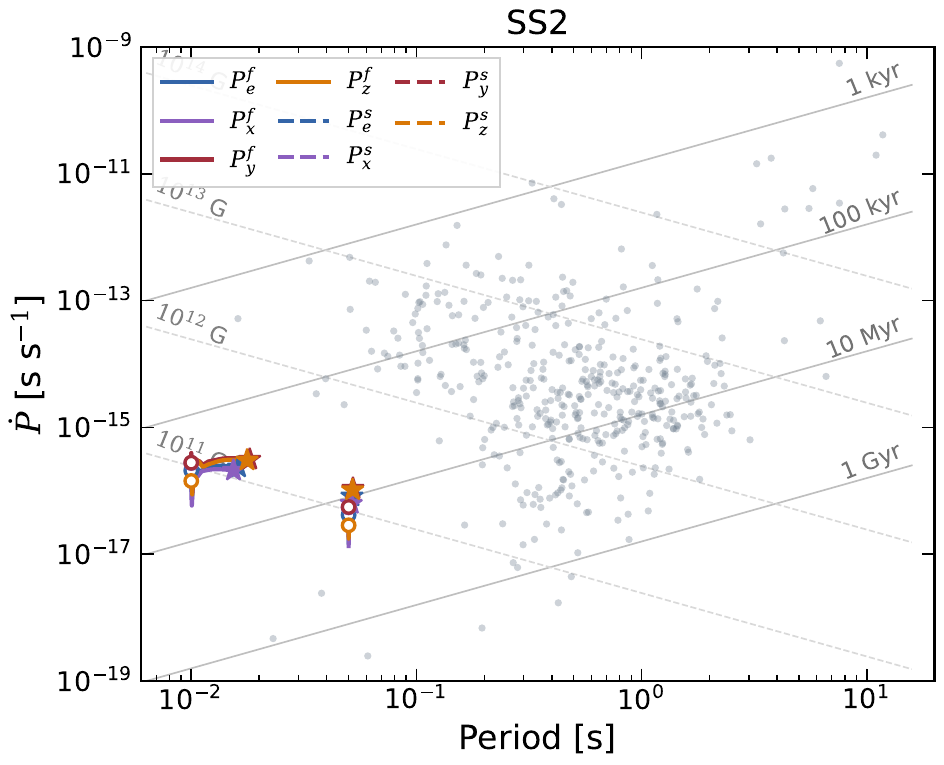} 
\includegraphics[width=0.37\linewidth]{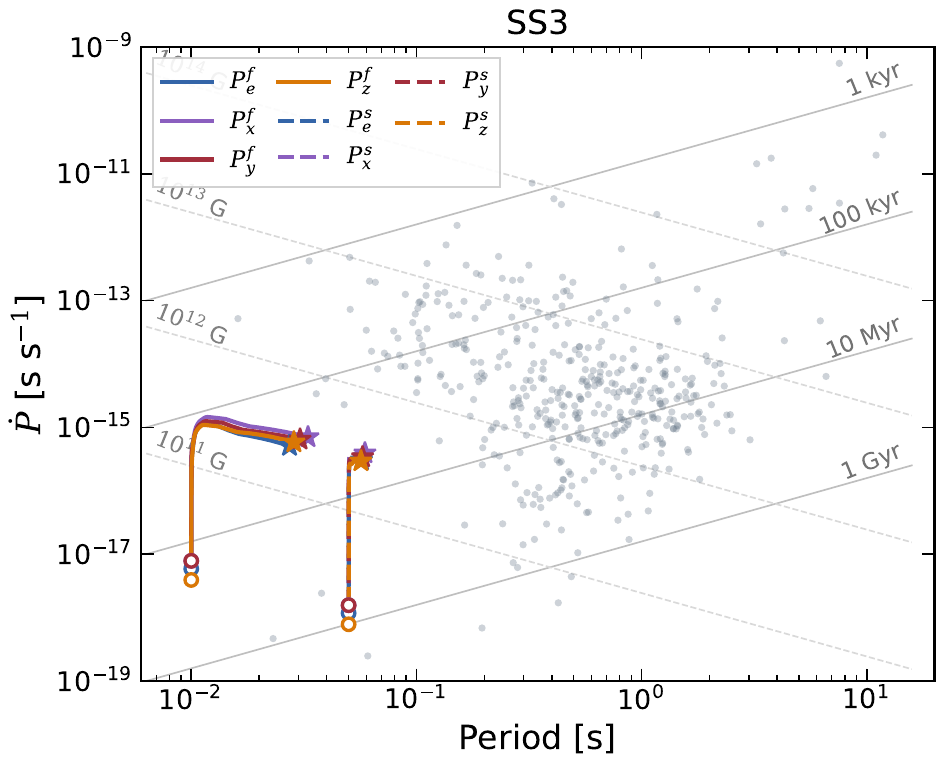} 
\caption{$P$--$\dot P$ tracks for the different models, indicated in each panel title, for each obliquity model ($e,x,y,z$; colours as in the previous figures) and for both initial periods (solid: $P_0^f=10$~ms; dashed: $P_0^s=50$~ms). Open circles mark $t=0$ and stars the end of each run. Grey solid lines mark constant characteristic age, and grey dashed lines constant inferred polar surface dipolar field, $B_d=6.4\times10^{19}\sqrt{P\dot P}$~G -- the expression commonly used in the literature, which corresponds to a vacuum orthogonal rotator with $R=10$~km and $I=10^{45}$~g~cm$^2$ and therefore does not coincide with the simulated $B_d$. Grey dots are the observed pulsars of the sample with $n$ measurements considered in Sect.~\ref{sec: Timing}.}
\label{fig: ppdot}
\end{figure*}

Figure~\ref{fig: ppdot} shows the corresponding tracks in the $P$--$\dot P$ diagram, superimposed on the observed values for our sample, which is representative of the general pulsar population at $\lesssim1$~Myr. Since $P\dot P\propto f_\alpha B_d^2$, a star whose field and obliquity are constant simply slides along a line of constant $B_d$, which in this plane has slope $-1$; any change in $f_\alpha B_d^2$ displaces it vertically from that line, downwards if the product decreases and upwards if it increases.

In \CCSN the star first moves to the right along a line of nearly constant $B_d$, starting just above the inferred $10^{12}$~G line, and turns downwards at $t\gtrsim100$~kyr, when the field decay makes $\dot P$ drop at almost constant $P$, approaching the $10^{11}$~G line. The two initial periods converge well before the end of the run, so the final position is set by the field rather than by $P_0$; the obliquity models are ordered as expected from $f_\alpha$, with the most orthogonal rotator ($\alpha_y$) always the rightmost track, but their spread is small.

\texttt{D12} is similar to \CCSN, ending only slightly further to the right. \texttt{D14} starts two orders of magnitude higher in $\dot P$, the two initial periods are already indistinguishable at the beginning of the plotted range, and the star crosses the observed population diagonally, ending among the majority of pulsars at $P$ of order seconds.

The small-scale models show a variety of behaviours. Their tracks are broadly horizontal, at least at late times, which indicates a mild increase of $B_d$ rather than a constant or decaying field. \texttt{SS1} and \texttt{SS2} barely move at all, remaining confined near their initial periods in the upper-left corner of the diagram, since their dipoles are never strong enough to brake the star appreciably: within that small excursion the \texttt{SS1} track first descends, as the dipole decays, and then flattens, whereas \texttt{SS2} stays nearly horizontal throughout. \texttt{SS3} is the extreme case: starting from a very weak dipole, its tracks first rise almost vertically as the dipolar
component is amplified by the transfer of energy from the small scales, and only afterwards turn to the right. Since the dipolar field is low, for a large enough initial period ($P_0^s$, dashed lines) the star ends the simulation with a negligible change in $P$.

The models that produce the most interesting $n$ behaviour therefore do not pass through the bulk of the observed population. This reflects the fact that, for a tangled configuration in which the dipolar component is subdominant, models reaching the observed values of $P$ and $\dot P$ are numerically much more challenging and computationally costly. The model \texttt{SS1-10}, for which the \texttt{SS1} $B_d(t)$ curve has been manually amplified by a factor of ten, serves as a first assessment of a more magnetised, tangled model: it moves the synthetic $P$--$\dot P$ track closer to the observed population while preserving the relevant features of $n(\tau_c)$. Ideally, one would fine-tune the initial configuration of more magnetised models so as to satisfy the $P$--$\dot P$ constraints while retaining the large spontaneous dipolar drift.

\end{appendix}

\end{document}